\documentclass[12pt]{article}
\usepackage{geometry}
\usepackage[onehalfspacing]{setspace}
\usepackage{amssymb}
\usepackage{amsfonts}
\usepackage{graphicx}
\usepackage{amsmath}
\usepackage{epstopdf}
\usepackage{natbib}
\usepackage{bibunits}
\usepackage{morefloats}
\usepackage{float}
\usepackage{comment}
\usepackage{pdflscape}
\usepackage{adjustbox}
\usepackage{subcaption}
\usepackage{longtable}
\usepackage[dvipsnames]{xcolor}

\usepackage[final]{pdfpages}

\usepackage{mathpazo} 
\usepackage[scaled]{helvet} 
\usepackage{courier} 
\normalfont
\usepackage[T1]{fontenc}

\usepackage[pdftex,bookmarks,colorlinks]{hyperref}
\hypersetup{
  colorlinks,
  citecolor=blue,
  linkcolor=red,
  urlcolor=Green}

\usepackage[compact]{titlesec}
\titlespacing{\section}{0pt}{0pt}{8pt}
\titlespacing{\subsection}{0pt}{2pt}{5pt}
\titlespacing{\subsubsection}{0pt}{3pt}{5pt}

\usepackage{enumitem}
\setlist[enumerate]{nosep,topsep={2pt},partopsep={2pt}}

\usepackage[bottom]{footmisc}

\begin{document}

\title{How is Machine Learning Useful for Macroeconomic Forecasting?\thanks{%
The third author acknowledges financial support from the Fonds de recherche sur
la soci\'{e}t\'{e} et la culture (Qu\'{e}bec) and the Social Sciences and
Humanities Research Council.}}
\author{Philippe Goulet Coulombe$^1$\thanks{%
Corresponding Author: \href{mailto:gouletc@sas.upenn.edu}{{gouletc@sas.upenn.edu}}. Department of Economics, UPenn.}  \and Maxime Leroux$^2$ \and Dalibor Stevanovic$^2$\thanks{%
Corresponding Author: \href{mailto:dstevanovic.econ@gmail.com}{{dstevanovic.econ@gmail.com}}. D\'{e}partement des sciences \'{e}conomiques, UQAM.} \and St\'{e}phane Surprenant$^2$ }
\date{%
    $^1$University of Pennsylvania\\%
    $^{2}$Universit\'{e} du Qu\'{e}bec \`{a} Montr\'{e}al\\[2ex]%
First version: October 2019 \\
This version: \today \\
\vspace{0.4cm}}
\maketitle

\vspace{-1.5cm}
\begin{abstract}

We move beyond \textit{Is Machine Learning Useful for Macroeconomic Forecasting?} by adding the \textit{how}.
The current forecasting literature has focused on matching specific variables and horizons with a particularly successful algorithm. To the contrary, we study the usefulness of the underlying features driving ML gains over standard macroeconometric methods. We distinguish four  so-called features  (nonlinearities, regularization, cross-validation and alternative loss function) and study their behavior in both the data-rich and data-poor environments. To do so, we  design experiments that allow  to identify the ``treatment'' effects of interest. 
We conclude that \textbf{(i)} nonlinearity is the true game changer for macroeconomic prediction, \textbf{(ii)} the standard factor model remains the best regularization, \textbf{(iii)} 
K-fold cross-validation is the best practice
and \textbf{(iv)} the  $L_2$ is preferred to the $\bar \epsilon$-insensitive in-sample loss. The forecasting gains of nonlinear techniques are associated with 
high macroeconomic uncertainty,  financial stress and housing bubble bursts. 
This suggests that Machine Learning is useful for macroeconomic forecasting by mostly capturing important nonlinearities that arise in the context of uncertainty and financial frictions.

\end{abstract}

\thispagestyle{empty}



\clearpage

\doublespace

\setlength{\abovedisplayskip}{2.2pt}
\setlength{\belowdisplayskip}{2.1pt}

\section{Introduction}\label{intro}

The intersection of Machine Learning (ML) with econometrics  has  become an important research
landscape in economics. ML has gained prominence due to the availability of large data sets, especially in microeconomic
applications \citep{Belloni2017, athey2019}.  
Despite the growing interest in ML, understanding the properties of ML  procedures when they are applied to predict  macroeconomic outcomes remains a difficult challenge.\footnote{The linear techniques have been extensively  
examined since \cite{Stock2002, Stock2002a}.
\cite{Kotchoni2019} compare  more than 30 forecasting models, including factor-augmented and regularized regressions. \cite{Giorgio2018} study the relevance of sparse modeling  in various economic prediction problems.} Nevertheless, that very understanding is an interesting econometric research endeavor \textit{per se}. It is more appealing to applied econometricians to upgrade a standard framework with a subset of specific insights rather than to drop everything altogether for an off-the-shelf ML model.

Despite appearances, ML has a  long history in macroeconometrics (see \cite{Lee1993,Kuan1994,Swanson1997,Stock1999,Trapletti2000,Medeiros2006}).  
However, only recently did the field of macroeconomic forecasting experience an overwhelming (and succesful) surge in the number of studies applying ML methods,\footnote{\cite{Moshiri2000, Nakamura2005, Marcellino2008} use neural networks to predict inflation and \cite{Cook2017} explore deep learning. \cite{Sermpinis2014} apply support vector regressions, while \cite{Diebold2019} propose a LASSO-based forecast combination technique. \cite{Ng2014}, \cite{Dopke2017} and \cite{Medeiros2019} improve forecast accuracy with random forests and boosting, while \cite{Yousuf2019} use boosting for high-dimensional predictive regressions with time varying parameters. Others compare machine learning methods in horse races \citep{Ahmed2010,Stock2012,Li2014,Kim2018,Smeekes2018,Chen2019,Milunovich2020}.} while works such as \cite{Joseph2019} and \cite{Zhao2019} contribute to their interpretability. However, the vast catalogue of tools, often evaluated with few models and forecasting targets, creates a large conceptual space, much of which remains to be explored. 
To map that large space without getting lost in it, we move beyond the coronation of a single winning model and its subsequent interpretation. Rather, we conduct a meta-analysis of many ML products by projecting them in their "characteristic" space. Then, we provide a direct assessment of which characteristics matter and which do not.



More precisely, we aim to answer the following question: 
What are the key features of ML modeling that improve the macroeconomic prediction? In particular, no clear attempt has been made at 
understanding why one algorithm might work while another does not. We address  this question by designing an \textit{experiment} to identify 
important characteristics of machine learning and big data techniques. The exercise consists of an extensive pseudo-out-of-sample forecasting 
horse race between many models that differ with respect to the four main features: nonlinearity, regularization, hyperparameter selection 
and loss function. To control for the big data aspect, we consider data-poor and data-rich models, and administer those \textit{patients} one particular ML \textit{treatment} or combinations of them. Monthly forecast errors are constructed for five important macroeconomic variables, five forecasting horizons and for almost 40 years.  Then, we provide a straightforward framework to identify which of them are
actual game changers for macroeconomic forecasting.

The main results can be summarized as follows. First, the ML nonparametric nonlinearities  constitute the most salient feature as they improve substantially the forecasting accuracy for all macroeconomic variables in our exercise, especially when predicting  at long horizons. Second, in the big data framework, alternative regularization methods (Lasso, Ridge, Elastic-net) do not improve over the factor model, suggesting that the factor representation of the macroeconomy is quite accurate as a means of dimensionality reduction.

Third, the hyperparameter selection by K-fold cross-validation (CV) and the standard BIC (when possible) do better on average than any other criterion. This suggests that ignoring information criteria when opting for more complicated ML models is not harmful. This is also quite convenient: K-fold is the built-in CV option in most standard ML packages. Fourth, replacing the standard in-sample quadratic loss function by the $\bar{\epsilon}$-insensitive loss function in Support Vector Regressions (SVR) is not useful, except in very rare cases. The latter finding is a direct by-product of our strategy to disentangle treatment effects. In accordance with other empirical results \citep{Sermpinis2014, Colombo2020}, in absolute terms, SVRs do perform well -- even if they use a loss at odds with the one used for evaluation. However, that performance is a mixture of the attributes of both nonlinearities (via the kernel trick) and an alternative loss function. Our results reveal that this change in the loss function has detrimental effects on performance in terms of both mean squared errors and absolute errors. Fifth, the marginal effect of big data is positive and significant, and improves as the forecast horizon grows. The robustness analysis shows that these results remain valid when: (i) the absolute loss is considered; (ii) quarterly targets are predicted; (iii) the exercise is re-conducted with a large Canadian data set. 

The evolution of economic uncertainty and financial conditions are important drivers of the NL treatment effect. ML nonlinearities are particularly useful: (i) when the level of macroeconomic uncertainty is high; (ii) when financial conditions are tight and (iii) during housing bubble bursts. The effects are bigger in the case of data-rich models, which suggests that combining nonlinearity with factors made of many predictors is an accurate way to capture complex macroeconomic relationships.  



These results give a clear recommendation for practitioners. For most cases, start by reducing the dimensionality 
with principal components and then augment the standard diffusion indices model by a ML nonlinear function approximator of your choice. 
That recommendation is conditional on being able to keep overfitting in check. To that end, if 
cross-validation must be applied to hyperparameter selection, the best practice is the standard K-fold. 

These novel empirical results also complement a growing theoretical literature on ML with dependent observations. As \cite{Alquier2013} points out, much of the work in statistical learning has focus on the cross-section setting where the assumption of independent draws is more plausible. Nevertheless, some theoretical guarantees exist in the time series context. \cite{Mohri2010} provide generalization bounds for Support Vector Machines and Regressions, and Kernel Ridge Regression under the assumption of a stationary joint distribution of predictors and target variable. \cite{Kuznetsov2015} generalize some of those results to non-stationary distributions and non-mixing processes. 
However, as the macroeconomic time series framework is characterized by short samples and structural instability, our exercise contributes to the general understanding of machine learning properties in the context of time series modeling and forecasting.

In the remainder of this paper, we first present the general prediction problem with machine learning and big data.  
Section \ref{features} describes the four important features of machine learning methods.  
Section \ref{empirics} presents the empirical setup, section \ref{results} discusses the main results, followed by section 
\ref{BlackBox} that aims to open the black box. Section \ref{conclusion} 
concludes. Appendices \ref{sec:rootMSE}, \ref{sec:robust}, \ref{sec:addgraphs} and \ref{sec:NNBT} contain respectively: tables with overall performance; robustness of 
treatment  analysis; additional results and robustness of nonlinearity analysis. The supplementary material contains the following appendices: results for absolute loss, results with quarterly 
US data, results with monthly Canadian data, description of CV techniques and technical details on  forecasting models.

\section{Making Predictions with Machine Learning and Big Data}\label{problem}

Machine learning methods  are meant to improve our predictive ability especially when the ``true'' model is unknown and complex. To illustrate this point, let $y_{t+h}$ be the variable to be predicted $h$ periods ahead (target) and $Z_t$ the $N_Z$-dimensional vector of predictors  made out of $H_t$, the set of all the inputs available at time $t$. 
Let $g^*(Z_t)$ be the true model and $g(Z_t)$ a functional (parametric or not) form selected by the practitioner. In addition, denote $\hat{g}(Z_t)$ and $\hat y_{t+h}$ the fitted model and its forecast. The forecast error can be decomposed as 
\begin{equation}\label{eq2}
		y_{t+h} - \hat y_{t+h} = \underbrace{ g^*(Z_t) - g(Z_t)}_{\text{approximation error}} + \underbrace{ g(Z_t) - \hat{g}(Z_t)}_{\text{estimation error}} + e_{t+h}.
\end{equation}
The intrinsic error $e_{t+h}$ is not shrinkable, while the estimation error can be reduced by adding more data. The approximation error is controlled by the functional estimator choice. While it can be potentially minimized by using flexible functions, it also rise the risk of overfitting and a judicious regularization is needed to control this risk. This problem can be embedded in the   
general prediction setup from \cite{Hastie2009}
\begin{equation}\label{eq1}
\min_{g \in \mathcal{G}} \{ \hat{L}(y_{t+h},g(Z_{t})) +pen(g;\tau) \}, \quad t=1,\ldots,T.
\end{equation}
This setup has four main features:
\begin{enumerate}
	\item $\mathcal{G}$ is the space of possible functions $g$ that combine the data to form the prediction. In particular,
the interest is how much nonlinearities can we allow for in order to reduce the approximation error in (\ref{eq2})? 
	\item $pen()$ is the regularization penalty limiting the flexibility of the function $g$ and hence controlling the overfitting risk. This is quite general and can accommodate Bridge-type penalties and dimension reduction techniques.  
	\item $\tau$ is the set of hyperparameters including those in the penalty and the approximator $g$. The usual problem is to choose the best data-driven method to optimize $\tau$. 
	\item $\hat{L}$ is the loss function that defines the optimal forecast. Some ML models  feature an in-sample loss function different from the standard $l_2$ norm.
\end{enumerate}
Most of (supervised) machine learning consists of a combination of those ingredients and popular methods like linear (penalized) regressions can be 
obtained as special cases of (\ref{eq1}).

\subsection{Predictive Modeling}

We consider the \emph{direct} predictive modeling
in which the target is projected on the information set, and the forecast
is made directly using the most recent observables. This is opposed to
\emph{iterative} approach where the model recursion is used to simulate the
future path of the variable.\footnote{\cite{Marcellino2006} conclude
that the direct approach provides slightly better
results but does not dominate uniformly across time and series. See \cite{Chevillon2007} for a  survey on multi-step forecasting.} Also, the direct approach 
is the standard practice for in ML applications.

We now define the forecast objective given the variable of interest $Y_t$.
If $Y_{t}$ is stationary, we forecast its level $h$ periods ahead: 
\begin{equation}
	y_{t+h}^{(h)}=y_{t+h},  \label{fcst0}
\end{equation}%
where $y_{t}\equiv \mathrm{ln}Y_{t}$ if $Y_{t}$ is strictly positive. If $Y_t$ is I(1), then we 
forecast the average  growth rate 
over the period $[t+1,t+h]$ \citep{Stock2002}. We shall therefore define $y_{t+h}^{(h)}$ as:
\begin{equation}
y_{t+h}^{(h)}=(1/h)\mathrm{ln}%
(Y_{t+h}/Y_{t}).  \label{fcst1}
\end{equation}%
In order to avoid a cumbersome notation, we use $y_{t+h}$ instead of $y_{t+h}^{(h)}$ in what follows. In addition, all the predictors in $Z_t$ are 
assumed to be covariance stationary.

\subsection{\emph{Data-Poor} versus \emph{Data-Rich} Environments}\label{patients}

Large time series panels are now widely constructed
and used for macroeconomic analysis. The most popular is FRED-MD monthly panel of US variables constructed by
\cite{McCracken2016}.\footnote{\cite{FLSS2020} have recently proposed similar data for Canada.} Unfortunately, the performance of standard econometric models tends to
deteriorate as the dimensionality of data increases. 
\cite{Stock2002}
first proposed to solve the problem by replacing the high-dimensional predictor set by common factors.\footnote{ Another way to approach the dimensionality problem is to use Bayesian methods. 
Indeed, some of our Ridge regressions will look  like a direct version of a Bayesian VAR with a \cite{Litterman1979} prior. \cite{Giannone2015} have shown that an hierarchical prior can lead the BVAR to perform as well as a factor model.}


On other hand, even though the machine
learning models do not require big data, they are useful to perform variable selection and digest 
large information sets to improve the prediction. Therefore, in addition to
treatment effects in terms of characteristics of forecasting models, we will also interact those with the width of the sample. 
The data-poor, defined as $H_t^{-}$, will only contain a finite number of lagged values of the target, while the data-rich
panel, defined as $H_t^{+}$ will also include a large number of exogenous predictors. Formally, 
\begin{align}
H_t^{-} \equiv {\lbrace y_{t-j} \rbrace}_{j=0}^{p_y} \quad \text{and} \quad
H_t^{+}  \equiv  \left[{\lbrace y_{t-j} \rbrace}_{j=0}^{p_y}, {\lbrace X_{t-j} \rbrace}_{j=0}^{p_f}  \right]. \label{Hs}
\end{align}

The analysis we propose can thus be summarized in the following way. We will consider two standard models for forecasting.
\begin{enumerate}
\item The $H_t^{-}$ model is the \textit{autoregressive direct} (AR)\ model, which is specified as:%
\begin{equation}
y_{t+h} = c + \rho(L)y_{t}+e_{t+h},\quad t=1,\ldots ,T,  \label{ard}
\end{equation}%
where $h\geq 1$ is the forecasting horizon. The only hyperparameter in
this model is $p_y$, the order of the lag polynomial  $\rho(L)$.

\item The $H_t^{+}$ workhorse model is the autoregression augmented with diffusion
indices (ARDI) from \cite{Stock2012}:
\begin{eqnarray}
y_{t+h} &=& c+ \rho(L)y_{t} + \beta(L)F_{t} +e_{t+h},\quad t=1,\ldots ,T  \label{swardi1} \\
X_{t} &=&\Lambda F_{t}+u_{t} \label{swardi2}
\end{eqnarray}%
where $F_{t}$ are $K$ \emph{consecutive} \emph{static} factors, and $\rho(L)$ and $\beta(L)$
are lag polynomials of orders $p_y$ and $p_f$ respectively. The feasible procedure requires an
estimate of $F_t$ that is usually obtained by principal component analysis (PCA).
\end{enumerate}
Then, we will take these models as two different types of ``patients'' and will administer them one particular ML treatment or combinations of them. That is, we will upgrade these models with one or many features of ML and evaluate the gains/losses in both environments.  From the perspective of the machine learning literature, equation (\ref{swardi2}) motivates the use of PCA as a form of feature engineering. Although more sophisticated methods have been used\footnote{The autoencoder method of \cite{Gu2020} can be seen as a form of feature engineering, just as the independent components used in conjunction with SVR in \cite{Lu2009}. The interested reader may also see \cite{Hastie2009} for a detailed discussion of the use of PCA and related method in machine learning.}, PCA remains popular \citep{Uddin2018}. As we insist on treating models as symmetrically as possible, we will use the same feature transformations throughout such that our nonlinear models, such as Kernel Ridge Regression, will introduce nonlinear transformations of lagged target values as well as of lagged values of the principal components.
Hence, our nonlinear models  postulate that a sparse set of latent variables impact the target  in a flexible way.\footnote{We omit considering a VAR  as an additional option. VAR iterative approach to produce $h$-step-ahead predictions is not comparable with the direct forecasting used with ML models.}

\subsection{Evaluation}\label{sec:eval}

The objective of this paper is to disentangle important characteristics of the ML prediction algorithms when forecasting macroeconomic
variables. To do so, we design an \textit{experiment} that consists of a pseudo-out-of-sample (POOS) forecasting horse race between many models that differ with respect to the
four main features above, i.e., nonlinearity, regularization, hyperparameter selection and loss function. To create variation around those \textit{treatments}, we will generate forecast errors from different models associated to each feature.

To test this paper's hypothesis, suppose the following model for forecasting errors
\begin{subequations}\label{exp_eq}
\begin{align}
	e^2_{t,h,v,m} = \alpha_m +  \psi_{t,v,h} + v_{t,h,v,m} \\ \alpha_m = \alpha_F'\boldsymbol{1}  + \eta_m
\end{align}
\end{subequations}
where $e^2_{t,h,v,m}$ are squared prediction errors of model $m$ for variable $v$ and horizon $h$ at time $t$. $\psi_{t,v,h}$ is a fixed effect term that demeans the dependent variable by ``forecasting target'', that is a combination of $t$, $v$ and $h$.  $\alpha_{F}$ is a vector of $\alpha_{\mathcal{G}}$, 
$\alpha_{pen()}$, $\alpha_{\tau}$ and $\alpha_{\hat{L}}$ terms associated to each feature. We re-arrange 
 equation (\ref{exp_eq}) to obtain 
\begin{equation}\label{log_eq}
	e^2_{t,h,v,m} = \alpha_F'\boldsymbol{1} +  \psi_{t,v,h} + u_{t,h,v,m}.
\end{equation}
$H_0$ is now $\alpha_f =0 \quad \forall f \in F = [\mathcal{G}, \ pen(), \ \tau, \ \hat{L}]$. In other words, the null is that there is no predictive accuracy gain with respect to a base model that does not have this particular feature.\footnote{If we consider two models that differ in one feature and run this regression for a specific $(h,v)$ pair, the t-test on   coefficients amounts to  \cite{Diebold1995} -- conditional on having the proper standard errors.} By interacting $\alpha_F$ with other fixed effects or  variables, we can test many hypotheses about the heterogeneity of the ``ML treatment effect.'' To get interpretable coefficients, we 
define $R^2_{t,h,v,m} \equiv 1 - \frac{e^2_{t,h,v,m}}{\frac{1}{T} \sum_{t=1}^T (y_{v,t+h} - \bar{y}_{v,h})^2}$ and run 
\begin{equation}\label{r2_eq}
	R^2_{t,h,v,m} = \dot{\alpha}_F'\boldsymbol{1} +  \dot{\psi}_{t,v,h} + \dot{u}_{t,h,v,m}.
\end{equation}
While \eqref{log_eq} has the benefit of connecting directly with the specification of a \cite{Diebold1995} test, the transformation of the regressand in \eqref{r2_eq} has two main advantages justifying its use. First and foremost, it provides standardized coefficients  $\dot{\alpha}_F$  interpretable as marginal improvements in OOS-$R^2$'s. In contrast, ${\alpha}_F$ are a unit- and series-dependant marginal increases in MSE. Second, the $R^2$ approach  has the advantage of standardizing \textit{ex-ante} the regressand and removing an obvious source of $(v,h)$-driven heteroskedasticity. 

While the generality of (\ref{log_eq}) and (\ref{r2_eq}) is appealing, when investigating the heterogeneity of specific partial effects, it will be much more convenient to run specific regressions for the  multiple hypothesis we wish to test. That is, to evaluate a feature $f$, we run
\begin{align}{\label{e_eq2}}
\forall m \in \mathcal{M}_f: \quad  R^2_{t,h,v,m} = \dot{\alpha}_f +  \dot{\phi}_{t,v,h} + \dot{u}_{t,h,v,m}
\end{align}
where $ \mathcal{M}_f$ is defined as the set of models that differs only by the feature under study $f$. An analogous evaluation setup has been considered in \cite{CARRIERO20191226}.

\section{Four Features of ML}\label{features}

In this section we detail the forecasting approaches that create variations for each characteristic of machine learning prediction
problem defined in (\ref{eq1}).

\subsection{Feature 1: Nonlinearity}

Although linearity is popular in practice, if the data generating process (DGP) is complex, using linear $g$ introduces approximation error as shown in (\ref{eq2}). As a solution, ML proposes an apparatus of nonlinear functions able to estimate the true DGP, and thus reduces  the approximation error. We  focus on applying the Kernel trick and random forests to our two baseline models to see if the nonlinearities they generate will lead to significant improvements.\footnote{A popular approach to model nonlinearity is deep learning. However, since we re-optimize our models recursively in a POOS, selecting an accurate network architecture by cross-validation is practically infeasible. In addition to optimize numerous neural net hyperparameters (such as the number of hidden layers and neurons, activation function, etc.), our forecasting models also require careful input selection (number of lags and number of factors in case of data-rich). An alternative is to fix ex-ante a variety of networks as in \cite{Gu2020a}, but this would potentially benefit other models that are optimized over time. Still, since few papers have found similar predictive ability of random forests and neural nets \citep{Gu2020,Joseph2019}, we believe that considering random forests and Kernel trick is enough to properly identify the ML nonlinear treatment. Nevertheless, we have conducted a robustness analysis with feed-forward neural networks and boosted trees. The results are presented in Appendix \ref{sec:NNBT}.} 


\subsubsection{Kernel Ridge Regression}\label{KTregression}

A simple way to make predictive regressions (\ref{ard}) and (\ref{swardi1}) nonlinear is to adopt a generalized linear model with multivariate functions of predictors (e.g. spline series expansions). However, this rapidly becomes overparameterized, so we opt for the Kernel trick (KT) to avoid computing all possible interactions and higher order terms.   
It is worth noting that Kernel Ridge Regression (KRR) has several implementation advantages. It has a closed-form solution that rules out convergence problems associated with models trained with gradient descent. It is also fast to implement since it implies inverting a $T$x$T$ matrix at each step. 

To show how KT is implemented in our benchmark models, suppose a Ridge regression direct forecast with generic regressors $Z_t$
\begin{align*}
\min_{\beta}{\sum_{t=1}^{T}}\left(y_{t+h}-Z_t \beta \right)^{2}+\lambda {\sum_{k=1}^{K}} \beta_{k}^{2}.
\end{align*}
The solution to that problem is $\hat{\beta}=(Z'Z+\lambda I_k)^{-1}Z'y$. By the representer theorem of \cite{SMOLA2004}, $\beta$ can also be obtained by solving the dual of the convex optimization problem above. The dual solution for $\beta$ is $\hat{\beta}=Z'(ZZ'+\lambda I_T)^{-1}y$. This equivalence allows to rewrite the conditional expectation in the following way:
\[
	\hat{E}(y_{t+h}|Z_t) = Z_t\hat{\beta} = \sum_{i=1}^t \hat{\alpha_i}\langle Z_i, Z_t \rangle
	\]
where $\hat{\alpha}= (ZZ'+\lambda I_T)^{-1}y$ is the solution to the dual Ridge Regression problem. 

Suppose now we  approximate a general nonlinear model $g(Z_t)$  with basis functions $\phi()$
\[
	y_{t+h} = g(Z_t) + \varepsilon_{t+h} = \phi(Z_t)'\gamma + \varepsilon_{t+h}.
\]	
The so-called Kernel trick is the fact that there exist a reproducing kernel $K()$ such that
\[
	\hat{E}(y_{t+h}|Z_t) = \sum_{i=1}^t \hat{\alpha_i}\langle \phi(Z_i), \phi(Z_t) \rangle = \sum_{i=1}^t \hat{\alpha_i}K( Z_i, Z_t ).
	\]
This means we do not need to specify the numerous basis functions, a well-chosen kernel implicitly replicates them. This paper will use the standard radial basis function (RBF) kernel

\[
{\displaystyle K_{\sigma}(\mathbf {x} ,\mathbf {x'} )=\exp \left(-{\frac {\|\mathbf {x} -\mathbf {x'} \|^{2}}{2\sigma ^{2}}}\right)}
\]
where $\sigma$ is a tuning parameter to be chosen by cross-validation. This choice of kernel is motivated by its good performance in macroeconomic forecasting as reported in \cite{Sermpinis2014} and  \cite{Exterkate2016}. The advantage of the kernel trick is that, by using the corresponding $Z_t$, we can easily make our data-rich or data-poor model nonlinear. For instance, in the case of the factor model, we can apply it to the regression equation to implicitly estimate
	\begin{align}
	y_{t+h} &= c + g(Z_t) + \varepsilon_{t+h},\label{KT1} \\
	Z_t &= \left[{\lbrace y_{t-j} \rbrace}_{j=0}^{p_y}, {\lbrace F_{t-j} \rbrace}_{j=0}^{p_f} \right], \label{KT2} \\
	X_t &= \Lambda F_t + u_t. \label{KT3}
\end{align}
In terms of implementation, this means extracting factors via PCA and then getting
\begin{equation}
	\hat{E}(y_{t+h}|Z_t)=K_{\sigma}(Z_t,Z)(K_{\sigma}(Z_t,Z)+\lambda I_T)^{-1}y_t. \label{KT4}
\end{equation}
The final set of tuning parameters for such a model is $\tau = \{\lambda,\sigma,p_y,p_f,n_f\}$.

\subsubsection{Random Forests} \label{RF}
Another way to introduce nonlinearity in the estimation of the predictive equation (\ref{swardi1}) is to use regression trees instead of OLS. 
The idea is to split sequentially the space of $Z_t$, as defined in (\ref{KT2}) into several regions and model the response by the mean of $y_{t+h}$ in each region. The process continues according to some stopping rule. The details of the recursive algorithm can be found in \cite{Hastie2009}.  Then, the tree regression forecast has the following form:
\begin{equation}
	\hat{f}(Z) =\sum_{m=1}^M c_m \mathrm{I}_{(Z \in R_m)},
\end{equation}
where $M$ is the number of terminal nodes, $c_m$ are node means and $R_1,...,R_M$ represents a partition of feature space. In the diffusion indices setup, the regression tree would estimate a nonlinear relationship linking factors and their lags to $y_{t+h}$. Once the tree structure is known, it can be related to a linear regression with dummy variables and their interactions. 

While the idea of obtaining nonlinearities via decision trees is intuitive and appealing  -- especially for its interpretability potential, the resulting prediction is usually plagued by high variance. The recursive tree fitting process is (i) unstable and (ii) prone to overfitting. The latter can be partially addressed by the use of pruning and related methodologies \citep{Hastie2009}. Notwithstanding, a much more successful (and hence popular) fix was proposed in \cite{Breiman2001}: Random Forests.  This consists in growing many trees on subsamples (or nonparametric bootstrap samples) of observations. Further randomization of underlying trees is obtained by considering a random subset of regressors for each potential split.\footnote{Only using a bootstrap sample of observations would be a procedure called Bagging -- for Bootstrap Aggregation. Also selecting randomly regressors has the effect of decorrelating the trees and hence boosting the variance reduction effect of averaging them.} The main hyperparameter to be selected is the number of variables to be considered at each split. The forecasts of the estimated regression trees are then averaged together to make one single "ensemble" prediction of the targeted variable.\footnote{In this paper, we consider 500 trees, which is usually more than enough to get a stabilized prediction (that will not change with the addition of another tree).} 

\subsection{Feature 2: Regularization}\label{sec:reg}


In this section we will only consider models where dimension reduction is needed, which are the models with $H_t^+$. The traditional shrinkage method used in macroeconomic forecasting is the ARDI model that consists of extracting principal components of $X_t$ and to use them as data in an ARDL model. Obviously, this is only one out of many ways to compress the information contained in $X_t$ to run a well-behaved regression of $y_{t+h}$ on it.\footnote{\cite{DeMol2008} compares Lasso, Ridge and ARDI and finds that forecasts are very much alike.} 

In order to create identifying variations for $pen()$ treatment, we need to generate multiple different shrinkage schemes. Some will also blend in selection, some will not. The alternative shrinkage methods will all be special cases of the Elastic Net (EN) problem:
\begin{align}
\min_{\beta}{\sum_{t=1}^{T}}\left(y_{t+h}-Z_t \beta \right)^{2}+\lambda {\sum_{k=1}^{K}} \left( \alpha |{\beta_k} | + (1- \alpha )  \beta_{k}^{2} \right) \label{en1}
\end{align}
where $Z_t=B(H_t)$ is some transformation of the original predictive set $X_t$. $\alpha \in [0,1]$ and $\lambda >0$ can either be fixed or found via CV. By using different $B$ operators, we can generate shrinkage schemes. Also, by setting $\alpha$ to either 1 or 0 we generate LASSO and Ridge Regression respectively. 
All these possibilities are reasonable alternatives to the traditional factor hard-thresholding procedure that is ARDI. 

Each type of shrinkage in this section will be defined by the tuple $S = \{\alpha,B()\}$. To begin with the most straightforward dimension, for a given $B$, we will evaluate the results for  $\alpha \in \{0,\hat{\alpha}_{CV},1\}$. For instance, if $B$ is the identity mapping, we get in turns the LASSO, EN and Ridge shrinkage. 
We now detail different $pen()$ resulting when we vary $B()$ for a fixed $\alpha$. 
\begin{enumerate}
\item \textbf{(Fat Regression)}: First, we consider the case $B_1()=I()$. That is, we use the entirety of the untransformed high-dimensional data set. The results of \cite{Giorgio2018} point in the direction that specifications with a higher $\alpha$ should do better, that is, sparse models do worse than models where every regressor is kept but shrunk to zero.
\item \textbf{(Big ARDI)} Second, $B_2()$ corresponds to first rotating $X_t \in {\rm I\!R}^N $ so that we get $N$-dimensional uncorrelated $F_t$. 
Note here that contrary to the ARDI approach, we do not select factors recursively, we keep them all.
Hence, $F_t$ has exactly the same span as $X_t$. 
Comparing LASSO and Ridge in this setup will allow to verify whether sparsity emerges in a rotated space. 
\item \textbf{(Principal Component Regression)} A third possibility is to rotate $H_t^+$ rather than $X_t$ and still keep all the factors. $H_t^+$ includes all the relevant preselected lags. If we were to just drop the $F_t$ using some hard-thresholding rule, this would correspond to Principal Component Regression (PCR). Note that $B_3()=B_2()$ only when no lags are included. 
\end{enumerate}
Hence, the tuple $S$ has a total of 9 elements. Since we will be considering both POOS-CV and K-fold CV for each of these models, this leads to a total of 18 models.\footnote{Adaptive versions (in the sense of \cite{Zou2006}) of the 9 models were also considered but gave either similar or deteriorated results with respect to their plain counterparts.}
 
To see clearly through all of this, we  describe where the benchmark ARDI model stands in this setup. Since it uses a hard thresholding rule that is based on the eigenvalues ordering, it cannot be a special case of the Elastic Net problem. While it uses $B_2$, we would need to set $\lambda=0$ and select $F_t$ \textit{a priori} with a hard-thresholding rule. 
The closest approximation in this EN setup would be to set $\alpha=1$ and fix the value of $\lambda$ to match the number of consecutive factors selected by an information criteria  directly in the predictive regression (\ref{swardi1}). 


\subsection{Feature 3: Hyperparameter Optimization}\label{sec:cv}

The conventional wisdom in macroeconomic forecasting is to either use AIC or BIC and compare results. 
The prime reason for the popularity of CV is that it can be applied to any model, including those for which the derivation of an information criterion is impossible.\footnote{\cite{Abadie2019} show that hyperparemeter tuning by CV performs uniformly well in high-dimensional context.} 

It is not  obvious that CV should work better only because it is ``out of sample'' while AIC and BIC are ''in sample''. All model selection methods are actually approximations to the OOS prediction error that relies on different assumptions that are sometime motivated by different theoretical goals. Also, it is well known that asymptotically, these methods have similar behavior.\footnote{\cite{Hansen2015} show equivalence between test statistics for OOS forecasting performance and in-sample Wald statistics. For instance,  one can show that Leave-one-out CV (a special case of K-fold) is asymptotically equivalent to the Takeuchi Information criterion (TIC), \cite{Cleaskens2008}. 
AIC is a special case of TIC where we need to assume in addition that all models being considered are at least correctly specified. Thus, under the latter assumption, Leave-one-out CV is asymptotically equivalent to AIC.} Hence, it is impossible \textit{a priori} to think of one model selection technique being the most appropriate for macroeconomic forecasting.

For samples of small to medium size encountered in macro, the question of which one is optimal in the forecasting sense is inevitably an empirical one. For instance, \cite{Granger2004} compared AIC and BIC in a generic forecasting exercise. In this paper, we will compare AIC, BIC and two types of CV for our two baseline models. The two types of CV are relatively standard. We will first use POOS CV and then K-fold CV. The first one will always behave correctly in the context of time series data, but may be quite inefficient by only using the end of the training set. The latter is known to be valid only if residual autocorrelation is absent from the models as shown in \cite{Bergmeir2018}. If it were not to be the case, then we should expect K-fold to underperform. 
The specific details of the implementation of both CVs is discussed in the section \ref{CVdetails} of the supplementary material.

The contributions of this section are twofold. First, it will shed light on which model selection method is most appropriate for typical macroeconomic data and models. Second, we will explore how much of the gains/losses of using ML can be attributed to widespread use of CV. Since most nonlinear ML models cannot be easily tuned by anything other than CV, it is hard for the researcher to disentangle between gains coming from the ML method itself or just the way it is tuned.\footnote{\cite{Zou2007} show that the number of remaining parameters in the LASSO is an unbiased estimator of the degrees of freedom and derive LASSO-BIC and LASSO-AIC criteria. Considering these as well would provide additional evidence on the empirical debate of CV vs IC.} Hence, it is worth asking the question whether some gains from ML are simply coming from selecting hyperparameters in a different fashion using a method whose assumptions are more in line with the data at hand. To investigate that, a natural first step is to look at our benchmark macro models, AR and ARDI, and see if  using CV to select hyperparameters gives different selected models and forecasting performances.

\subsection{Feature 4:  Loss Function}\label{sec:lf}

Until now, all of our estimators  use a quadratic loss function. Of course, it is very natural for them to do so: the quadratic loss is the measure used for out-of-sample evaluation. Thus, someone may legitimately wonder if the fate of the SVR is not sealed in advance as it uses an in-sample loss function which is 
inconsistent with the out-of-sample performance metric. As we will discuss later after the explanation of the SVR, there are reasons to believe the alternative (and mismatched) loss function can help. As a matter of fact, SVR has been successfully applied to forecasting financial and macroeconomic time series.\footnote{See for example, \cite{Lu2009}, \cite{Choudhury2014}, \cite{Patel2015}, \cite{Patel2015a}, \cite{Yeh2011} and \cite{Qu2016} for financial forecasting. See \cite{Sermpinis2014} and \cite{Zhang2010} macroeconomic forecasting.} An important question remains unanswered: are the good results due to kernel-based non-linearities or to the use of an alternative loss-function?

We provide a strategy to isolate the marginal effect of the SVR's $\bar{\epsilon}$-insensitive loss function which consists in, perhaps unsurprisingly by now, estimating different variants of the same model. 
We considered the Kernel Ridge Regression earlier. The latter only differs from the Kernel-SVR by the use of different in-sample loss functions. This identifies directly the effect of the loss function, for nonlinear models. Furthermore, we  do the same exercise for linear models: comparing a linear SVR to the plain ARDI. To sum up, to isolate the ``treatment effect'' of a different in-sample loss function, we consider: (1) the linear SVR with $H_t^-$; (2) the linear SVR with $H_t^+$; (3) the RBF Kernel SVR with $H_t^-$; and (4) the RBF Kernel SVR with $H_t^+$.
	
What follows is a bird's-eye overview of the underlying mechanics of the SVR. As it was the case for the Kernel Ridge regression, the SVR estimator approximates the function $g \in G$ with basis functions.  We opted to use the $\epsilon$-SVR variant which implicitly defines the size $2\bar{\epsilon}$ of the insensitivity tube of the loss function. The $\epsilon$-SVR is defined by:
	\begin{align*}
	\underset{\gamma}{min} \; \frac{1}{2}\gamma'\gamma + C \left[\sum_{t=1}^{T} (\xi_t + \xi_t^*) \right]\\
	s.t. \begin{cases}
	y_{t+h} - \gamma'\phi(Z_t) - \alpha \leq \bar{\epsilon} + \xi_t \\
	\gamma'\phi(Z_t) + \alpha - y_{t+h} \leq \bar{\epsilon} + \xi_t^* \\
	\xi_t, \xi_t^* \geq 0.
	\end{cases}
	\end{align*}
	Where $\xi_t, \xi_t^*$ are slack variables, $\phi()$ is the basis function of the feature space implicitly defined by the kernel used and $T$ is the size of the sample used for estimation. $C$ and $\bar{\epsilon}$ are hyperparameters. Additional hyperparameters vary depending on the choice of a kernel. In case of the RBF kernel, a scale parameter $\sigma$ also has to be cross-validated.
Associating Lagrange multipliers $\lambda_j, \lambda_j^*$ to the first two types of constraints, \cite{SMOLA2004} show that we can derive the dual problem out of which we would find the optimal weights $\gamma = \sum_{j=1}^{T} (\lambda_j - \lambda_j^*)\phi(Z_j)$ and the forecasted values
	\begin{equation}
	\hat{E}(y_{t+h}|Z_t) = \hat{c} + \sum_{j=1}^{T} (\lambda_j - \lambda_j^*)\phi(Z_j)\phi(Z_j) = \hat{c} + \sum_{j=1}^{T} (\lambda_j - \lambda_j^*)K(Z_j,Z_t). \label{svr1}
	\end{equation}
	
	Let us now turn to the resulting loss function of such a problem. 
	For the $\epsilon$-SVR, the penalty is given by:
	\begin{equation*}
	P_{\bar{\epsilon}}(\epsilon_{t+h|t}) := \begin{cases}
	0 \; \; \;  \; \; \;  \; \; \; \; \;  \; \; \;  \; \; \; \; \; \; if \quad |e_{t+h}| \leq \bar{\epsilon} \\
	|e_{t+h}| - \bar{\epsilon}  \; \; \; \; \; \;  otherwise
	\end{cases}.
	\end{equation*}
	
	For other estimators, the penalty function is quadratic $P(e_{t+h}) := e_{t+h}^2$. Hence, for our other estimators, the rate of the penalty increases with the size of the forecasting error, whereas it is constant and only applies to excess errors in the case of the $\epsilon$-SVR. Note that this insensitivity has a nontrivial consequence for the forecasting values. The Karush-Kuhn-Tucker conditions imply that only support vectors, i.e. points lying outside the insensitivity tube, will have nonzero Lagrange multipliers and contribute to the weight vector. 
	
As discussed briefly earlier, given that SVR forecasts will eventually be evaluated according to a quadratic loss, it is reasonable to ask why 
this alternative loss function isn't trivially suboptimal. \cite{Smola1998} show that the optimal size of $\bar{\epsilon}$ is a linear function of the underlying noise, with the exact relationship depending on the nature of the data generating process. This idea is not at odds with \cite{Gu2020} using the Huber Loss for asset pricing with ML (where outliers seldomly happen in-sample) or \cite{Colombo2020} successfully using SVR to forecast (notoriously noisy) exchange rates.
Thus, while SVR can work well in macroeconomic forecasting, it is unclear which feature between the nonlinearity and $\bar{\epsilon}$-insensitive loss  has the primary influence on its performance. 


	

	
To sum up, the table \ref{tab:models} shows a list of all  forecasting models and highlights their relationship with each of four features discussed above. The computational details for every model in this list are available in section \ref{sec:models} in the supplementary material.

\begin{table}[!h]
\caption{List of all forecasting models}
\label{tab:models}\vspace{-0.4cm}
\par
\begin{center}
{\scriptsize
\begin{tabular}{lllll}
\hline\hline
Models & Feature 1: selecting  & Feature 2: selecting  & Feature 3: optimizing  & Feature 4: selecting  \\
             & the function $g$ &  the regularization  & hyperparameters $\tau$ & the loss function \\
\hline
\multicolumn{1}{c}{Data-poor models} \\
AR,BIC & Linear &  & BIC & Quadratic \\
AR,AIC & Linear &  & AIC & Quadratic \\
AR,POOS-CV & Linear &  & POOS CV & Quadratic \\
AR,K-fold & Linear &  & K-fold CV & Quadratic \\
RRAR,POOS-CV & Linear & Ridge & POOS CV & Quadratic \\
RRAR,K-fold & Lineal & Ridge & K-fold CV & Quadratic \\
RFAR,POOS-CV & Nonlinear &  & POOS CV & Quadratic \\
RFAR,K-fold & Nonlinear &  & K-fold CV & Quadratic \\
KRRAR,POOS-CV & Nonlinear & Ridge & POOS CV & Quadratic \\
KRRAR,K-fold & Nonlinear & Ridge & K-fold CV & Quadratic \\
SVR-AR,Lin,POOS-CV & Linear &  & POOS CV & $\bar{\epsilon}$-insensitive \\
SVR-AR,Lin,K-fold & Linear &  & K-fold CV & $\bar{\epsilon}$-insensitive \\
SVR-AR,RBF,POOS-CV & Nonlinear &  & POOS CV & $\bar{\epsilon}$-insensitive \\
SVR-AR,RBF,K-fold & Nonlinear &  & K-fold CV & $\bar{\epsilon}$-insensitive \\
\hline
\multicolumn{1}{c}{Data-rich models} \\
ARDI,BIC & Linear & PCA & BIC & Quadratic \\
ARDI,AIC & Linear & PCA & AIC & Quadratic \\
ARDI,POOS-CV & Linear & PCA & POOS CV & Quadratic \\
ARDI,K-fold & Linear & PCA & K-fold CV & Quadratic \\
RRARDI,POOS-CV & Linear & Ridge-PCA & POOS CV & Quadratic \\
RRARDI,K-fold & Linear & Ridge-PCA & K-fold CV & Quadratic \\
RFARDI,POOS-CV & Nonlinear & PCA & POOS CV & Quadratic \\
RFARDI,K-fold & Nonlinear & PCA & K-fold CV & Quadratic \\
KRRARDI,POOS-CV & Nonlinear & Ridge-PCR & POOS CV & Quadratic \\
KRRARDI,K-fold & Nonlinear & Ridge-PCR & K-fold CV & Quadratic \\
($B_1,\alpha=\hat{\alpha}$),POOS-CV & Linear & EN & POOS CV & Quadratic \\
($B_1,\alpha=\hat{\alpha}$),K-fold & Linear & EN & K-fold CV & Quadratic \\
($B_1,\alpha=1$),POOS-CV & Linear & Lasso & POOS CV & Quadratic \\
($B_1,\alpha=1$),K-fold & Linear & Lasso & K-fold CV & Quadratic \\
($B_1,\alpha=0$),POOS-CV & Linear & Ridge & POOS CV & Quadratic \\
($B_1,\alpha=0$),K-fold & Linear & Ridge & K-fold CV & Quadratic \\
($B_2,\alpha=\hat{\alpha}$),POOS-CV & Linear & EN-PCA & POOS CV & Quadratic \\
($B_2,\alpha=\hat{\alpha}$),K-fold & Linear & EN-PCA & K-fold CV & Quadratic \\
($B_2,\alpha=1$),POOS-CV & Linear & Lasso-PCA & POOS CV & Quadratic \\
($B_2,\alpha=1$),K-fold & Linear & Lasso-PCA & K-fold CV & Quadratic \\
($B_2,\alpha=0$),POOS-CV & Linear & Ridge-PCA & POOS CV & Quadratic \\
($B_2,\alpha=0$),K-fold & Linear & Ridge-PCA & K-fold CV & Quadratic \\
($B_3,\alpha=\hat{\alpha}$),POOS-CV & Linear & EN-PCR & POOS CV & Quadratic \\
($B_3,\alpha=\hat{\alpha}$),K-fold & Linear & EN-PCR & K-fold CV & Quadratic \\
($B_3,\alpha=1$),POOS-CV & Linear & Lasso-PCR & POOS CV & Quadratic \\
($B_3,\alpha=1$),K-fold & Linear & Lasso-PCR & K-fold CV & Quadratic \\
($B_3,\alpha=0$),POOS-CV & Linear & Ridge-PCR & POOS CV & Quadratic \\
($B_3,\alpha=0$),K-fold & Linear & Ridge-PCR & K-fold CV & Quadratic \\
SVR-ARDI,Lin,POOS-CV & Linear & PCA & POOS CV & $\bar{\epsilon}$-insensitive \\
SVR-ARDI,Lin,K-fold & Linear & PCA & K-fold CV & $\bar{\epsilon}$-insensitive \\
SVR-ARDI,RBF,POOS-CV & Nonlinear & PCA & POOS CV & $\bar{\epsilon}$-insensitive \\
SVR-ARDI,RBF,K-fold & Nonlinear & PCA & K-fold CV & $\bar{\epsilon}$-insensitive \\
\hline\hline
\end{tabular}
}
\end{center}
\vspace{-0.5cm}
{\scriptsize \emph{%
\singlespacing{Note: PCA stands for Principal Component Analysis, EN for Elastic Net regularizer, 
PCR for Principal Component Regression.}}}
\end{table}

\section{Empirical setup}\label{empirics}

This section presents the data and the design of the pseudo-of-sample
experiment used to generate the treatment effects above.

\subsection{Data}

We use historical data to evaluate and compare the performance of all the
forecasting models described previously. The dataset is FRED-MD,  available
at the Federal Reserve of St-Louis's web site. It contains 134 monthly
US macroeconomic and financial indicators observed from 1960M01 to 2017M12.
Since many of them are usually very persistent or
not stationary, we follow \cite{McCracken2016}
in the choice of transformations in order to achieve stationarity.\footnote{Alternative data transformations 
in the context of ML modeling are used in \cite{GouletMDTM2020}.} 
Even though the universe of time series available at FRED is huge, we stick to FRED-MD 
for several reasons. First, we want to have the test 
set as long as possible since most of the variables do not start early 
enough.
Second, most of the timely available series are  
disaggregated components of the variables in FRED-MD. Hence, adding them alters 
the estimation of common factors \citep{Boivin2006}, and induces too much collinearity 
for Lasso performance \citep{Fan2010}. Third, it is the standard high-dimensional dataset 
that has been extensively used in the macroeconomic literature. 

\subsection{Variables of Interest}

We focus on predicting five representative macroeconomic indicators of the US economy: Industrial
Production (INDPRO), Unemployment rate (UNRATE), Consumer Price Index
(INF), difference between 10-year Treasury Constant Maturity rate and Federal funds rate (SPREAD) 
and housing starts (HOUST). INDPRO, CPI and HOUST are assumed $I(1)$ so we forecast the average growth rate  as in equation (\ref{fcst1}). UNRATE is considered $I(1)$ and we target the average change as in 
(\ref{fcst1}) but without logs. SPREAD is  $I(0)$ and the 
target is as in (\ref{fcst0}).\footnote{The US CPI is sometimes modeled as $I(2)$ due to the possible stochastic trend in inflation 
rate in 70's and 80's, see \citep{Stock2002}. Since in our test set the  the inflation is mostly stationary, we treat the price index as $I(1)$, as in \cite{Medeiros2019}.  
We have compared the mean squared predictive errors of best models under $I(1)$ and $I(2)$ alternatives, and 
found that errors are minimized when predicting the inflation rate directly.}

\subsection{Pseudo-Out-of-Sample Experiment Design}

The pseudo-out-of-sample period is 1980M01 - 2017M12. The forecasting
horizons considered are 1, 3, 9, 12 and 24 months. Hence, there are 456 evaluation periods for
each horizon. All models are estimated recursively with an expanding window as means of erring on the side of including more data so as to potentially reduce the variance of more flexible models.\footnote{The alternative is obviously that of a rolling window, which could be more robust to issues of model instability. These are valid concerns and have motivated tests and methods for taking them into account (see for example, \cite{Pesaran2007,Pesaran2013,Inoue2017,Boot2020}), an adequate evaluation lies beyond the scope of this paper. Moreover, as noted in \cite{Boot2020}, the number of relevant breaks may be much smaller than previously thought.}


Hyperparameter optimization is done with in-sample criteria (AIC and BIC) and
 two types of CV (POOS and K-fold). The in-sample
 selection is standard, we fix the upper bounds for the set
of HPs. 
For the POOS CV, 
the validation set consists
of last 25\% of the in-sample. In case of K-fold CV, we set $k=5$. We re-optimize hyperparameters every two years. This isn't uncommon for computationally demanding studies.\footnote{\cite{Sermpinis2014}, for example, split their out-of-sample into four year periods and update both hyperparameters and model parameter estimates every 4 years. Likewise, \cite{Terasvirta2006} selected the number of lagged values to be included in nonlinear autoregressive models once and for all at the start of the POOS.} It is also reasonable to assume that optimal hyperparameters would not be terribly affected by expanding the training set with observations that account for 2-3\% of the new training set size. The information on upper / lower bounds and grid search for HPs for every model is available in section \ref{sec:models} in the supplementary material.


\subsection{Forecast Evaluation Metrics}

Following a standard practice in the forecasting literature, we evaluate the
quality of our point forecasts using the root Mean Square Prediction Error
(MSPE).  \cite{Diebold1995} (DM)  procedure is used to test
the predictive accuracy of each model against the reference (ARDI,BIC).
We also implement the Model Confidence Set (MCS), \citep{Hansen2011}, 
that selects the subset of best models at a given confidence level. 
These metrics 
measure the overall predictive performance and classify models according to 
DM and MCS tests. Regression analysis from section \ref{sec:eval} is used to 
estimate the treatment effect 
of each ML ingredient.

\section{Results}\label{results}

We present the results in several ways. First, for each variable, we summarize tables containing the relative 
root MSPEs (to AR,BIC model) with DM and MCS outputs, for the whole pseudo-out-of-sample and NBER recession
periods. Second, we evaluate the marginal effect of important features of ML using regressions described in 
section \ref{sec:eval}. 

\subsection{Overall Predictive Performance}\label{overall}

Tables \ref{tab:RelMSPE_INDPRO} - \ref{tab:RelMSPE_Hous}, in the appendix \ref{sec:rootMSE}, summarize the overall predictive performance  
in terms of root MSPE relative to the reference model AR,BIC. The analysis is done for the full out-of-sample as well as for NBER recessions 
 (i.e., when the target belongs to a recession episode). This address two  
questions: is ML already useful for macroeconomic forecasting and when?\footnote{ The knowledge of the models that 
have performed best historically during recessions is of interest for practitioners. If the probability of recession is high 
enough at a given period, our results can provide an ex-ante guidance on which model is likely to perform 
best in such circumstances.}

In case of industrial production, table \ref{tab:RelMSPE_INDPRO} shows that principal component regressions $B_2$ and $B_3$ with 
Ridge and Lasso penalty respectively are the best at short-run horizons of 1 and 3 months. 
The kernel ridge ARDI with POOS CV is best for $h=9$, while its autoregressive counterpart with K-fold minimizes the MSPE at 
the one-year horizon. Random forest ARDI, the alternative nonlinear approximator, outperforms the reference model by 11\% for $h=24$.  
During recessions, the ARDI with CV is the best for 1, 3 and 9 months ahead, while the nonlinear SVR-ARDI minimizes the MSPE at 
the one-year horizon. The ridge regression ARDI is the best for $h=24$.  Ameliorations with respect to 
AR,BIC are much larger during economic downturns, and the MCS selects fewer models.

Results for the unemployment rate, table \ref{tab:RelMSPE_UNRATE}, highlight the performance of nonlinear models especially for longer horizons. Improvements with respect to the AR,BIC model 
are bigger for both full OOS and recessions. MCSs are narrower than in case of INDPRO. A similar pattern is observed during NBER recessions.  
Table \ref{tab:RelMSPE_GS1} summarizes results for the Spread. Nonlinear models are generally the best, combined with data-rich 
predictors' set. 

For inflation, table \ref{tab:RelMSPE_CPI} shows that the kernel ridge autoregressive model with K-fold CV is the best for 3, 9 
and 12 months ahead, while the nonlinear SVR-ARDI optimized with K-fold CV reduces the MSPE by more than 20\% at two-year 
horizon. Random forest models are very resilient, as in \cite{Medeiros2019}, but generally outperformed by KRR form of nonlinearity. During recessions, the fat 
regression models ($B_1$) are the best at short horizons, while the ridge regression ARDI with K-fold 
dominates for $h={9,12,24}$.
Housing starts, in table \ref{tab:RelMSPE_Hous}, are best predicted with nonlinear data-rich models for almost all horizons.

Overall, using data-rich models and nonlinear $g$ functions improve macroeconomic prediction. 
Their marginal contribution depends on the state of the economy.

\subsection{Disentangling ML Treatment Effects}

\begin{figure}
\centering
\includegraphics[scale=.4, trim= 0mm 10mm 0mm 10mm, clip]{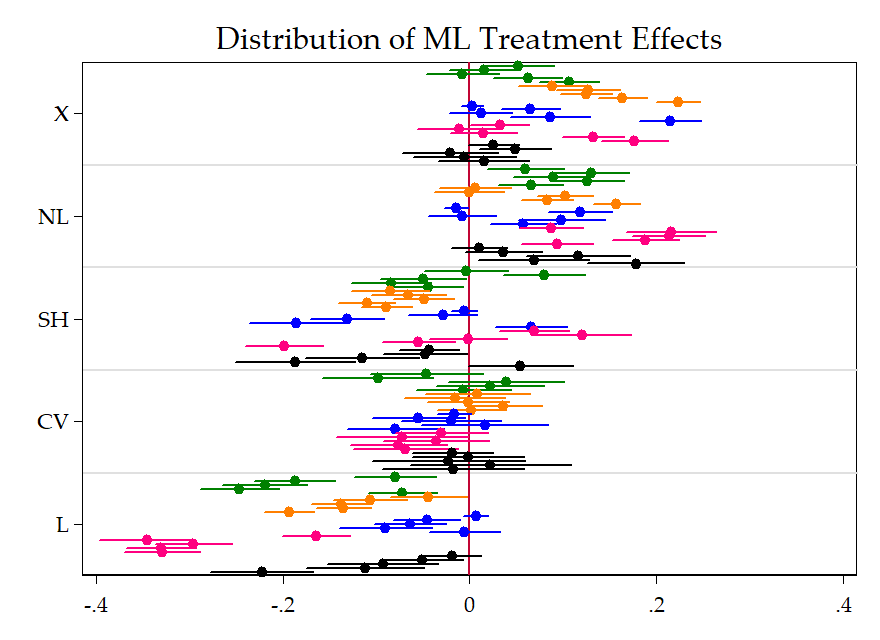}
\vspace{-0cm}
	\caption{\footnotesize{This figure plots the distribution of $\dot{\alpha}_F^{(h,v)}$ from equation (\ref{r2_eq}) done by $(h,v)$ subsets. That is, we are looking at the average partial effect on the pseudo-OOS $R^2$ from augmenting the model with ML features, keeping everything else fixed. $X$ is making the switch from data-poor to data-rich. Finally, variables are \textcolor{Green}{INDPRO}, \textcolor{orange}{UNRATE}, \textcolor{blue}{SPREAD}, \textcolor{magenta}{INF} and HOUST. Within a specific color block, the horizon increases from $h=1$ to $h=24$ as we are going down. As an example, we clearly see that the partial effect of $X$ on the $R^2$ of \textcolor{magenta}{INF} increases drastically with the forecasted horizon $h$. SEs are HAC. These are the 95\% confidence bands.}} 
	\label{dist_all}
\end{figure}

The results in the previous section does not easily allow to disentangle the marginal effects of important ML features -- as presented in section \ref{features}. Therefore, we turn to the regression analysis described in section \ref{sec:eval}. In what follows, [X, NL, SH, CV and LF] stand for data-rich, nonlinearity, alternative shrinkage, cross-validation and loss function features respectively.

Figure \ref{dist_all} shows the distribution of $\dot{\alpha}_F^{(h,v)}$ from equation (\ref{r2_eq}) done by $(h,v)$ subsets. Hence, here we allow for heterogeneous treatment effects according to 25 different targets. This figure highlights by itself the main findings of this paper.  \textbf{First}, ML nonlinearities improve substantially the forecasting accuracy in almost all situations. The effects are positive and significant for all horizons in case of INDPRO and SPREAD, and for most of the cases when predicting UNRATE, INF and HOUST. 
The improvements of the nonlinearity treatment reach up to 23\% in terms of pseudo-$R^2$. This is in contrast with previous literature that did not find substantial forecasting power 
from nonlinear methods, see for example \cite{Stock1999}. In fact, the ML  nonlinearity is highly flexible and well disciplined by a careful regularization, and thus can solve the general overfitting problem of standard nonlinear models \citep{Terasvirta2006}. This is also in line with the finding in \cite{Gu2020a} that nonlinearities (from ML models) can help predicting financial returns.

\textbf{Second},  alternative regularization means of dimensionality reduction do not improve on average over the standard factor model, except few cases. Choosing sparse modeling can decrease the forecast accuracy by up to 20\% of the pseudo-$R^2$ which is not negligible. Interestingly, \cite{Gu2020a} also reach similar conclusions that dense outperforms sparse in the context of applying ML to returns.
 
\textbf{Third}, the average effect of CV appears not significant. However, as we will see in section \ref{CVeffect}, the averaging in this case hides some interesting and relevant differences between K-fold and POOS CVs. 
\textbf{Fourth}, on average, dropping the standard in-sample squared-loss function for what the SVR proposes is not useful, except in very rare cases. 
\textbf{Fifth} and lastly, the marginal benefits of data-rich models ($X$) seems roughly to increase with horizons for every variable-horizon pair, except for few cases with spread and housing. Note that this is almost exactly like the picture we described for NL. Indeed, visually, it seems like the results for $X$ are a compressed-range version of NL that was translated to the right. Seeing NL models as data augmentation via  basis expansions, we  conclude that for predicting macroeconomic variables, we need to augment the AR($p$) model with more regressors either created from the lags of the dependent variable itself or coming from additional data. The possibility of joining these two forces to create a ``data-filthy-rich'' model is studied in section \ref{NLeffect}.

It turns out these findings are somewhat robust as graphs included in the appendix section \ref{sec:robust} show. ML treatment effects plots of very similar shapes are obtained for data-poor models only (figure \ref{dist_datap}), data-rich models only (figure \ref{dist_datar}) and recessions / expansions periods (figures \ref{dist_rec} and \ref{dist_exp}). It is important to notice that nonlinearity 
effect is not only present during recession periods, but it is even more important during expansions.\footnote{This suggests that our models behave relatively similarly over the business cycle and that our analysis does not suffer from undesirable forecast ranking due to extreme events as pointed out in \cite{Lerch2017}.} The 
only exception is the data-rich feature that has negative and significant effects for  housing starts prediction when we condition  on the last 20 years of the forecasting exercise (figure \ref{dist_last20yrs}). 

\begin{figure}
\centering
\includegraphics[width=0.8\textwidth, height=0.38\textheight, trim= 0mm 10mm 0mm 0mm, clip]{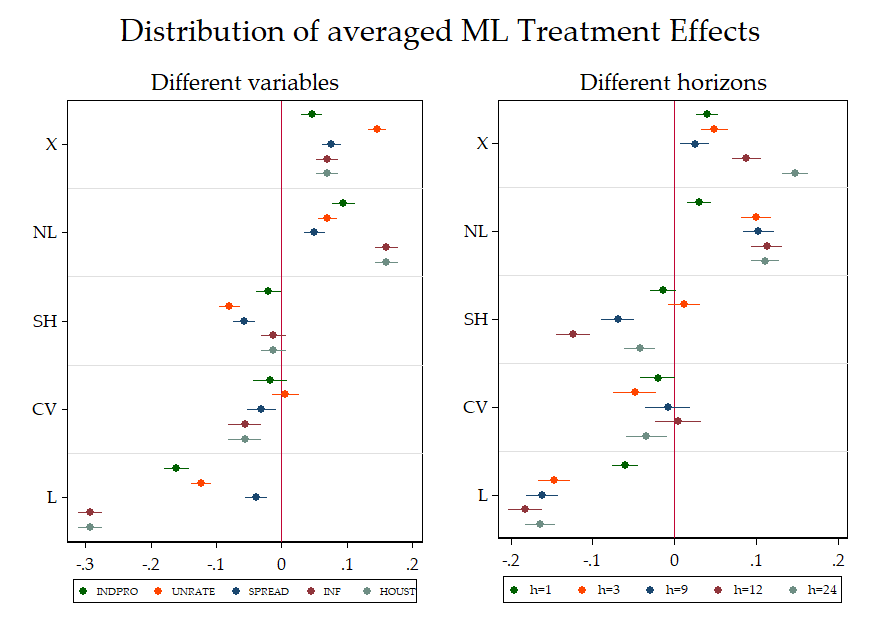}
	\caption{\footnotesize{This figure plots the distribution of $\dot{\alpha}_F^{(v)}$ and $\dot{\alpha}_F^{(h)}$ from equation (\ref{r2_eq}) done by $h$ and $v$ subsets. That is, we are looking at the average partial effect on the pseudo-OOS $R^2$ from augmenting the model with ML features, keeping everything else fixed. $X$ is making the switch from data-poor to data-rich. However, in this graph, $v-$specific heterogeneity and $h-$specific heterogeneity have been integrated out in turns. SEs are HAC. These are the 95\% confidence bands.}}
	\label{dist_all_hv}
\end{figure}

Figure \ref{dist_all_hv} aggregates by $h$ and $v$ in order to clarify whether variable or horizon heterogeneity matters most. Two facts detailed earlier are now quite easy to see. For both $X$ and NL, the average marginal effects roughly increase in $h$. In addition, it is now clear that all the variables benefit from both additional information and nonlinearities. Alternative shrinkage is least harmful for inflation and housing, and at short horizons. Cross-validation has negative and sometimes significant impacts, while the SVR loss function is often damaging. 

Supplementary material contains additional results. Section \ref{sec:mae} shows the results obtained using the absolute loss. 
The importance of each feature and the way it behaves according to the variable/horizon pair is the same. 
Finally, sections \ref{sec:quart} and \ref{sec:can} show results for two similar exercises. The first consider quarterly US data where we forecast the average growth rate of GDP, consumption, 
investment and disposable income, and the PCE inflation. The results are consistent with the findings obtained in the main body of this paper. In the second, we use a large Canadian 
monthly dataset and forecast the same target variables for Canada. Results are qualitatively in line with those on US data, except that NL  effect is smaller in size. 


In what follows we break down averages and run specific regressions as in (\ref{e_eq2}) to study how homogeneous are the $\dot{\alpha}_F$'s reported above.

\subsubsection{Nonlinearities}\label{NLeffect}

	\begin{figure}
			\centering
\includegraphics[scale=.4]{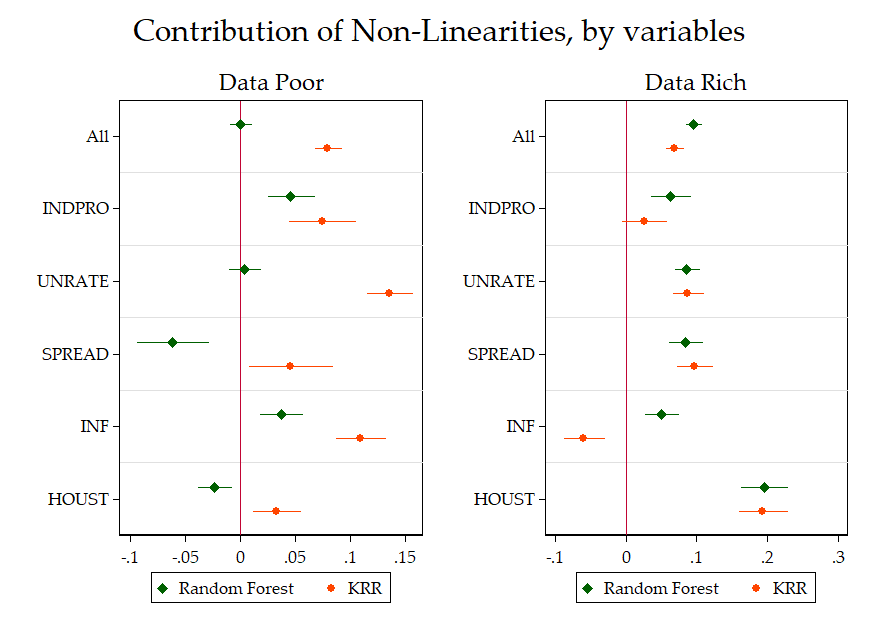}  
		\caption{\footnotesize{This figure compares the two NL models averaged over all horizons. The unit of the x-axis are  improvements in OOS $R^2$ over the basis model. SEs are HAC. These are the 95\% confidence bands.}}
		\label{g_nl_v}
		\end{figure}

		\begin{figure}
			\centering
\includegraphics[scale=.4]{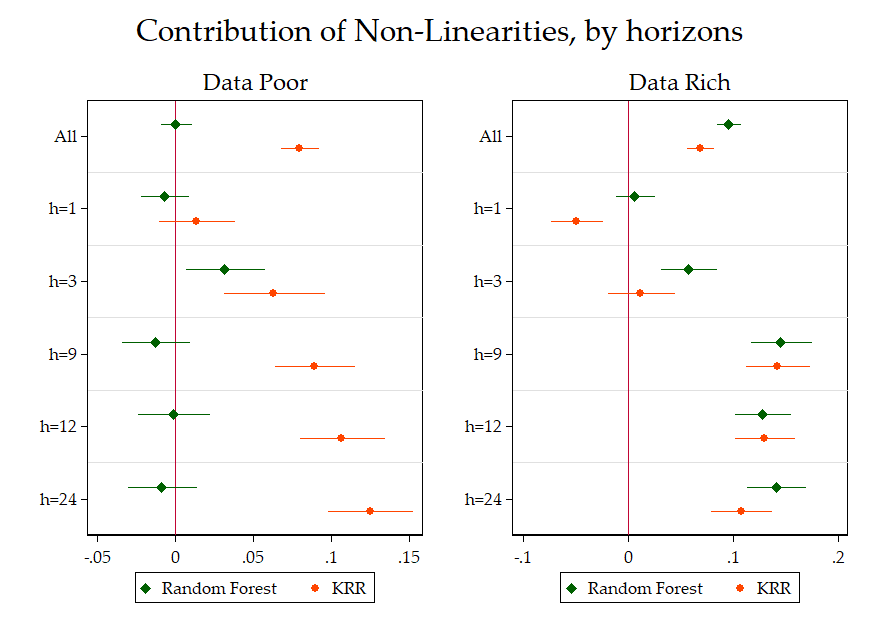}  
	\caption{\footnotesize{This figure compares the two NL models averaged over all variables. The unit of the x-axis are improvements in OOS $R^2$ over the basis model.  SEs are HAC. These are the 95\% confidence bands.}}
	\label{g_nl_h}
		\end{figure}

Figure \ref{g_nl_v} suggests that nonlinearities can be very helpful at forecasting all the five variables in the data-rich environment. The marginal effects of random forests and KRR are almost never statistically different for data-rich models, except for inflation combined with data-rich, suggesting that the common NL feature is the driving force. However, this is not the case for data-poor models where the kernel-type nonlinearity shows  significant improvements for all variables, while the random forests have positive impact on predicting INDPRO and inflation, but decrease forecasting accuracy for the rest of the variables.  

Figure \ref{g_nl_h} suggests that nonlinearities are in general more useful for longer horizons in data-rich environment while the KRR can be harmful for a very short horizon. Note again that both nonlinear models follow the same pattern for data-rich models with random forest often being better (but never statistically different from KRR). For data-poor models, it is KRR that has a (statistically significant) growing advantage as $h$ increases.
Seeing NL models as data augmentation via some basis expansions, we can join the two facts together to conclude that the need for a complex and ``data-filthy-rich'' model arises for predicting macroeconomic variables at longer horizons. Similar conclusions are obtained with neural networks and boosted trees as shown in figures \ref{g_nl_v_NNBT} and \ref{g_nl_h_NNBT} in Appendix \ref{sec:NNBT}. 

Figure \ref{TV_MSPE} in the appendix \ref{sec:addgraphs} plots the cumulative and 3-year rolling window root MSPE for linear and nonlinear data-poor and data-rich models, for $h=12$, as well as \cite{Giacomini2010} fluctuation test for those alternatives. The cumulative root MSPE clearly shows the positive impact on forecast accuracy of both nonlinearities and data-rich environment for all series except INF.  
The rolling window depicts the changing level of forecast accuracy. For all series except the SPREAD, there is a common cyclical behavior with two relatively similar peaks (1981 and 2008 recessions), as well as a drop in MSPE during the Great Moderation period. Fluctuation tests confirm the important role of nonlinear and data-rich models. 

For CPI inflation at horizons of 3, 9 and 12 months, Random Forests perform distinctively well. In both its data-poor and data-rich incarnations, the algorithm is included in the superior model set of \cite{Hansen2011} and significantly outperforms the AR-BIC benchmark according to the DM test. This result can help shed some light on long standing issues in the inflation forecasting literature.  A consensus emerged that nonlinear models in-sample good performance does not materialize out-of-sample \citep{Marcellino2008, StockWatson2009}.\footnote{Concurrently, simple benchmarks such as a random walk or moving averages emerged as surprisingly hard to beat  \citep{Atkeson2001,StockWatson2009,Kotchoni2019}.}  In contrast, we found -- as in \cite{Medeiros2019}, that  Random Forests are a particularly potent tool to forecast CPI inflation. One possible explanation is that previous studies suffer from overfitting \citep{Marcellino2008} while Random Forests are arguably completely immune from it \citep{MSoRF}, all this while retaining relevant nonlinearities. In that regard, it is noted that INF is the only target where KRR performance does not match that of Random Forests in the data rich environment. In the data-poor case, roles are reversed. Unlike most other  targets, it seems the type of NL being used matters for inflation. Nonetheless, ML generally appears to be useful for inflation forecasting by providing better-behaved non-parametric nonlinearities than what was considered by the older literature.


\subsubsection{Regularization}\label{SHeffect}

		\begin{figure}
			\centering
					
\includegraphics[scale=.35]{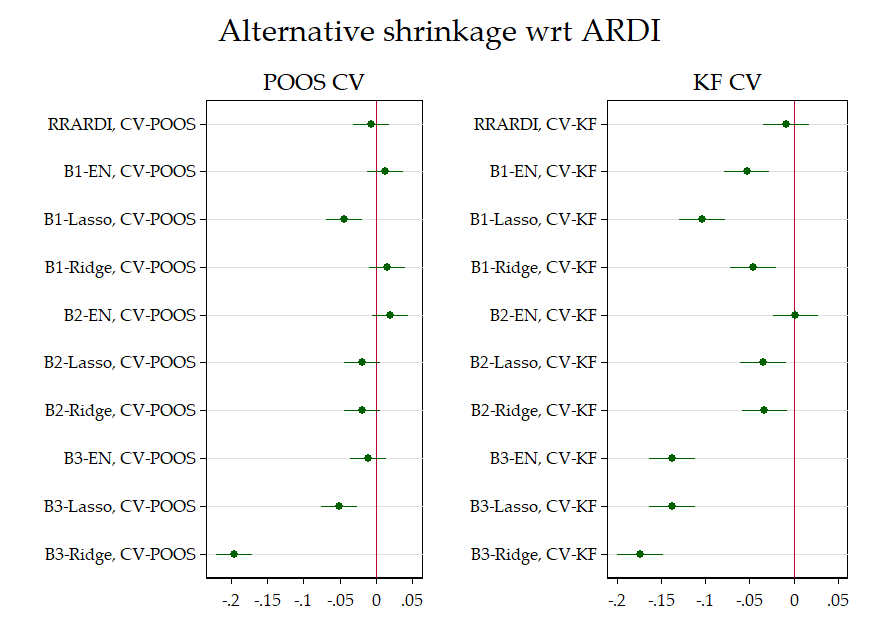}  
\caption{\footnotesize{This figure compares models of section \ref{sec:reg} averaged over all variables and horizons. The unit of the x-axis are improvements in OOS $R^2$ over the basis model.  The base models are ARDIs specified with POOS-CV and KF-CV respectively. SEs are HAC. These are the 95\% confidence bands.}}\label{SHgraph}
		\end{figure}
	
Figure \ref{SHgraph} shows that the ARDI reduces dimensionality in a way that certainly works well with economic data: all competing schemes do at most as good on average. It is overall safe to say that on average, all shrinkage schemes give similar or lower performance,  
which is in line with conclusions from \cite{Stock2012} and \cite{Kim2018}, but contrary to \cite{Smeekes2018}. No clear superiority for the Bayesian versions of some of these models was also documented in \cite{DeMol2008}. This suggests that the factor model view of the macroeconomy is quite accurate in the sense that when we use it as a means of dimensionality reduction, it extracts the most relevant information to forecast the relevant time series. This is good news. The ARDI is the simplest model to run and results from the preceding section tells us that adding nonlinearities to an ARDI can be quite helpful. 

Obviously, the deceiving  behavior of alternative shrinkage methods does not mean there are no interesting $(h,v)$ cases where using a different dimensionality reduction has significant benefits as discussed in section \ref{overall} and \cite{Smeekes2018}. Furthermore, LASSO and Ridge can still be useful to tackle specific time series  problems (other than dimensionality reduction), as shown with time-varying parameters in \cite{Coulombe2019}.

\subsubsection{Hyperparameter Optimization}\label{CVeffect}

\begin{figure}[h!]
\centering
\includegraphics[scale=.8,trim= 0mm 0mm 0mm 20mm, clip]{{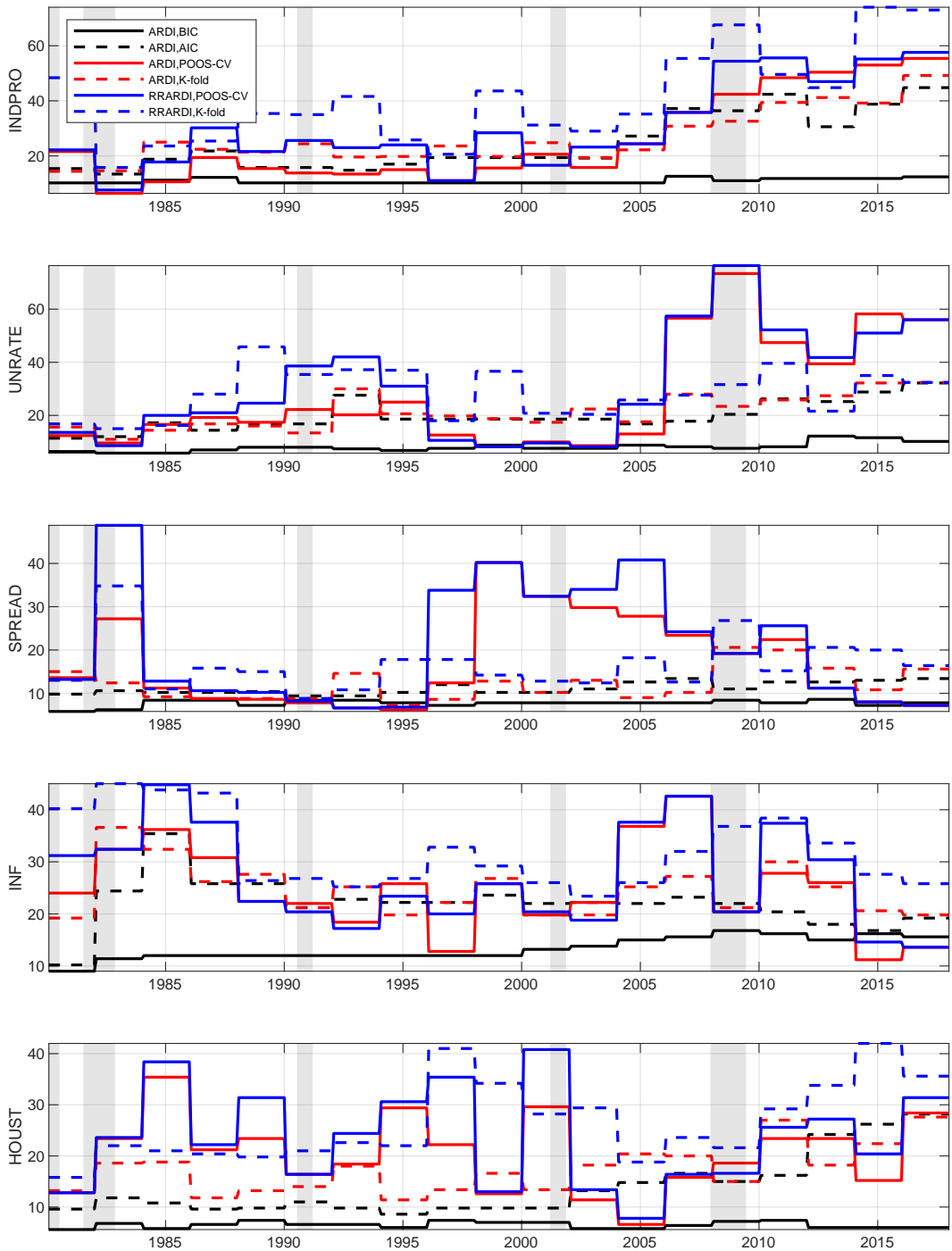}} 
\vspace{-2.5cm} 
\caption{\footnotesize{This figure shows the number of regressors in linear  ARDI models. Results averaged across horizons.}}
\label{ARDI_HP}
\end{figure}

Figure \ref{ARDI_HP} shows how many regressors are kept by different  selection methods in the case of ARDI. As expected, BIC is  in general the lower envelope of each of these graphs. 
Both cross-validations favor larger models, especially when combined with Ridge regression. 
We remark a common upward trend for all model selection methods in case of INDPRO and UNRATE. This is not the case for inflation where large models have been selected in 80's and most recently since 2005. In case of HOUST, there is a downward trend since 2000's which is consistent with the finding in Figure \ref{dist_last20yrs} that data-poor models do better in last 20 years.  POOS CV 
selection is  more volatile and 
selects bigger models for unemployment rate, spread and housing. While K-fold also selects models of considerable size, it does so in a more slowly growing fashion. This is not surprising because K-fold samples from all available data to build the CV criterion: adding new data points only gradually change the average. POOS CV is a shorter  window approach that offers flexibility against structural hyperparameters change at the cost of greater variance and vulnerability of rapid regime changes  in the data.

We know that different model selection methods  lead to quite different models, but what about their predictions? First, let us note that changes in OOS-$R^2$ are much smaller in magnitude for CV (as can be seen easily in figures \ref{dist_all} and \ref{dist_all_hv}) than for other studied ML treatment effects. Nevertheless, table \ref{CVreg_main} tells many interesting tales. The models included in the regressions are the standard linear  ARs and ARDIs (that is, excluding the Ridge versions) that have all been tuned using BIC, AIC, POOS CV and CV-KF. First, we see that overall, only POOS CV is distinctively worse, especially in data-rich environment, and that AIC and CV-KF are not significantly 
different from BIC on average. 
For data-poor models and during recessions, AIC and CV-KF are being significantly better than BIC  in downturns, while CV-KF seems harmless. The state-dependent effects are not significant in data-rich environment.  
Hence, for that class of models, we can safely opt for either BIC or CV-KF. Assuming some degree of external validity beyond that model class, we can be reassured that the quasi-necessity of leaving ICs behind when opting for more complicated ML models is not harmful.

\begin{table}[htbp]\centering
\def\sym#1{\ifmmode^{#1}\else\(^{#1}\)\fi}
\caption{CV comparison}\label{CVreg_main}
\begin{tabular}{l*{5}{c}}
\hline\hline
                    &\multicolumn{1}{c}{(1)}&\multicolumn{1}{c}{(2)}&\multicolumn{1}{c}{(3)}&\multicolumn{1}{c}{(4)}&\multicolumn{1}{c}{(5)}\\
                    &\multicolumn{1}{c}{All}&\multicolumn{1}{c}{Data-rich}&\multicolumn{1}{c}{Data-poor}&\multicolumn{1}{c}{Data-rich}&\multicolumn{1}{c}{Data-poor}\\
\hline
CV-KF               &     -0.0380         &      -0.314         &       0.237         &      -0.494         &      -0.181         \\
                    &     (0.800)         &     (0.711)         &     (0.411)         &     (0.759)         &     (0.438)         \\
CV-POOS             &      -1.351         &      -1.440\sym{*}  &      -1.262\sym{**} &      -1.069         &      -1.454\sym{***}\\
                    &     (0.800)         &     (0.711)         &     (0.411)         &     (0.759)         &     (0.438)         \\
AIC                 &      -0.509         &      -0.648         &      -0.370         &      -0.580         &      -0.812         \\
                    &     (0.800)         &     (0.711)         &     (0.411)         &     (0.759)         &     (0.438)         \\
CV-KF * Recessions  &                     &                     &                     &       1.473         &       3.405\sym{**} \\
                    &                     &                     &                     &     (2.166)         &     (1.251)         \\
CV-POOS * Recessions&                     &                     &                     &      -3.020         &       1.562         \\
                    &                     &                     &                     &     (2.166)         &     (1.251)         \\
AIC * Recessions    &                     &                     &                     &      -0.550         &       3.606\sym{**} \\
                    &                     &                     &                     &     (2.166)         &     (1.251)         \\
\hline
Observations        &       91200         &       45600         &       45600         &       45600         &       45600         \\
\hline\hline
\multicolumn{6}{l}{\footnotesize Standard errors in parentheses. \sym{*} \(p<0.05\), \sym{**} \(p<0.01\), \sym{***} \(p<0.001\)}
\end{tabular}
\end{table}

		\begin{figure}
			\centering
\includegraphics[scale=.4]{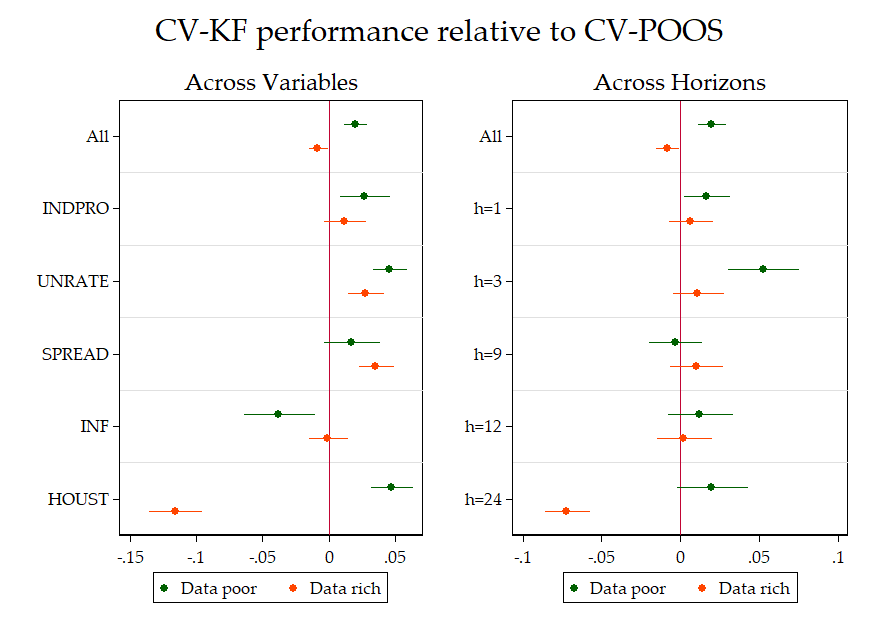}
		\caption{\footnotesize{This figure compares the two CVs procedure averaged over all the models that use them. The unit of the x-axis are improvements in OOS $R^2$ over the basis model.  SEs are HAC. These are the 95\% confidence bands.}}\label{CVbyX}
		\end{figure}
\begin{figure}
			\centering
\includegraphics[scale=.4]{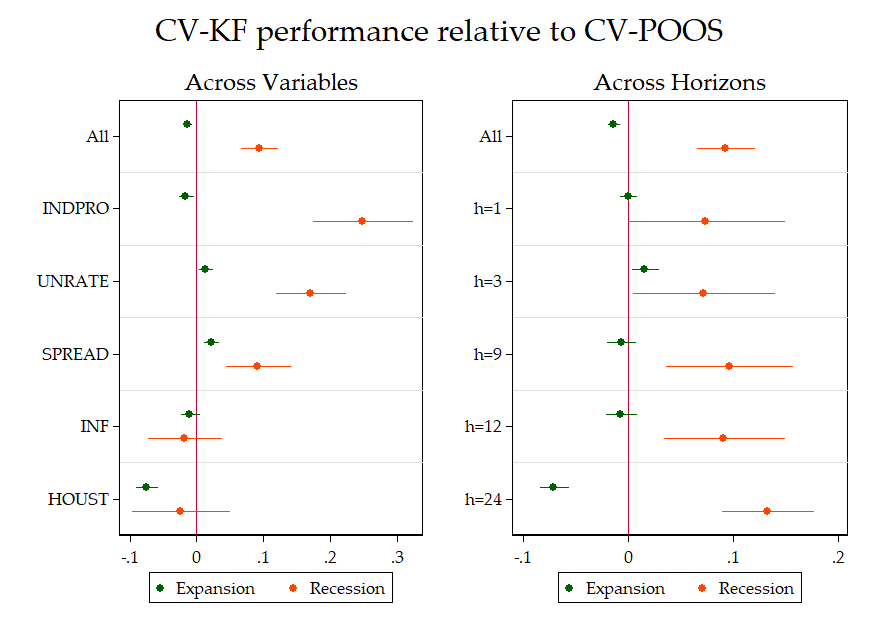}
\vspace{-0.5cm}
		\caption{\footnotesize{This figure compares the two CVs procedure averaged over all the models that use them. The unit of the x-axis are improvements in OOS $R^2$ over the basis model.  SEs are HAC. These are the 95\% confidence bands.}}\label{CVbyrec}
		\end{figure}

We now consider models that are usually tuned by CV and compare the performance of the two CVs by horizon and variables.		
Since we are now pooling multiple models, including all the alternative shrinkage models, if a clear pattern only attributable to a certain CV existed, it would most likely appear in figure \ref{CVbyX}. What we see are two things. First, CV-KF is at least as good as POOS CV on average for almost all variables and horizons, irrespective of the informational content of the regression. The exceptions are HOUST in data-rich and INF in data-poor frameworks, and the two-year horizon with large data. 
Figure \ref{CVbyrec}'s message has the virtue of clarity. POOS CV's failure is mostly attributable to its poor record in recessions periods  for the first three variables at any horizon. Note that this is the same subset of variables that benefits from adding in more data ($X$) and nonlinearities as discussed in \ref{NLeffect}.

By using only recent data, POOS CV will be more robust to gradual structural change but will perhaps have an Achilles heel in regime switching behavior. If the optimal hyperparameters are state-dependent, then a switch from expansion to recession at time $t$ can be quite harmful. K-fold, by taking the average over the whole sample, is less immune to such problems. Since results in \ref{overall} point in the direction that smaller models are better in expansions and bigger models in recessions, the behavior of CV and how it picks the effective complexity of the model can have an  effect on overall predictive ability. This is exactly what we see in figure \ref{CVbyrec}: POOS CV is having a hard time in recessions with respect to K-fold.

\subsubsection{Loss Function}\label{Leffectlin}

In this section, we investigate whether replacing the $l_2$ norm as an in-sample loss function for the SVR machinery helps in forecasting. We again use as baseline models ARs and ARDIs trained by the same corresponding CVs. The very nature of this ML feature is that the model is less sensible to extreme residuals, thanks to the $l_1$ norm outside of the $\bar \epsilon$-insensitivity tube. We first compare linear models  in figure \ref{Leffectlin}. Clearly, changing the loss function is generally harmful and that is mostly due to recessions period. However, in expansions, the linear SVR is better on average than a standard ARDI for UNRATE and SPREAD, but these small gains are clearly offset (on average) by the huge recession losses. 

\begin{figure}
			\centering
\includegraphics[scale=.4]{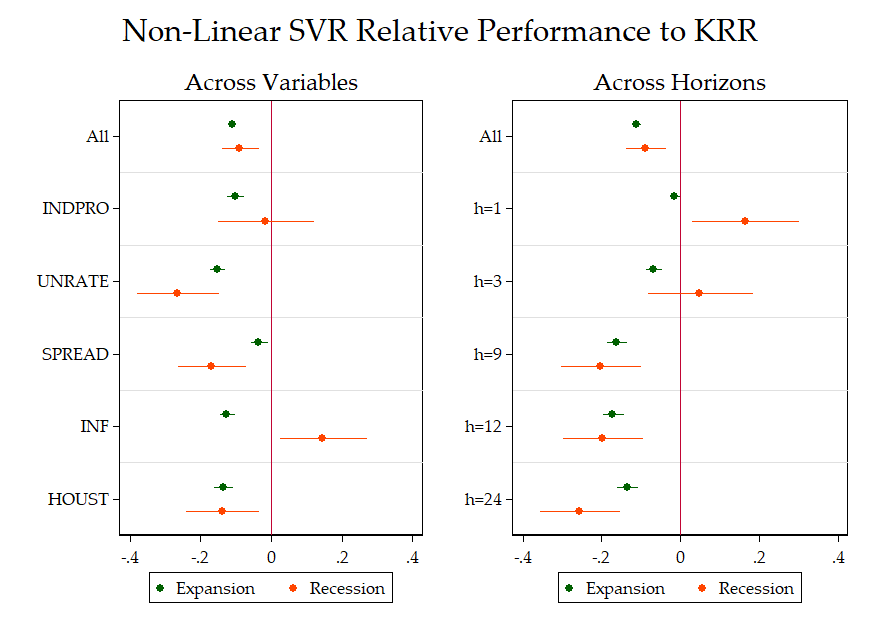}
\vspace{-0.3cm}
		\caption{\footnotesize{This graph displays the marginal (un)improvements by variables and horizons to opt for the SVR in-sample loss function in both recession and expansion periods. The unit of the x-axis are improvements in OOS $R^2$ over the basis model.  SEs are HAC. These are the 95\% confidence bands.}}\label{LeffectNL}
		\end{figure}

The SVR is usually used in its nonlinear form. We hereby compare KRR and SVR-NL to study whether the loss function effect could reverse when a nonlinear model is considered. Comparing these models makes sense since they both use the same kernel trick (with an RBF kernel). Hence, like linear models of figure \ref{Leffectlin}, models in figure \ref{LeffectNL} only differ by the use of a different loss function $\hat{L}$. It turns out conclusions are exactly the same as for linear models with the negative effects being slightly smaller in nonlinear world. There are few exceptions: inflation rate and one month ahead horizon during recessions. Furthermore, figures \ref{SVRlin_byX} and \ref{SVRNL_byX} in the appendix \ref{sec:addgraphs} confirm that these findings are valid for both the data-rich and the data-poor environments. 

By investigating these results more in depth using tables \ref{tab:RelMSPE_INDPRO} - \ref{tab:RelMSPE_Hous}, we see an emerging pattern. First, SVR sometimes does very good (best model for UNRATE at horizon 3 months) but underperforms for many targets -- in its AR or ARDI form. When it does perform well compared to the benchmark, it is more often than not outshined marginally by the KRR version. For instance, in table \ref{tab:RelMSPE_UNRATE}, linear and nonlinear SVR-Kfold provide respectively reductions of 17\% and 13\% in RMSPE over the benchmark for UNRATE at horizon 9 months. However, analogous KRR and Random Forest similarly do so. Moreover, for targets for which SVR fails, the two models it is compared to in order to extract $\alpha_{\hat{L}}$, KRR or the AR/ARDI, have a more stable (good) record. Hence, on average nonlinear SVR is much worse than KRR and the linear SVR is also inferior to the plain ARDI. This explains the clear-cut results reported in this section: if the SVR wins, it is rather for its use of the kernel trick (nonlinearities) than an alternative in-sample loss function.

		\begin{figure}
			\centering
\includegraphics[scale=.4]{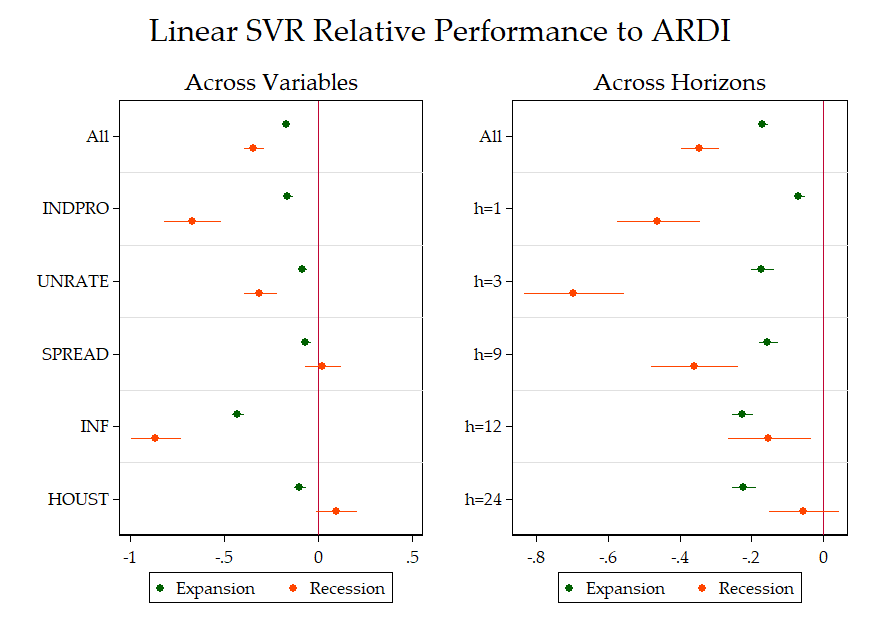}
\vspace{-0.3cm}
		\caption{\footnotesize{This graph displays the marginal (un)improvements by variables and horizons to opt for the SVR in-sample loss function in both recession and expansion periods. The unit of the x-axis are improvements in OOS $R^2$ over the basis model.  SEs are HAC. These are the 95\% confidence bands.}}\label{Leffectlin}
		\end{figure}

These results point out that an alternative $\hat{L}$ like the $\bar{\epsilon}$-insensitive loss function is not the most salient feature ML has to offer for macroeconomic forecasting. From a practical point of view, our results indicate that, on average, one can obtain the benefits of SVR and more by considering the much simpler KRR. This is convenient since obtaining the KRR forecast is a matter of less than 10 lines of codes implying the most straightforward form of linear algebra. In contrast, obtaining the SVR solution can be a serious numerical enterprise. 

\section{When are the ML Nonlinearities Important?}\label{BlackBox}

In this section we aim to explain some of the heterogeneity of ML treatment effects by interacting them in equation (\ref{e_eq2}) with few macroeconomic variables $\xi_t$ that have been used to explain main sources of 
observed nonlinear macroeconomic fluctuations. 
We focus  on NL feature only given its importance for both macroeconomic 
prediction and modeling. 

The first element in $\xi_t$ is  the Chicago Fed adjusted national financial conditions 
index (ANFCI). 
\cite{Adrian2019} find  that lower quantiles of GDP growth are time varying and are predictable by tighter financial conditions, suggesting  that higher order approximations are needed in general equilibrium models with 
financial frictions. 
In addition, \cite{Beaudry2018} build on the observation that recessions are preceded by accumulations of business, consumer and housing capital, while  \cite{Beaudry2020} add nonlinearities in the estimation part of a model with financial frictions and household capital accumulation.   
Therefore, we add to the list the house price growth (HOUSPRICE), measured by the S\&P/Case-Shiller U.S. National Home Price Index. 
The goal is then to test whether financial conditions and capital buildups interact with the nonlinear ML feature, and if they could explain its superior performance in macroeconomic forecasting. 

Uncertainty is also related to nonlinearity in macroeconomic modeling \citep{Bloom2009}. 
\cite{Benigno2013} provide a second-order approximation solution for a model with time-varying risk that has its own effect on endogenous 
variables. \cite{Gorodnichenko2017} find evidence on volatility factors that are persistent and load on the housing sector, while \cite{Carriero2018} 
estimate uncertainty and its effects in a large nonlinear VAR model. Hence, we include the Macro Uncertainty from \cite{Jurado2015} (MACROUNCERT).\footnote{We 
did not consider the Economic Policy Uncertainty from \cite{Baker2016} as it starts only from 1985.}  

Then we add measures of sentiments: University of Michigan Consumer Expectations (UMCSENT) and Purchasing Managers Index (PMI). \cite{Angelatos2013} 
and \cite{Benhabib2015} have  suggested that waves of pessimism and optimism play an important role in generating (nonlinear) macroeconomic 
fluctuations. In the case of \cite{Benhabib2015}, optimal decisions based on sentiments produce multiple self-fulfilling rational expectations equilibria. 
Consequently, including measures of sentiment in $\xi_t$ aims to test if this channel plays a role for nonlinearities in macro forecasting.  
Standard monetary VAR series are used as controls: UNRATE, PCE inflation (PCEPI) and one-year treasury rate (GS1).\footnote{We consider GS1 instead of the federal funds rate because of the long zero lower bound period. Time series  of elements in $\xi_t$ are plotted in figure \ref{xi_series}.}

Interactions are formed with $\xi_{t-h}$ to measure its impact  when the 
forecast is made. This  is of interest for practitioners as it indicates which macroeconomic conditions favor nonlinear ML forecast modeling. Hence, this expands the  equation (\ref{e_eq2}) to
\begin{align*} 
\forall m \in \mathcal{M}_{NL}: \quad  R^2_{t,h,v,m} = \dot{\alpha}_{NL} + \dot{\gamma} I(m \in NL) \xi_{t-h}  +  \dot{\phi}_{t,v,h} + \dot{u}_{t,h,v,m}
\end{align*}
where $ \mathcal{M}_{NL}$ is defined as the set of models that differs only by the use of NL. 

\begin{table}[t!]\centering
\def\sym#1{\ifmmode^{#1}\else\(^{#1}\)\fi}
\caption{Heterogeneity of NL treatment effect}\label{tab:level_0829}
\begin{tabular}{l*{4}{c}}
\hline\hline
                    &\multicolumn{1}{c}{(1)}&\multicolumn{1}{c}{(2)}&\multicolumn{1}{c}{(3)}&\multicolumn{1}{c}{(4)}\\
                    &\multicolumn{1}{c}{Base}&\multicolumn{1}{c}{All Horizons}&\multicolumn{1}{c}{Data-Rich}&\multicolumn{1}{c}{Last 20 years}\\
\hline
NL                  &       8.998\sym{***}&       5.808\sym{***}&       13.48\sym{***}&       19.87\sym{***}\\
                    &     (0.748)         &     (0.528)         &     (1.012)         &     (1.565)                \\
HOUSPRICE           &      -9.668\sym{***}&      -4.491\sym{***}&      -11.56\sym{***}&      -1.219         \\
                    &     (1.269)         &     (0.871)         &     (1.715)         &     (1.596)                 \\
ANFCI               &       7.244\sym{***}&       2.625         &       6.803\sym{**} &       20.29\sym{***}\\
                    &     (1.881)         &     (1.379)         &     (2.439)         &     (4.891)                \\
MACROUNCERT         &       17.98\sym{***}&       10.28\sym{***}&       34.87\sym{***}&       9.660\sym{***}\\
                    &     (1.875)         &     (1.414)         &     (2.745)         &     (2.038)                 \\
UMCSENT             &       4.695\sym{**} &       3.853\sym{**} &       10.29\sym{***}&      -3.625                 \\
                    &     (1.768)         &     (1.315)         &     (2.294)         &     (1.922)                 \\
PMI                 &      0.0787         &      -1.443         &      -2.048         &      -1.919                 \\
                    &     (1.179)         &     (0.879)         &     (1.643)         &     (1.288)               \\
UNRATE              &       0.834         &       2.517\sym{**} &       5.732\sym{***}&       8.526\sym{***}\\
                    &     (1.353)         &     (0.938)         &     (1.734)         &     (2.199)                 \\
GS1                 &      -14.24\sym{***}&      -9.500\sym{***}&      -17.30\sym{***}&       2.081         \\
                    &     (2.288)         &     (1.682)         &     (3.208)         &     (3.390)                 \\
PCEPI               &       5.953\sym{*}  &       6.814\sym{**} &      -1.142         &      -6.242             \\
                    &     (2.828)         &     (2.180)         &     (4.093)         &     (3.888)                 \\
\hline
Observations        &      136800         &      228000         &       68400         &       72300                \\
\hline\hline
\multicolumn{5}{l}{\footnotesize Standard errors in parentheses. \sym{*} \(p<0.05\), \sym{**} \(p<0.01\), \sym{***} \(p<0.001\)} 
\end{tabular}
\end{table}

The results are presented in table \ref{tab:level_0829}. The first column shows regression 
coefficients for $h=\{9, 12, 24 \}$, since nonlinearity has been found more important for longer horizons. The second column average across all horizons, while the third presents the results for data-rich models only. The last column shows the heterogeneity of NL treatments during last 20 years.  

Results show that macroeconomic uncertainty is a true game changer for ML nonlinearity as it improves its forecast accuracy by 34\% in the case of data-rich 
models. This means that if the macro uncertainty goes from -1 standard deviation to +1 standard deviation from its mean, the expected NL treatment effect (in terms OOS-$R^2$ difference) is 2*34=+68\%. Tighter financial conditions and a decrease in house prices are also positively correlated with a higher NL treatment, which 
supports the findings in  \cite{Adrian2019} and \cite{Beaudry2020}. It is particularly interesting that the effect of ANFCI reaches 20\% during last 20 years, while 
the impact of uncertainty decreases to less than 10\%, emphasizing that the determinant role of financial conditions in recent US macro history is also reflected in our results.  Waves of consumer optimism positively affect nonlinearities, especially with data-rich models. 

Among control variables, unemployment rate has a positive effect on nonlinearity. As expected, this suggests that the importance of nonlinearities is a cyclical feature. 
Lower interest rates also improve NL treatment by as much as 17\% in the data-rich setup. Higher inflation also leads to stronger gains from
ML nonlinearities, but mainly at shorter horizons and for data-poor models, as suggested by comparing specifications (2) and (3).  

These results document clear historical situations where NL consistently helps: (i) when the level of macroeconomic uncertainty is high and (ii)  during episodes of tighter financial conditions and housing bubble bursts.\footnote{\cite{Granziera2019} have exploited similar regression setup for model selection and found that `economic' forecasting models, AR augmented by few  macroeconomic indicators, outperform the time series models during turbulent times (recessions, tight financial conditions and high uncertainty).} Also, we note that effects are often bigger in the case of data-rich models. Hence, allowing nonlinear relationship between factors made of many predictors can capture better the complex relationships that characterize the episodes above.

These findings suggest that ML captures important macroeconomic nonlinearities, especially in the context of 
financial frictions and high macroeconomic uncertainty. They  can also serve as guidance for forecasters that use a portfolio of predictive models: one should put more weight on nonlinear specifications if economic 
conditions evolve as described above.


\section{Conclusion}\label{conclusion}

In this paper we have studied important  features driving the performance of machine learning techniques in the context of macroeconomic forecasting. 
We have considered many ML methods in a substantive POOS setup over 38 years for 5 key variables and 5  horizons.    
We have classified these models by ``features'' of machine learning: nonlinearities, regularization, cross-validation and alternative loss function. The data-rich and data-poor environments were 
considered. In order to recover their marginal effects on forecasting performance, we designed a series of experiments that easily allow  to identify the 
treatment effects of interest.  

The first result indicates that nonlinearities are the true game changer for the data-rich environment, as they improve substantially the 
forecasting accuracy for all macroeconomic variables in our exercise and especially when predicting  at long horizons. This gives a stark recommendation for practitioners. It recommends for most variables and horizons what is in the 
end a partially nonlinear factor model -- that is, factors are still obtained by PCA. The best of ML (at least of what considered here) can be obtained by 
simply generating the data for a standard ARDI model and then feed it into a ML nonlinear function of choice. 
The performance of nonlinear models is magnified  during  periods of high macroeconomic uncertainty, financial stress and housing bubble bursts.  
These findings suggest that Machine Learning is useful for macroeconomic forecasting by mostly capturing important nonlinearities that arise in the context of uncertainty and financial frictions.

The second result is that the standard factor model remains the best regularization. Alternative regularization schemes are most of the time harmful. 
Third, if cross-validation has to be applied to select models' features, 
the best practice is the standard K-fold. Finally, the standard $L_2$ is preferred to the $\bar{\epsilon}$-insensitive loss function for 
macroeconomic predictions. We found that most (if not all) the benefits from the use of SVR in fact comes from the nonlinearities it creates via the kernel trick rather than its use of an alternative loss function.


\onehalfspace

\setlength\bibsep{5pt}
\bibliographystyle{apalike}
\bibliography{references}

\clearpage

\doublespace

\appendix
\section{Detailed Overall Predictive Performance}\label{sec:rootMSE}

\begin{table}[!h]
\caption{Industrial Production: Relative Root MSPE}
\label{tab:RelMSPE_INDPRO}\vspace{-0.3cm}
\par
\begin{center}
{\scriptsize
\begin{tabular}{llllll|lllll}
\hline\hline
 & \multicolumn{5}{c}{Full Out-of-Sample} & \multicolumn{5}{|c}{NBER Recessions Periods} \\
Models & h=1 & h=3 & h=9 & h=12 & h=24 & \multicolumn{1}{|l}{h=1} & h=3 & h=9 & h=12 & h=24 \\
\hline
\multicolumn{3}{l}{Data-poor ($H_t^-$) models} \\
AR,BIC (RMSPE) & 0.0765 & \textbf{0.0515} & \textbf{0.0451} & \textbf{0.0428} & \textbf{0.0344} & 0.127 & 0.1014 & 0.0973 & 0.0898 & 0.0571 \\
AR,AIC & 0.991* & \textbf{1.000} & \textbf{0.999} & \textbf{1.000} & \textbf{1.000} & 0.987* & 1.000 & 1.000 & 1.000 & 1.000 \\
AR,POOS-CV & 0.999 & 1.021*** & \textbf{0.985}* & \textbf{1.001} & 1.032* & 1.01 & 1.023*** & 0.988* & 1.000 & 1.076** \\
AR,K-fold & 0.991* & \textbf{1.000} & \textbf{0.987}* & \textbf{1.000} & 1.033* & 0.987* & 1.000 & 0.992* & 1.000 & 1.078** \\
RRAR,POOS-CV & 1.003 & 1.041** & \textbf{0.989} & \textbf{0.993}* & \textbf{1.002} & 1.039** & 1.083** & 0.991 & 0.993 & 1.016** \\
RRAR,K-fold & 0.988** & \textbf{1.000} & \textbf{0.991} & \textbf{1.001} & \textbf{1.027} & 0.992 & 1.007** & 0.995 & 1.001** & 1.074** \\
RFAR,POOS-CV & 0.995 & 1.045 & \textbf{0.985} & \textbf{0.955} & \textbf{0.991} & 1.009 & 1.073 & 0.902*** & 0.890** & 0.983 \\
RFAR,K-fold & 0.995 & \textbf{1.020} & \textbf{0.960} & \textbf{0.930}** & \textbf{0.983} & 0.999 & 1.013 & 0.894*** & 0.887*** & 0.970* \\
KRR-AR,POOS-CV & 1.023 & 1.09 & \textbf{0.980} & \textbf{0.944} & \textbf{0.982} & 1.117 & 1.166* & 0.896** & 0.853*** & 0.903*** \\
KRR,AR,K-fold & \textbf{0.947}*** & \textbf{0.937}** & \textbf{0.936} & \underline{\textbf{0.910}}* & \textbf{0.959} & 0.922** & 0.902** & 0.835*** & \textbf{0.799}*** & 0.864*** \\
SVR-AR,Lin,POOS-CV & 1.134*** & 1.226*** & 1.114*** & 1.132*** & \textbf{0.952}* & 1.186** & 1.285*** & 1.079** & 1.034*** & 0.893*** \\
SVR-AR,Lin,K-fold & 1.069* & 1.159** & 1.055** & 1.042*** & 1.016*** & 1.268*** & 1.319*** & 1.067*** & 1.035*** & 1.013*** \\
SVR-AR,RBF,POOS-CV & 0.999 & 1.061*** & \textbf{1.020} & \textbf{1.048} & \textbf{0.980} & 1.062* & 1.082*** & 0.876*** & 0.941*** & 0.930*** \\
SVR-AR,RBF,K-fold & \textbf{0.978}* & \textbf{1.004} & 1.080* & 1.193** & 1.017*** & 0.992 & 1.009 & 0.989 & 1.016*** & 1.012*** \\
\hline
\multicolumn{3}{l}{Data-rich ($H_t^+$) models} \\
ARDI,BIC & \textbf{0.946}* & \textbf{0.991} & 1.037 & \textbf{1.004} & \textbf{0.968} & \textbf{0.801}*** & \textbf{0.807}*** & 0.887** & 0.833*** & 0.784*** \\
ARDI,AIC & \textbf{0.959}* & \textbf{0.968} & \textbf{1.017} & \textbf{0.998} & \textbf{0.943} & 0.840*** & \textbf{0.803}*** & 0.844** & \textbf{0.798}** & \textbf{0.768}*** \\
ARDI,POOS-CV & 0.994 & \textbf{1.015} & \textbf{0.984} & \textbf{0.968} & \textbf{0.966} & 0.896*** & \underline{\textbf{0.698}}*** & \underline{\textbf{0.773}}*** & \textbf{0.777}*** & 0.812*** \\
ARDI,K-fold & \textbf{0.940}* & \textbf{0.977} & \textbf{1.013} & \textbf{0.982} & \textbf{0.912}* & \underline{\textbf{0.787}}*** & \textbf{0.812}*** & 0.841** & \textbf{0.808}** & \textbf{0.762}*** \\
RRARDI,POOS-CV & \textbf{0.994} & \textbf{1.032} & \textbf{0.987} & \textbf{0.973} & \textbf{0.948} & 0.908** & \textbf{0.725}*** & 0.793*** & \textbf{0.778}*** & 0.861** \\
RRARDI,K-fold & \textbf{0.943}** & \textbf{0.977} & \textbf{0.986} & \textbf{0.990} & \textbf{0.921} & 0.847** & \textbf{0.718}*** & \textbf{0.794}*** & \textbf{0.796}*** & \underline{\textbf{0.702}}*** \\
RFARDI,POOS-CV & \textbf{0.948}** & \textbf{0.991} & \textbf{0.951} & \textbf{0.919}* & \textbf{0.899}** & 0.865** & \textbf{0.802}*** & 0.837*** & \textbf{0.782}*** & 0.819*** \\
RFARDI,K-fold & \textbf{0.953}** & \textbf{1.016} & \textbf{0.957} & \textbf{0.924}* & \underline{\textbf{0.890}}** & 0.889*** & \textbf{0.864}* & 0.846*** & \textbf{0.803}*** & \textbf{0.767}*** \\
KRR-ARDI,POOS-CV & 1.038 & \textbf{1.016} & \underline{\textbf{0.921}}* & \textbf{0.934} & \textbf{0.959} & 1.152* & 1.021 & 0.847*** & 0.814*** & 0.886** \\
KRR,ARDI,K-fold & \textbf{0.971} & \textbf{0.983} & \textbf{0.923}* & \textbf{0.914}* & \textbf{0.959} & 1.006 & 0.983 & 0.827*** & \textbf{0.793}*** & 0.848*** \\
$(B_1,\alpha=\hat{\alpha})$,POOS-CV & 1.014 & \textbf{1.001} & \textbf{1.023} & \textbf{0.996} & \textbf{0.946} & 1.067 & 0.956 & 0.979 & 0.916** & 0.855*** \\
$(B_1,\alpha=\hat{\alpha})$,K-fold & \textbf{0.957}** & \textbf{0.952} & \textbf{1.029} & 1.046 & 1.051 & 0.908** & 0.856*** & 0.874** & \textbf{0.816}*** & 0.890* \\
$(B_1,\alpha=1)$,POOS-CV & \textbf{0.971}* & \textbf{1.013} & 1.067* & \textbf{1.020} & \textbf{0.955} & 0.991 & 0.889 & 1.01 & 0.935* & 0.880** \\
$(B_1,\alpha=1)$,K-fold & \textbf{0.957}** & \textbf{0.952} & \textbf{1.029} & \textbf{1.046} & 1.051 & 0.908** & 0.856*** & 0.874** & \textbf{0.816}*** & 0.890* \\
$(B_1,\alpha=0)$,POOS-CV & 1.047 & 1.112** & \textbf{1.021} & 1.051 & \textbf{0.969} & 1.134* & 1.182** & 0.997 & 1.005 & 0.821*** \\
$(B_1,\alpha=0)$,K-fold & 1.025 & 1.056* & 1.065 & 1.082 & 1.052 & 1.032 & 0.974 & 0.923 & 0.929 & 0.847*** \\
$(B_2,\alpha=\hat{\alpha})$,POOS-CV & 1.061 & \textbf{0.968} & \textbf{0.975} & \textbf{0.999} & \textbf{0.923}** & 1.237 & 0.810*** & 0.889*** & 0.904** & 0.869** \\
$(B_2,\alpha=\hat{\alpha})$,K-fold & 1.098 & \textbf{0.949} & \textbf{0.993} & \textbf{0.974} & \textbf{0.970} & 1.332 & \textbf{0.801}*** & 0.896** & 0.851*** & \textbf{0.756}*** \\
$(B_2,\alpha=1)$,POOS-CV & \textbf{0.973} & 1.045 & \textbf{1.012} & \textbf{1.023} & \textbf{0.920}** & 1.034 & 1.033 & 0.997 & 0.957 & 0.839*** \\
$(B_2,\alpha=1)$,K-fold & \textbf{0.956}** & 1.022 & 1.032 & \textbf{1.025} & \textbf{0.990} & 0.961 & 0.935 & 0.959 & 0.913** & 0.809*** \\
$(B_2,\alpha=0)$,POOS-CV & \underline{\textbf{0.933}}*** & \textbf{0.955} & \textbf{0.972} & \textbf{0.937} & \textbf{0.913}** & 0.902** & \textbf{0.781}*** & 0.904** & 0.840*** & 0.807*** \\
$(B_2,\alpha=0)$,K-fold & \textbf{0.937}** & \textbf{0.927}** & \textbf{0.961} & \textbf{0.927} & \textbf{0.959} & 0.871*** & \textbf{0.787}*** & 0.858*** & \textbf{0.775}*** & 0.776*** \\
$(B_3,\alpha=\hat{\alpha})$,POOS-CV & \textbf{0.980} & \textbf{0.994} & \textbf{1.016} & 1.05 & \textbf{0.952} & 1.032 & 0.95 & 0.957 & 0.97 & 0.861*** \\
$(B_3,\alpha=\hat{\alpha})$,K-fold & \textbf{0.973}** & \textbf{0.946}** & 1.042 & \textbf{0.948} & \textbf{0.997} & 1.016 & 0.916** & 0.938 & 0.825*** & 0.827*** \\
$(B_3,\alpha=1)$,POOS-CV & \textbf{0.969}* & 1.053 & 1.053 & 1.080* & \textbf{0.956} & 0.972 & 0.946 & 1.002 & 1.014 & 0.906** \\
$(B_3,\alpha=1)$,K-fold & \textbf{0.946}*** & \underline{\textbf{0.913}}** & \textbf{0.994} & \textbf{0.976} & 1.01 & 0.924** & 0.829*** & 0.888* & \textbf{0.803}*** & 0.822*** \\
$(B_3,\alpha=0)$,POOS-CV & \textbf{0.976} & 1.049 & 1.04 & 1.063 & \textbf{0.973} & 1.034 & 1.061 & 0.997 & 0.932* & 0.846*** \\
$(B_3,\alpha=0)$,K-fold & 0.981 & 1.01 & 1.03 & \textbf{1.011} & \textbf{0.985} & 1.002 & 0.997 & 0.95 & \textbf{0.826}*** & 0.787*** \\
SVR-ARDI,Lin,POOS-CV & \textbf{0.989} & 1.165** & 1.216** & 1.193** & \textbf{1.034} & 0.915* & 0.900** & 1.006 & 0.862** & \textbf{0.778}*** \\
SVR-ARDI,Lin,K-fold & 1.109** & 1.367*** & \textbf{1.024} & \textbf{1.038} & \textbf{1.028} & 1.129 & 1.133 & \textbf{0.776}*** & \textbf{0.808}*** & \textbf{0.726}*** \\
SVR-ARDI,RBF,POOS-CV & \textbf{0.968}* & \textbf{0.986} & 1.100* & \textbf{0.960} & \textbf{0.936}* & 0.958 & 0.900* & 0.873** & \underline{\textbf{0.760}}*** & 0.820*** \\
SVR-ARDI,RBF,K-fold & \textbf{0.951}* & \textbf{0.946} & \textbf{0.993} & \textbf{0.952} & \textbf{1.001} & 0.860** & \textbf{0.793}*** & \textbf{0.806}*** & \textbf{0.777}*** & 0.791*** \\
\hline\hline
\end{tabular}
}
\end{center}
\vspace{-0.4cm}
{\scriptsize \emph{%
\singlespacing{Note: The numbers represent the relative. with respect to AR,BIC model. root MSPE. Models retained in model confidence set are in bold. the
minimum values are underlined. while
$^{***}$. $^{**}$. $^{*}$ stand for 1\%. 5\% and 10\% significance of
Diebold-Mariano test.}}}
\end{table}

\clearpage

\begin{table}[!h]
\caption{Unemployment rate: Relative Root MSPE}
\label{tab:RelMSPE_UNRATE}\vspace{-0.3cm}
\par
\begin{center}
{\scriptsize
\begin{tabular}{llllll|lllll}
\hline\hline
 & \multicolumn{5}{c}{Full Out-of-Sample} & \multicolumn{5}{|c}{NBER Recessions Periods} \\
Models & h=1 & h=3 & h=9 & h=12 & h=24 & \multicolumn{1}{|l}{h=1} & h=3 & h=9 & h=12 & h=24 \\
\hline
\multicolumn{3}{l}{Data-poor ($H_t^-$) models} \\
AR,BIC (RMSPE) & 1.9578 & 1.1905 & 1.0169 & 1.0058 & 0.869 & 2.5318 & 2.0826 & 1.8823 & 1.7276 & 1.0562 \\
AR,AIC & 0.991 & 0.984 & 0.988 & 0.993*** & 1.000 & 0.958 & 0.960** & 0.984* & 1.000 & 1.000 \\
AR,POOS-CV & 0.988 & 0.999 & 1.002 & 0.995 & 0.987 & 0.978 & 0.980** & 0.996 & 0.998 & 1.04 \\
AR,K-fold & 0.994 & 0.984 & 0.989 & 0.986*** & 0.991 & 0.956* & 0.960** & 0.998 & 1.000 & 1.038 \\
RRAR,POOS-CV & 0.989 & 1.000 & 1.002 & 0.990* & 0.972** & 0.984 & 0.988* & 0.997 & 0.991* & 1.001 \\
RRAR,K-fold & 0.988 & 0.982* & 0.983* & 0.989** & 0.999 & 0.963 & 0.971* & 0.992 & 0.995 & 1.033 \\
RFAR,POOS-CV & 0.983 & 0.995 & 0.968 & 1.000 & 1.002 & 0.989 & 1.003 & 0.929** & 0.951** & 0.994 \\
RFAR,K-fold & 0.98 & 0.985 & 0.979 & 1.006 & 0.99 & 0.985 & 0.972 & 0.896*** & 0.943* & 0.983 \\
KRR-AR,POOS-CV & 0.99 & 1.04 & \textbf{0.882}*** & \textbf{0.889}*** & 0.876*** & 1.04 & 1.116 & 0.843*** & 0.883*** & 0.904** \\
KRR,AR,K-fold & \textbf{0.940}*** & \textbf{0.910}*** & \textbf{0.878}*** & \textbf{0.869}*** & 0.852*** & 0.847*** & 0.838*** & \textbf{0.788}*** & \textbf{0.798}*** & 0.908** \\
SVR-AR,Lin,POOS-CV & 1.028 & 1.133** & 1.130*** & 1.108*** & 1.174*** & 1.065* & 1.274*** & 1.137*** & 1.094*** & 1.185*** \\
SVR-AR,Lin,K-fold & 0.993 & 1.061** & 1.068*** & 1.045*** & 1.013*** & 1.062** & 1.108*** & 1.032** & 1.011 & 1.018*** \\
SVR-AR,RBF,POOS-CV & 1.019 & 1.094* & 1.029 & 1.076** & 1.01 & 1.097** & 1.247** & 1.047* & 1.034*** & 1.112* \\
SVR-AR,RBF,K-fold & 0.997 & 1.011 & 1.078** & 1.053* & 0.993 & 1.026 & 1.009 & 1.058 & 1.023 & 0.985 \\
\hline
\multicolumn{3}{l}{Data-rich ($H_t^+$) models} \\
ARDI,BIC & \textbf{0.937}** & \textbf{0.893}** & 0.938 & 0.939 & 0.875*** & \textbf{0.690}*** & 0.715*** & \textbf{0.798}*** & 0.782*** & \textbf{0.783}*** \\
ARDI,AIC & \textbf{0.933}** & \textbf{0.878}*** & 0.928 & 0.953 & 0.893** & \textbf{0.720}*** & 0.719*** & \textbf{0.798}*** & 0.799*** & \textbf{0.787}*** \\
ARDI,POOS-CV & \textbf{0.924}*** & \textbf{0.913}* & 0.957 & 0.925* & \textbf{0.856}*** & \underline{\textbf{0.686}}*** & \textbf{0.676}*** & 0.840** & \textbf{0.737}*** & \textbf{0.777}*** \\
ARDI,K-fold & \textbf{0.935}** & \textbf{0.895}** & 0.929 & 0.93 & 0.915** & \textbf{0.696}*** & 0.697*** & \textbf{0.801}*** & 0.807*** & \textbf{0.787}*** \\
RRARDI,POOS-CV & \textbf{0.924}*** & \textbf{0.896}* & 0.968 & 0.946 & 0.870*** & \textbf{0.711}*** & \underline{\textbf{0.635}}*** & 0.849** & \textbf{0.768}*** & \textbf{0.767}*** \\
RRARDI,K-fold & \textbf{0.940}** & \textbf{0.899}** & 0.946 & 0.931* & 0.908** & \textbf{0.755}** & \textbf{0.681}*** & \textbf{0.803}*** & 0.790*** & \textbf{0.753}*** \\
RFARDI,POOS-CV & \textbf{0.934}*** & 0.945 & \textbf{0.857}*** & \textbf{0.842}*** & \underline{\textbf{0.763}}*** & \textbf{0.724}*** & 0.769*** & \underline{\textbf{0.718}}*** & \textbf{0.734}*** & \textbf{0.722}*** \\
RFARDI,K-fold & \textbf{0.932}*** & \textbf{0.897}*** & \textbf{0.873}** & \textbf{0.854}*** & \textbf{0.785}*** & \textbf{0.749}*** & 0.742*** & \textbf{0.731}*** & \underline{\textbf{0.720}}*** & \underline{\textbf{0.710}}*** \\
KRR-ARDI,POOS-CV & \textbf{0.959}* & \textbf{0.961} & \textbf{0.839}*** & \underline{\textbf{0.813}}*** & \textbf{0.804}*** & 1.01 & 1.017 & \textbf{0.748}*** & \textbf{0.732}*** & 0.828*** \\
KRR,ARDI,K-fold & \textbf{0.938}*** & \textbf{0.907}** & \underline{\textbf{0.827}}*** & \textbf{0.817}*** & \textbf{0.795}*** & 0.925 & 0.933 & \textbf{0.785}*** & \textbf{0.729}*** & \textbf{0.814}*** \\
$(B_1.\alpha=\hat{\alpha})$,POOS-CV & 0.979 & \textbf{0.945} & 0.976 & 0.953 & 0.913*** & 1.049 & 0.899* & 0.933 & 0.910* & 0.871*** \\
$(B_1.\alpha=\hat{\alpha})$,K-fold & 0.971 & \textbf{0.925}** & \textbf{0.867}*** & 0.919* & 0.925* & 0.787*** & 0.848*** & 0.840*** & 0.839*** & 0.829** \\
$(B_1.\alpha=1)$,POOS-CV & 0.947*** & \textbf{0.937}* & 0.962 & 0.922** & 0.889*** & 0.857** & 0.789*** & 0.888** & 0.860*** & 0.915* \\
$(B_1.\alpha=1)$,K-fold & 0.971 & \textbf{0.925}** & \textbf{0.867}*** & 0.919* & 0.925* & 0.787*** & 0.848*** & 0.840*** & 0.839*** & 0.829** \\
$(B_1.\alpha=0)$,POOS-CV & 1.238** & 1.319** & 1.021 & 1.07 & 1.01 & 1.393* & 1.476* & 0.979 & 0.972 & \textbf{0.764}*** \\
$(B_1.\alpha=0)$,K-fold & 1.246** & 0.994 & 1.062* & 1.077* & 1.018 & 1.322 & 0.963 & 0.991 & 0.933 & 0.802*** \\
$(B_2,\alpha=\hat{\alpha})$,POOS-CV & \underline{\textbf{0.907}}*** & \textbf{0.918}** & 0.926* & 0.936* & 0.911** & \textbf{0.756}*** & 0.767*** & 0.869** & 0.832*** & 0.808*** \\
$(B_2,\alpha=\hat{\alpha})$,K-fold & \textbf{0.917}*** & \textbf{0.900}*** & 0.915* & 0.931 & 0.974 & \textbf{0.728}*** & 0.777*** & 0.829*** & \textbf{0.738}*** & \textbf{0.713}*** \\
$(B_2,\alpha=1)$,POOS-CV & \textbf{0.914}*** & 0.955 & 1.057 & 1.011 & 0.883*** & 0.810*** & 0.830*** & 1.029 & 0.952 & 0.795*** \\
$(B_2,\alpha=1)$,K-fold & 0.97 & \textbf{0.901}** & 0.991 & 0.983 & 0.918** & 0.837** & 0.754*** & 0.903 & 0.833*** & \textbf{0.753}*** \\
$(B_2,\alpha=0)$,POOS-CV & \textbf{0.908}*** & \textbf{0.893}*** & 0.991 & 0.922* & 0.889*** & 0.781** & 0.769*** & 0.915 & 0.786*** & \textbf{0.788}*** \\
$(B_2,\alpha=0)$,K-fold & \textbf{0.949}** & \textbf{0.898}*** & 0.908** & 0.906** & 0.967 & 0.875 & 0.777*** & 0.817*** & \textbf{0.756}*** & \textbf{0.741}*** \\
$(B_3,\alpha=\hat{\alpha})$,POOS-CV & \textbf{0.949}** & \textbf{0.888}*** & 0.952 & 0.943 & 0.874*** & 0.933 & 0.843*** & 0.886** & 0.829*** & 0.827*** \\
$(B_3,\alpha=\hat{\alpha})$,K-fold & \textbf{0.937}** & \textbf{0.910}*** & \textbf{0.882}** & 0.923* & 0.921** & 0.836* & 0.831*** & 0.868*** & 0.839*** & \textbf{0.795}*** \\
$(B_3,\alpha=1)$,POOS-CV & \textbf{0.929}*** & \textbf{0.921}** & 0.958 & 0.983 & 0.884*** & 0.812** & 0.771*** & 0.864** & 0.851** & 0.845*** \\
$(B_3,\alpha=1)$,K-fold & 0.968 & 0.941* & \textbf{0.861}*** & 0.907* & 0.943 & 0.808** & 0.806*** & 0.832*** & 0.873** & \textbf{0.736}*** \\
$(B_3,\alpha=0)$,POOS-CV & \textbf{0.948}** & 0.974 & 0.994 & 1.066 & 0.946* & 0.979 & 1.03 & 0.956 & 0.877** & \textbf{0.799}*** \\
$(B_3,\alpha=0)$,K-fold & 0.969 & \textbf{0.918}*** & 0.983 & 0.998 & 0.945* & 0.963 & 0.901* & 0.957 & 0.912* & \textbf{0.730}*** \\
SVR-ARDI,Lin,POOS-CV & \textbf{0.960}* & 1.041 & 1.072 & 0.929 & 1.028 & 0.872 & 0.858* & 0.941 & 0.809*** & \textbf{0.779}*** \\
SVR-ARDI,Lin,K-fold & 0.959* & \underline{\textbf{0.873}}*** & \textbf{0.838}*** & 0.926 & 0.946 & 0.801** & 0.791*** & \textbf{0.756}*** & \textbf{0.800}** & 0.872* \\
SVR-ARDI,RBF,POOS-CV & \textbf{0.966} & \textbf{0.995} & 1.016 & 0.957 & 0.872*** & 0.938 & 0.859* & 0.937 & 0.786*** & \textbf{0.777}** \\
SVR-ARDI,RBF,K-fold & \textbf{0.943}** & 0.958 & \textbf{0.871}** & 0.911* & 0.930* & \textbf{0.769}*** & 0.796*** & \textbf{0.770}*** & \textbf{0.763}*** & \textbf{0.787}*** \\
\hline\hline
\end{tabular}
}
\end{center}
\vspace{-0.4cm}
{\scriptsize \emph{%
\singlespacing{Note: The numbers represent the relative, with respect to AR,BIC model, root MSPE. Models retained in model confidence set are in bold, the
minimum values are underlined, while
$^{***}$, $^{**}$, $^{*}$ stand for 1\%, 5\% and 10\% significance of
Diebold-Mariano test.}}}
\end{table}

\clearpage

\begin{table}[!h]
\caption{Term spread: Relative Root MSPE}
\label{tab:RelMSPE_GS1}\vspace{-0.3cm}
\par
\begin{center}
{\scriptsize
\begin{tabular}{llllll|lllll}
\hline\hline
 & \multicolumn{5}{c}{Full Out-of-Sample} & \multicolumn{5}{|c}{NBER Recessions Periods} \\
Models & h=1 & h=3 & h=9 & h=12 & h=24 & \multicolumn{1}{|l}{h=1} & h=3 & h=9 & h=12 & h=24 \\
\hline
\multicolumn{3}{l}{Data-poor ($H_t^-$) models} \\
AR,BIC (RMSPE) & \textbf{6.4792} & 12.8246 & \textbf{16.3575} & 20.0828 & 22.2091 & \textbf{13.3702} & 23.16 & 23.5697 & 31.597 & 23.0842 \\
AR,AIC & \textbf{1.002}* & 0.998 & \textbf{1.053}* & 1.034** & 1.041** & \textbf{1.002} & 1.001 & 1.034 & 0.993 & 0.972 \\
AR,POOS-CV & \textbf{1.055}* & 1.139* & \textbf{1.000} & 0.969 & 1.040** & \textbf{1.041} & 1.017 & 0.895* & 0.857* & 0.972 \\
AR,K-fold & \textbf{1.001} & 1.000 & \textbf{1.003} & 0.979 & 1.038* & \textbf{1.002} & 0.998 & 0.911 & 0.890* & 0.983 \\
RRAR,POOS-CV & \textbf{1.055}** & 1.142* & \textbf{1.004} & 0.998 & 1.016 & \textbf{1.036} & 1.014 & 0.899 & 0.966 & 0.945** \\
RRAR,K-fold & \textbf{1.044}* & 0.992 & \textbf{1.027} & 0.96 & 1.015 & \textbf{1.024} & 0.982 & 0.959 & 0.795** & 0.957* \\
RFAR,POOS-CV & \textbf{0.997} & 0.886 & 1.125*** & 1.019 & 1.107** & \textbf{0.906} & \textbf{0.816} & 1.039 & 0.747** & 1.077** \\
RFAR,K-fold & \textbf{0.991} & 0.941 & 1.136*** & 1.011 & 1.084** & \textbf{0.909} & 0.823 & 1.023 & 0.764* & 1.038 \\
KRR-AR,POOS-CV & 1.223** & 0.881 & \underline{\textbf{0.949}} & \textbf{0.888}** & 0.945* & \textbf{1.083} & \textbf{0.702} & \textbf{0.788}*** & 0.758*** & 0.948 \\
KRR,AR,K-fold & \textbf{1.141} & 0.983 & \textbf{1.098}** & 0.999 & 1.048 & \textbf{0.999} & \textbf{0.737} & 0.833* & \textbf{0.663}** & \textbf{0.924} \\
SVR-AR,Lin,POOS-CV & 1.158** & 1.326*** & \textbf{1.071}* & 1.045 & 1.045 & 1.111* & 1.072 & 0.894* & 0.828* & 0.967 \\
SVR-AR,Lin,K-fold & 1.191** & 1.056 & \textbf{1.018} & 0.963 & 0.993 & 1.061 & 1.009 & 0.886** & 0.845** & 0.916*** \\
SVR-AR,RBF,POOS-CV & \textbf{1.006} & 1.039 & \textbf{1.050}* & 0.951 & 0.969 & \textbf{0.964} & 0.902 & 0.876* & 0.761** & \textbf{0.864}*** \\
SVR-AR,RBF,K-fold & \textbf{0.985} & 0.911 & \textbf{1.038} & 0.946 & 0.933** & \textbf{0.990} & \textbf{0.737} & 0.851** & 0.747* & 0.968 \\
\hline
\multicolumn{3}{l}{Data-rich ($H_t^+$) models} \\
ARDI,BIC & \textbf{0.953} & 0.971 & \textbf{0.979} & 0.93 & \textbf{0.892}*** & \textbf{0.921} & 0.9 & \textbf{0.790}*** & \textbf{0.633}*** & 1.049 \\
ARDI,AIC & \textbf{0.970} & 0.956 & \textbf{1.019} & 0.944 & 0.917** & \textbf{0.929} & 0.867 & \textbf{0.814}*** & \textbf{0.647}*** & 1.076 \\
ARDI,POOS-CV & \textbf{0.954} & 1.015 & \textbf{1.067} & 0.991 & \textbf{0.915}** & \textbf{0.912} & 0.92 & 0.958 & 0.769** & 1.087 \\
ARDI,K-fold & \textbf{0.991} & 1.026 & \textbf{1.001} & 0.928 & 0.939 & \textbf{0.958} & 0.967 & 0.812*** & \textbf{0.662}*** & 1.041 \\
RRARDI,POOS-CV & \underline{\textbf{0.936}} & 0.994 & \textbf{1.078} & 0.991 & 0.964 & \textbf{0.896} & \textbf{0.850} & 0.952 & 0.784** & 1.092 \\
RRARDI,K-fold & \textbf{1.015} & 0.992 & \textbf{1.018} & 0.934 & 0.981 & \textbf{0.978} & 0.899 & 0.881* & \textbf{0.635}*** & 1.163* \\
RFARDI,POOS-CV & \textbf{0.988} & \underline{\textbf{0.830}}* & \textbf{0.957} & \textbf{0.873}** & \textbf{0.921}** & \underline{\textbf{0.804}} & \textbf{0.691} & \textbf{0.785}*** & \textbf{0.606}*** & 0.985 \\
RFARDI,K-fold & \textbf{1.010} & 0.883 & \textbf{0.997} & 0.909 & 0.935** & \textbf{0.808} & \textbf{0.778} & 0.827** & \textbf{0.626}*** & 0.97 \\
KRR-ARDI,POOS-CV & 1.355** & 0.898 & \textbf{0.993} & \textbf{0.856}** & \textbf{0.884}*** & \textbf{0.861} & \underline{\textbf{0.682}}* & \textbf{0.772}*** & \textbf{0.621}** & \textbf{0.905}* \\
KRR,ARDI,K-fold & 1.382*** & 0.96 & \textbf{0.974} & \underline{\textbf{0.827}}** & \textbf{0.862}*** & \textbf{0.858} & \textbf{0.684}* & \underline{\textbf{0.754}}*** & \underline{\textbf{0.569}}*** & 0.912* \\
$(B_1,\alpha=\hat{\alpha})$,POOS-CV & 1.114 & 1.06 & 1.126*** & 1.021 & \textbf{0.866}*** & \textbf{1.009} & 0.981 & 1.02 & \textbf{0.701}** & 1.012 \\
$(B_1,\alpha=\hat{\alpha})$,K-fold & \textbf{1.089} & 1.149** & 1.199** & 1.106* & 0.969 & \textbf{1.001} & 1.041 & 0.885 & 0.767** & 0.941 \\
$(B_1,\alpha=1)$,POOS-CV & 1.125* & 1.115 & 1.172*** & 1.072 & \underline{\textbf{0.844}}*** & 1.071 & 1.006 & 1.033 & 0.833 & 0.96 \\
$(B_1,\alpha=1)$,K-fold & \textbf{1.089} & 1.149** & 1.199** & 1.106* & 0.969 & \textbf{1.001} & 1.041 & 0.885 & 0.767** & 0.941 \\
$(B_1,\alpha=0)$,POOS-CV & 1.173** & 1.312** & 1.176*** & 1.088 & 0.978 & 1.089 & 1.065 & 0.981 & 0.799 & 0.966 \\
$(B_1,\alpha=0)$,K-fold & 1.163* & 1.059 & \textbf{1.069} & 0.929 & \textbf{0.921}** & \textbf{1.041} & 0.869 & \textbf{0.810}** & 0.729** & \textbf{0.880}* \\
$(B_2,\alpha=\hat{\alpha})$,POOS-CV & \textbf{1.025} & 0.993 & \textbf{1.101}** & 1.028 & \textbf{0.897}*** & \textbf{0.918} & 0.908 & 1.02 & \textbf{0.651}*** & 0.989 \\
$(B_2,\alpha=\hat{\alpha})$,K-fold & \textbf{0.976} & 0.954 & \textbf{1.098}* & 1.059 & 0.935* & \textbf{0.931} & 0.875 & 0.938 & 0.779* & 0.952 \\
$(B_2,\alpha=1)$,POOS-CV & 1.062 & 0.968 & \textbf{1.125}** & 1.049 & 0.926*** & \textbf{0.897} & 0.855 & 1.058 & 0.79 & 1.001 \\
$(B_2,\alpha=1)$,K-fold & \textbf{0.980} & 0.938 & \textbf{1.130}** & 1.01 & 0.950* & \textbf{0.948} & 0.858 & 0.976 & 0.679** & 1.001 \\
$(B_2,\alpha=0)$,POOS-CV & 1.118* & 1.082 & \textbf{1.097}** & 1.008 & \textbf{0.901}*** & \textbf{1.004} & 0.919 & 1.008 & \textbf{0.669}*** & 1.016 \\
$(B_2,\alpha=0)$,K-fold & 1.102 & 0.988 & \textbf{1.047} & 1.041 & \textbf{0.919}** & \textbf{0.985} & 0.909 & 0.870* & 0.757* & 0.986 \\
$(B_3,\alpha=\hat{\alpha})$,POOS-CV & \textbf{0.971} & 0.964 & \textbf{1.089}** & 1.076 & 0.933* & \textbf{0.887} & \textbf{0.837} & 0.908 & 0.783* & 0.904** \\
$(B_3,\alpha=\hat{\alpha})$,K-fold & \textbf{0.968} & 0.944 & \textbf{1.009} & 0.999 & \textbf{0.898}*** & \textbf{0.895} & 0.872 & 0.883** & 0.744** & \textbf{0.907}*** \\
$(B_3,\alpha=1)$,POOS-CV & \textbf{1.006} & 1.066 & \textbf{1.059}* & 1.039 & \textbf{0.896}*** & \textbf{0.894} & 1.131 & 0.974 & 0.764* & 0.987 \\
$(B_3,\alpha=1)$,K-fold & \textbf{0.994} & 0.924 & \textbf{1.037} & 0.96 & 0.975 & \textbf{0.934} & 0.852 & 0.834** & 0.712** & 1.01 \\
$(B_3,\alpha=0)$,POOS-CV & 1.181* & 0.961 & \textbf{1.104}** & 1.056 & 0.937** & 1.215 & 0.901 & 1.013 & 0.825 & 0.919* \\
$(B_3,\alpha=0)$,K-fold & \textbf{0.999} & 0.953 & \textbf{1.036} & 0.94 & 0.97 & \textbf{0.897} & 0.845 & 0.923 & 0.735** & 0.925** \\
SVR-ARDI,Lin,POOS-CV & \textbf{1.062} & 0.967 & \textbf{1.164}** & 1.113* & 1.065 & 1.016 & \textbf{0.762}* & 1.117 & 0.714** & 1.097 \\
SVR-ARDI,Lin,K-fold & \textbf{0.990} & 0.98 & \textbf{1.011} & \textbf{0.922} & \textbf{0.909}** & \textbf{0.935} & 0.885 & 0.825** & \textbf{0.667}** & 0.994 \\
SVR-ARDI,RBF,POOS-CV & \textbf{0.972} & 0.937 & \textbf{1.069} & 1.039 & 1.068 & \textbf{0.875} & \textbf{0.741} & \textbf{0.796}*** & 0.707*** & 1.204* \\
SVR-ARDI,RBF,K-fold & \textbf{1.018} & 0.938 & \textbf{1.123} & \textbf{0.914}* & \textbf{0.882}*** & \textbf{0.931} & \textbf{0.781} & 0.858** & 0.778** & \underline{\textbf{0.858}}** \\
\hline \hline
\end{tabular}
}
\end{center}
\vspace{-0.4cm}
{\scriptsize \emph{%
\singlespacing{Note: The numbers represent the relative, with respect to AR,BIC model, root MSPE. Models retained in model confidence set are in bold, the
minimum values are underlined, while
$^{***}$, $^{**}$, $^{*}$ stand for 1\%, 5\% and 10\% significance of
Diebold-Mariano test.}}}
\end{table}

\clearpage

\begin{table}[!h]
\caption{CPI Inflation: Relative Root MSPE}
\label{tab:RelMSPE_CPI}\vspace{-0.3cm}
\par
\begin{center}
{\scriptsize
\begin{tabular}{llllll|lllll}
\hline\hline
 & \multicolumn{5}{c}{Full Out-of-Sample} & \multicolumn{5}{|c}{NBER Recessions Periods} \\
Models & h=1 & h=3 & h=9 & h=12 & h=24 & \multicolumn{1}{|l}{h=1} & h=3 & h=9 & h=12 & h=24 \\
\hline
\multicolumn{3}{l}{Data-poor ($H_t^-$) models} \\
AR,BIC (RMSPE) & 0.0312 & 0.0257 & 0.0194 & 0.0187 & 0.0188 & 0.0556 & 0.0484 & 0.032 & 0.0277 & 0.0221 \\
AR,AIC & 0.969*** & 0.984 & 0.976* & 0.988 & 0.995 & 1.000 & 0.970** & 0.999 & 0.992 & 1.005 \\
AR,POOS-CV & 0.966** & 0.988 & 0.997 & 0.992 & 1.009 & 0.961** & 0.981 & 0.995 & 0.978 & 1.003 \\
AR,K-fold & 0.972** & 0.976** & 0.975* & 0.988 & 0.987 & 1.002 & 0.965*** & 0.998 & 0.992 & 1.005 \\
RRAR,POOS-CV & 0.969** & 0.984 & 0.99 & 0.993 & 1.006 & 0.961** & 0.982 & 0.995 & 0.963* & 0.998 \\
RRAR,K-fold & 0.964*** & 0.979** & 0.970* & 0.980* & 0.989 & 0.989 & 0.973** & 0.996 & 0.992 & 0.997 \\
RFAR,POOS-CV & 0.983 & \textbf{0.944}* & \textbf{0.909}* & \textbf{0.930} & 1.022 & 1.018 & 0.998 & 1.063 & 1.047 & 0.998 \\
RFAR,K-fold & \textbf{0.975} & \textbf{0.927}** & \textbf{0.909}* & \textbf{0.956} & 0.998 & 1.032 & 0.972 & 1.065 & 1.103 & 1.019 \\
KRR-AR,POOS-CV & \textbf{0.972} & \textbf{0.905}** & \textbf{0.872}** & \textbf{0.872}** & 0.907** & 1.023 & \textbf{0.930}** & 0.927 & 0.91 & 0.852* \\
KRR,AR,K-fold & \textbf{0.931}*** & \underline{\textbf{0.888}}*** & \underline{\textbf{0.836}}** & \underline{\textbf{0.827}}*** & 0.942 & 0.965 & \textbf{0.920}** & 0.92 & 0.915 & 0.975 \\
SVR-AR,Lin,POOS-CV & 1.119** & 1.291** & 1.210*** & 1.438*** & 1.417*** & 1.116 & 1.196** & 1.204** & 1.055 & 1.613*** \\
SVR-AR,Lin,K-fold & 1.239*** & 1.369** & 1.518*** & 1.606*** & 1.411*** & 1.159* & 1.326* & 1.459** & 1.501* & 1.016 \\
SVR-AR,RBF,POOS-CV & 0.988 & 1.004 & 1.086* & 1.068** & 1.127** & 0.999 & 1.004 & 0.969 & 1.091** & 1.501*** \\
SVR-AR,RBF,K-fold & 0.99 & 1.025 & 1.025 & 1.003 & 1.370*** & 0.965 & 0.979 & 0.996 & 0.896** & 1.553** \\
\hline
\multicolumn{3}{l}{Data-rich ($H_t^+$) models} \\
ARDI,BIC & 0.96 & \textbf{0.973} & 1.024 & \textbf{0.895}* & 0.880* & 0.919* & \textbf{0.906}* & 0.779* & 0.755** & 0.713** \\
ARDI,AIC & \textbf{0.954} & \textbf{0.990} & 1.034 & \textbf{0.895} & 0.884 & 0.925 & \textbf{0.898} & 0.778* & \textbf{0.736}** & 0.676** \\
ARDI,POOS-CV & \textbf{0.950} & \textbf{0.984} & 1.017 & \textbf{0.910} & 0.916 & 0.916* & \textbf{0.913}* & 0.832** & 0.781*** & \textbf{0.669}** \\
ARDI,K-fold & \textbf{0.941}* & \textbf{0.990} & 1.028 & \textbf{0.873}* & \textbf{0.858}* & 0.891** & \textbf{0.900} & 0.784* & \textbf{0.709}*** & \textbf{0.635}** \\
RRARDI,POOS-CV & \textbf{0.943}* & \textbf{0.975} & 1.001 & \textbf{0.917} & 0.914 & 0.905* & \textbf{0.912}* & 0.828** & \textbf{0.780}*** & \textbf{0.666}** \\
RRARDI,K-fold & \textbf{0.943}** & \textbf{0.983} & 1.022 & \textbf{0.875}* & \textbf{0.882} & 0.927* & \textbf{0.901} & \underline{\textbf{0.744}}** & \underline{\textbf{0.664}}*** & \underline{\textbf{0.613}}** \\
RFARDI,POOS-CV & \textbf{0.947}** & \textbf{0.908}*** & \textbf{0.853}** & \textbf{0.914}* & 0.979 & 0.976 & \textbf{0.939}** & 0.988 & 1.051 & 0.964 \\
RFARDI,K-fold & \textbf{0.936}*** & \textbf{0.907}*** & \textbf{0.854}** & \textbf{0.868}** & 0.909* & 0.962 & \textbf{0.933}** & 0.979 & 0.93 & 1.003 \\
KRR-ARDI,POOS-CV & 1.006 & 1.043 & 0.959 & 0.972 & 1.067 & 1.046 & 1.093 & 0.952 & 0.948 & 0.946 \\
KRR,ARDI,K-fold & \textbf{0.985} & 0.999 & 0.983 & 0.977 & 0.938 & 0.998 & 0.99 & 1.023 & 1.022 & 0.986 \\
$(B_1,\alpha=\hat{\alpha})$,POOS-CV & \textbf{0.918}** & \textbf{0.916}* & 0.976 & 0.96 & 1.026 & \underline{\textbf{0.803}}*** & \textbf{0.900}* & 0.8 & 0.848 & 0.974 \\
$(B_1,\alpha=\hat{\alpha})$,K-fold & \underline{\textbf{0.908}}** & \textbf{0.921}* & 1.012 & 1.056 & 1.092* & \textbf{0.823}** & \underline{\textbf{0.873}}* & \textbf{0.774} & 0.836 & 1.069 \\
$(B_1,\alpha=1)$,POOS-CV & \textbf{0.960} & \textbf{0.908}** & 1.11 & 1.03 & 1.076 & \textbf{0.813}** & \textbf{0.889}* & 0.794 & 0.825 & 0.989 \\
$(B_1,\alpha=1)$,K-fold & \textbf{0.908}** & \textbf{0.921}* & 1.012 & 1.056 & 1.092* & \textbf{0.823}** & \textbf{0.873}* & \textbf{0.774} & 0.836 & 1.069 \\
$(B_1,\alpha=0)$,POOS-CV & 0.971 & 1.035 & 1.114* & 1.048 & 1.263** & 0.848** & \textbf{0.906} & 0.935 & 0.881 & 0.99 \\
$(B_1,\alpha=0)$,K-fold & \textbf{0.945}* & 1.057 & 1.246** & 1.289** & 1.260*** & \textbf{0.850}*** & \textbf{0.939} & 0.954 & 0.944 & 1.095 \\
$(B_2,\alpha=\hat{\alpha})$,POOS-CV & \textbf{0.923}** & \textbf{0.956}** & \textbf{0.940} & 0.934 & 0.945 & 0.871* & 0.959 & 0.803* & 0.802* & 0.822* \\
$(B_2,\alpha=\hat{\alpha})$,K-fold & \textbf{0.921}** & \textbf{0.963}* & 0.995 & 0.956 & 1.037 & 0.868* & 0.957* & 0.817* & 0.778** & 0.861 \\
$(B_2,\alpha=1)$,POOS-CV & \textbf{0.942} & \textbf{0.959} & 1.158* & 1.174** & 1.151** & 0.877 & 0.927 & 0.799 & 0.907 & 1.087 \\
$(B_2,\alpha=1)$,K-fold & \textbf{0.922}** & \textbf{0.970} & 1.066 & 0.995 & 1.168* & 0.879 & 0.929 & 0.853 & 0.816* & 1.009 \\
$(B_2,\alpha=0)$,POOS-CV & \textbf{0.921}** & \textbf{0.940} & 1.079 & 0.959 & 1.071 & 0.857* & \textbf{0.881} & 1.129 & 0.883 & 0.851 \\
$(B_2,\alpha=0)$,K-fold & \textbf{0.919}** & \textbf{0.929}* & 0.997 & 1.011 & 1.212** & 0.865* & \textbf{0.883} & 0.825 & 0.961 & 0.853 \\
$(B_3,\alpha=\hat{\alpha})$,POOS-CV & \textbf{0.935}* & \textbf{0.941}*** & \textbf{0.961} & \textbf{0.849}** & 0.901* & 0.889* & \textbf{0.947}** & 0.791** & 0.785** & 0.808** \\
$(B_3,\alpha=\hat{\alpha})$,K-fold & \textbf{0.938}* & \textbf{0.952}** & \textbf{0.937} & \textbf{0.915} & 0.952 & 0.891* & 0.958* & 0.801* & 0.784** & 0.91 \\
$(B_3,\alpha=1)$,POOS-CV & \textbf{0.933}* & \textbf{0.960} & 1.076 & 1.000 & 1.017 & \textbf{0.856}* & \textbf{0.917}* & \textbf{0.755}* & \textbf{0.769}** & 0.86 \\
$(B_3,\alpha=1)$,K-fold & \textbf{0.943} & 0.978 & 1.006 & \textbf{0.894} & 1.002 & 0.889 & 0.946 & 0.805 & 0.806* & 0.879 \\
$(B_3,\alpha=0)$,POOS-CV & \textbf{0.946}* & \textbf{0.939}** & \textbf{0.896}* & \textbf{0.871}** & 1.022 & 0.894* & \textbf{0.931}** & 0.865 & 0.875 & 0.896 \\
$(B_3,\alpha=0)$,K-fold & \textbf{0.921}*** & \textbf{0.975} & \textbf{0.926} & \textbf{0.920} & 1.106 & 0.877*** & \textbf{0.936} & 0.839 & 0.892 & 1.147 \\
SVR-ARDI,Lin,POOS-CV & 1.148*** & 1.202* & 1.251*** & 1.209*** & 1.219** & 1.068 & 1.053 & 0.969 & 0.969 & 0.943 \\
SVR-ARDI,Lin,K-fold & 1.115*** & 1.390** & 1.197** & 1.114 & 1.177* & 1.058 & 1.295* & 0.944 & 0.954 & 1.036 \\
SVR-ARDI,RBF,POOS-CV & 0.963 & 1.031 & 1.002 & 0.962 & 0.951 & 0.922 & \textbf{0.915} & 0.848 & 0.861 & 0.996 \\
SVR-ARDI,RBF,K-fold & \textbf{0.951}** & 1.002 & 0.997 & 0.945 & \underline{\textbf{0.797}}*** & 0.927* & 0.964 & 0.816** & 0.826** & \textbf{0.659}** \\
\hline \hline
\end{tabular}
}
\end{center}
\vspace{-0.4cm}
{\scriptsize \emph{%
\singlespacing{Note: The numbers represent the relative, with respect to AR,BIC model, root MSPE. Models retained in model confidence set are in bold, the
minimum values are underlined, while
$^{***}$, $^{**}$, $^{*}$ stand for 1\%, 5\% and 10\% significance of
Diebold-Mariano test.}}}
\end{table}

\clearpage

\begin{table}[!h]
\caption{Housing starts: Relative Root MSPE}
\label{tab:RelMSPE_Hous}\vspace{-0.3cm}
\par
\begin{center}
{\scriptsize
\begin{tabular}{llllll|lllll}
\hline\hline
 & \multicolumn{5}{c}{Full Out-of-Sample} & \multicolumn{5}{|c}{NBER Recessions Periods} \\
Models & h=1 & h=3 & h=9 & h=12 & h=24 & \multicolumn{1}{|l}{h=1} & h=3 & h=9 & h=12 & h=24 \\
\hline
\multicolumn{3}{l}{Data-poor ($H_t^-$) models} \\
AR,BIC (RMSPE) & \textbf{0.9040} & \textbf{0.4142} & \textbf{0.2499} & 0.2198 & 0.1671 & \textbf{1.2526} & 0.6658 & 0.4897 & 0.4158 & 0.2954 \\
AR,AIC & \textbf{0.998} & \textbf{1.019} & \textbf{1.000} & \textbf{1.000} & 1.000 & 1.01 & 0.965* & 1.000 & 1.000 & 1.000 \\
AR,POOS-CV & \textbf{1.001} & \textbf{1.012} & \textbf{1.019}* & 1.01 & 1.036** & 1.015 & \underline{\textbf{0.936}}** & 1.011* & 1.013 & 1.057** \\
AR,K-fold & \textbf{0.993} & \textbf{1.017} & \textbf{1.001} & 1.000 & 1.02 & 1.01 & \textbf{0.951}** & 1.000 & 1.000 & 1.036 \\
RRAR,POOS-CV & \textbf{1.007} & \textbf{1.007} & \textbf{1.008} & 1.009 & 1.031** & 1.027* & \textbf{0.939}** & 1.001 & 1.013 & 1.050** \\
RRAR,K-fold & \textbf{0.999} & \textbf{1.014} & \textbf{0.998} & \textbf{0.998} & 1.024* & 1.013 & \textbf{0.941}** & 1.000** & 0.999 & 1.042** \\
RFAR,POOS-CV & 1.030*** & \textbf{1.026}* & \textbf{1.028}* & 1.045** & 1.018 & 1.023 & \textbf{0.941}* & \textbf{0.992} & 1.048* & 1.013 \\
RFAR,K-fold & 1.017* & \textbf{1.022} & \textbf{1.007} & 1.031** & 1.008 & 1.02 & \textbf{0.942}* & \textbf{0.990} & 1.026 & 1.01 \\
KRR-AR,POOS-CV & \textbf{0.995} & \textbf{0.999} & \textbf{0.969}* & 1.044* & 1.037* & \textbf{0.990} & \textbf{0.972} & \textbf{0.971} & 1.050** & 0.993 \\
KRR,AR,K-fold & \textbf{0.977}* & \textbf{0.975} & \textbf{0.957}** & \textbf{0.989} & 1.001 & \textbf{0.985} & \textbf{0.976} & 1.01 & 1.006 & 1.004 \\
SVR-AR,Lin,POOS-CV & 1.032*** & \textbf{0.997} & \textbf{1.044}*** & 1.064*** & 1.223** & 1.024* & \textbf{0.962}* & \textbf{0.986}* & 0.984 & \textbf{0.957}*** \\
SVR-AR,Lin,K-fold & 1.036*** & 1.031 & \textbf{1.002} & 1.006 & 1.002 & 1.013 & 0.976 & 1.002 & 1.009 & 1.004 \\
SVR-AR,RBF,POOS-CV & \textbf{1.008} & 1.047** & \textbf{1.023} & 1.035*** & 1.060*** & \textbf{1.014} & 0.981 & \textbf{0.947}*** & 1.015 & 1.017 \\
SVR-AR,RBF,K-fold & 1.009 & \textbf{1.011} & \textbf{1.012}** & 1.020*** & 1.034** & 1.021* & 0.969* & 1.010*** & 1.017** & 1.001 \\
\hline
\multicolumn{3}{l}{Data-rich ($H_t^+$) models} \\
ARDI,BIC & \underline{\textbf{0.973}}* & \textbf{0.989} & \textbf{1.031} & 1.051 & 1.05 & \textbf{0.946} & 1.139 & 1.048 & 0.988 & 0.944 \\
ARDI,AIC & \textbf{0.992} & \textbf{0.995} & \textbf{1.018} & 1.06 & 1.078 & \textbf{1.000} & 1.113 & 1.025 & 1.025 & 0.96 \\
ARDI,POOS-CV & 1.01 & \textbf{1.007} & \textbf{1.080} & 1.027 & 0.998 & 1.023 & 1.128 & 1.054 & 1.015 & 1.021 \\
ARDI,K-fold & \textbf{0.992} & \textbf{0.984} & \textbf{1.026} & 1.061 & 1.094 & \textbf{1.011} & 1.093 & 1.027 & 1.027 & 0.958 \\
RRARDI,POOS-CV & \textbf{0.998} & \textbf{1.007} & \textbf{1.043} & \textbf{0.996} & 1.082 & 1.008 & 1.119 & 1.041 & 0.991 & 1.022 \\
RRARDI,K-fold & \textbf{0.998} & \textbf{0.988} & \textbf{1.051} & 1.064 & 1.089 & 1.017 & 1.118 & 1.033 & 0.998 & 0.941 \\
RFARDI,POOS-CV & \textbf{0.997} & \textbf{0.944}** & \textbf{0.930}** & \textbf{0.920}* & 0.899** & \textbf{0.982} & \textbf{0.971} & \textbf{0.965} & 0.957 & 0.972 \\
RFARDI,K-fold & \textbf{0.994} & \textbf{0.962} & \textbf{0.939}* & \underline{\textbf{0.914}}* & \underline{\textbf{0.838}}*** & \textbf{0.993} & 0.985 & \textbf{0.986} & \textbf{0.943} & 0.902* \\
KRR-ARDI,POOS-CV & \textbf{0.980} & \underline{\textbf{0.943}}*** & \underline{\textbf{0.915}}** & \textbf{0.942}** & \textbf{0.884}*** & \underline{\textbf{0.941}}* & \textbf{0.952}* & \textbf{0.949} & 0.964** & 0.986 \\
KRR,ARDI,K-fold & \textbf{0.982}** & \textbf{0.949}** & \textbf{0.928} & \textbf{0.933} & \textbf{0.889}** & \textbf{0.973} & \textbf{0.973} & \textbf{1.003} & 1.022 & 0.994 \\
$(B_1.\alpha=\hat{\alpha})$,POOS-CV & \textbf{1.006} & \textbf{1.000} & \textbf{1.063} & 1.016 & \textbf{0.895}** & 1.023 & 1.099 & \textbf{0.985} & 1.026 & 1.022 \\
$(B_1.\alpha=\hat{\alpha})$,K-fold & 1.040* & 1.095** & 1.250** & 1.335** & 1.151* & 1.096* & 1.152** & 1.021 & 1.127 & \textbf{0.890} \\
$(B_1.\alpha=1)$,POOS-CV & 1.032** & 1.039 & 1.155 & 1.045 & 0.949 & 1.013 & 1.063 & \textbf{0.961} & 1.025 & 1.062 \\
$(B_1.\alpha=1)$,K-fold & 1.040* & 1.095** & 1.250** & 1.335** & 1.151* & 1.096* & 1.152** & 1.021 & 1.127 & \textbf{0.890} \\
$(B_1.\alpha=0)$,POOS-CV & \textbf{0.982} & \textbf{0.977} & 1.084 & 1.337** & 0.959 & \textbf{0.999} & 1.017 & 1.014 & 1.152** & 0.964 \\
$(B_1.\alpha=0)$,K-fold & \textbf{0.982} & \textbf{1.006} & 1.137* & 1.158** & 1.007 & \textbf{0.994} & 1.03 & \textbf{1.017} & 1.067 & \underline{\textbf{0.809}}** \\
$(B_2.\alpha=\hat{\alpha})$,POOS-CV & 1.044 & \textbf{0.992} & \textbf{0.975} & 0.988 & 0.969 & 1.177 & 1.126* & 1.034 & 0.989 & 0.972 \\
$(B_2.\alpha=\hat{\alpha})$,K-fold & \textbf{0.988} & \textbf{1.003} & \textbf{1.069} & 1.193** & 1.069 & 1.11 & 1.188* & 1.085 & 1.133* & 0.917 \\
$(B_2.\alpha=1)$,POOS-CV & \textbf{1.001} & \textbf{1.000} & \textbf{0.967} & 1.02 & 0.940* & \textbf{0.961} & 1.047 & \textbf{0.943} & 0.985 & 1.006 \\
$(B_2.\alpha=1)$,K-fold & \textbf{0.989} & 1.095 & 1.245** & 1.203* & 1.093 & 1.007 & 1.322*** & 1.1 & \textbf{0.919} & \textbf{0.848}** \\
$(B_2.\alpha=0)$,POOS-CV & 1.091* & \textbf{0.949} & \textbf{0.987} & \textbf{0.971} & 0.939 & 1.255 & 1.027 & \textbf{0.992} & \textbf{0.956} & 0.994 \\
$(B_2.\alpha=0)$,K-fold & 1.066 & \textbf{1.068} & 1.19 & 1.044 & 1.064 & 1.248 & 1.332** & 1.057 & \underline{\textbf{0.896}}*** & 0.917 \\
$(B_3,\alpha=\hat{\alpha})$,POOS-CV & 1.009 & \textbf{0.951}* & \textbf{0.935} & 0.99 & \textbf{0.891}** & 1.028 & 1.019 & \textbf{0.958} & \textbf{0.963} & 0.987 \\
$(B_3,\alpha=\hat{\alpha})$,K-fold & \textbf{0.998} & \textbf{0.977} & \textbf{1.007} & 1.055 & 1.044 & 1.019 & 1.115 & 1.017 & \textbf{0.979} & \textbf{0.882}* \\
$(B_3,\alpha=1)$,POOS-CV & \textbf{0.997} & \textbf{0.975} & \textbf{1.024} & 0.996 & 0.928* & \textbf{0.976} & 1.001 & 1.021 & \textbf{0.940} & 1.001 \\
$(B_3,\alpha=1)$,K-fold & 1.013 & \textbf{1.040} & 1.071 & 1.106 & 1.145 & 1.042 & 1.219* & 1.036 & 0.992 & 1.009 \\
$(B_3,\alpha=0)$,POOS-CV & 1.022* & \textbf{0.951}* & \textbf{0.962} & \textbf{0.944} & 0.932* & 1.022 & 0.981 & \underline{\textbf{0.930}} & \textbf{0.915}** & 1.001 \\
$(B_3,\alpha=0)$,K-fold & 1.030** & \textbf{1.003} & \textbf{1.005} & 1.011 & 1.029 & \textbf{0.986} & 1.114 & \textbf{0.998} & \textbf{0.955} & 0.934 \\
SVR-ARDI,Lin,POOS-CV & \textbf{0.998} & 1.078* & 1.154* & 1.137* & 1.142 & 1.047 & 1.111 & \textbf{0.989} & 1.009 & 1.111 \\
SVR-ARDI,Lin,K-fold & \textbf{0.992} & \textbf{0.971} & \textbf{1.017} & 1.038 & 1.11 & 1.007 & 1.021 & \textbf{0.988} & \textbf{0.937} & 0.959 \\
SVR-ARDI,RBF,POOS-CV & \textbf{0.991} & \textbf{1.004} & \textbf{1.010} & 1.044 & 1.034 & \textbf{0.987} & 1.095 & \textbf{0.981} & 0.969 & 1.096 \\
SVR-ARDI,RBF,K-fold & \textbf{1.003} & \textbf{0.998} & \textbf{1.045} & 1.078 & 1.162* & 1.022 & 1.081 & 1.03 & 0.984 & 1.026 \\
\hline \hline
\end{tabular}
}
\end{center}
\vspace{-0.4cm}
{\scriptsize \emph{%
\singlespacing{Note: The numbers represent the relative, with respect to AR,BIC model, root MSPE. Models retained in model confidence set are in bold, the
minimum values are underlined, while
$^{***}$, $^{**}$, $^{*}$ stand for 1\%, 5\% and 10\% significance of
Diebold-Mariano test.}}}
\end{table}

\clearpage

\section{Robustness of Treatment Effects Graphs}\label{sec:robust}

\begin{figure}[!h]
\centering
\includegraphics[scale=.35]{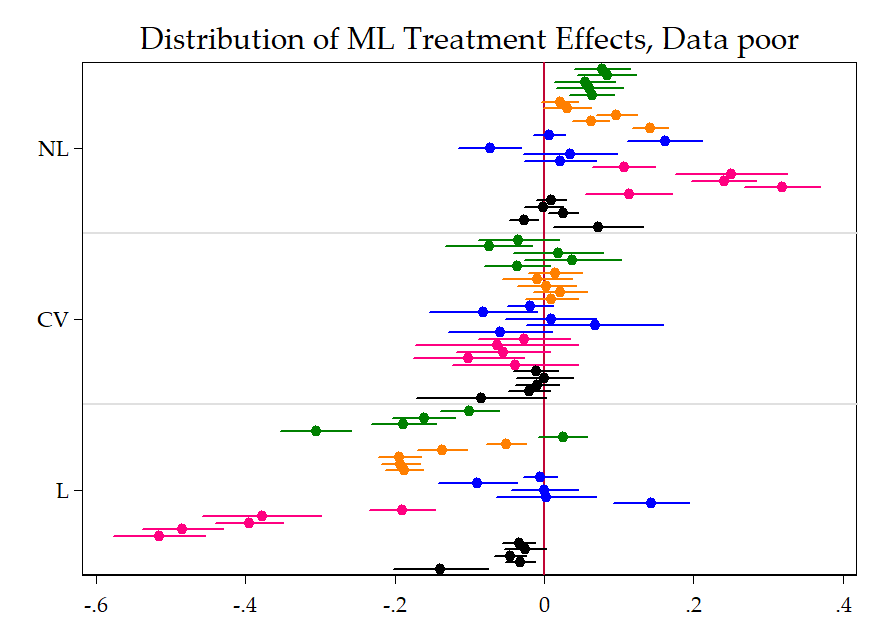}
\vspace{-0.5cm}
	\caption{\footnotesize{This figure plots the distribution of $\dot{\alpha}_F^{(h,v)}$ from equation \ref{r2_eq} done by $(h,v)$ subsets. The subsample under consideration here is \textbf{data-poor models}. The unit of the x-axis are improvements in OOS $R^2$ over the basis model. Variables are \textcolor{Green}{INDPRO}, \textcolor{orange}{UNRATE}, \textcolor{blue}{SPREAD}, \textcolor{magenta}{INF} and HOUST. Within a specific color block, the horizon increases from $h=1$ to $h=24$ as we are going down.  SEs are HAC. These are the 95\% confidence bands.}} 
	\label{dist_datap}
\end{figure}

\begin{figure}[!h]
\centering
\includegraphics[scale=.35]{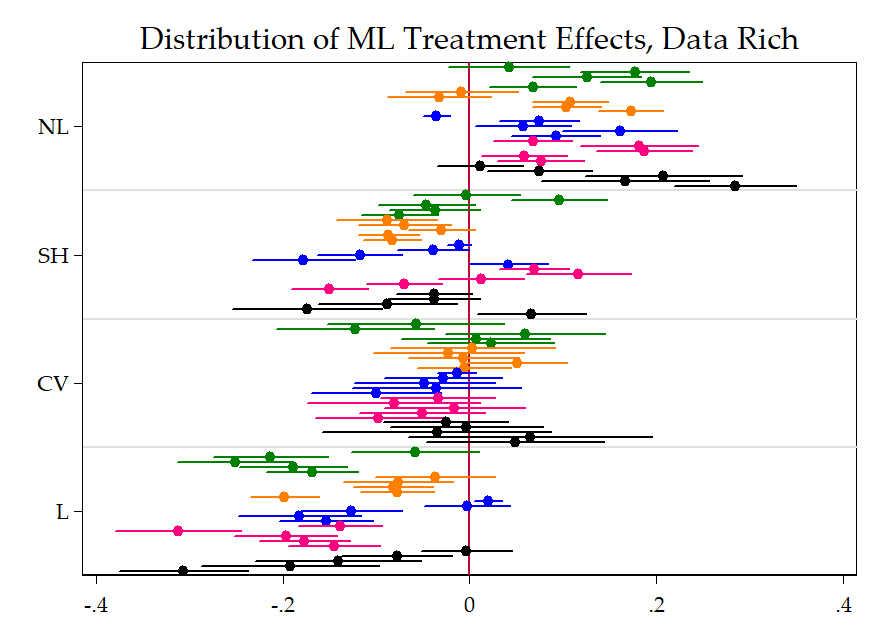}
\vspace{-0.5cm}
	\caption{\footnotesize{This figure plots the distribution of $\dot{\alpha}_F^{(h,v)}$ from equation \ref{r2_eq} done by $(h,v)$ subsets. The subsample under consideration here is \textbf{data-rich models}. The unit of the x-axis are improvements in OOS $R^2$ over the basis model. Variables are \textcolor{Green}{INDPRO}, \textcolor{orange}{UNRATE}, \textcolor{blue}{SPREAD}, \textcolor{magenta}{INF} and HOUST. Within a specific color block, the horizon increases from $h=1$ to $h=24$ as we are going down.  SEs are HAC. These are the 95\% confidence bands.}} 
		\label{dist_datar}
\end{figure}

\begin{figure}[!h]
\centering
\includegraphics[scale=.35]{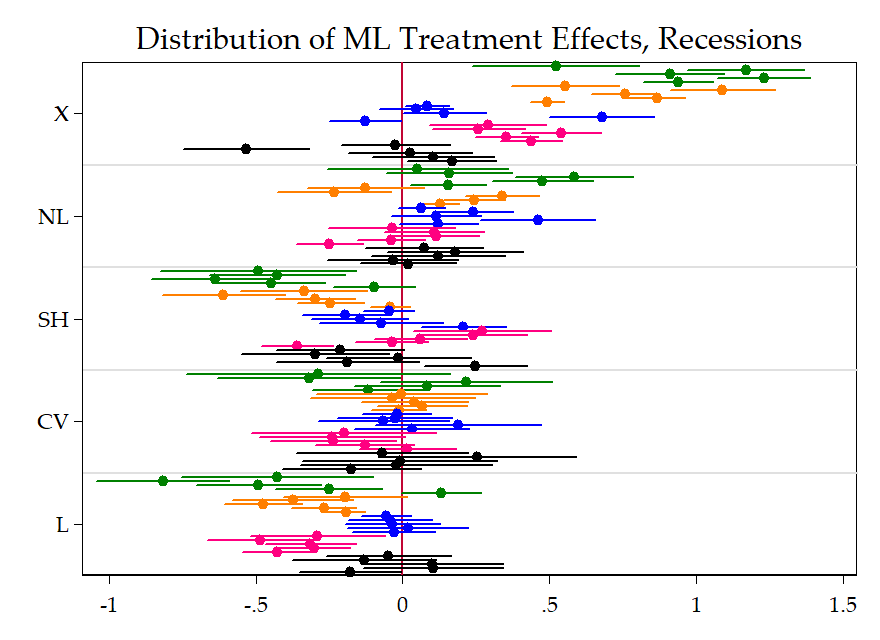}
\vspace{-0.5cm}
	\caption{\footnotesize{This figure plots the distribution of $\dot{\alpha}_F^{(h,v)}$ from equation \ref{r2_eq} done by $(h,v)$ subsets. The subsample under consideration here are \textbf{recessions}. The unit of the x-axis are improvements in OOS $R^2$ over the basis model. Variables are \textcolor{Green}{INDPRO}, \textcolor{orange}{UNRATE}, \textcolor{blue}{SPREAD}, \textcolor{magenta}{INF} and HOUST. Within a specific color block, the horizon increases from $h=1$ to $h=24$ as we are going down.  SEs are HAC. These are the 95\% confidence bands.}} 
	\label{dist_rec}
\end{figure}

\begin{figure}[!h]
\centering
\includegraphics[scale=.35]{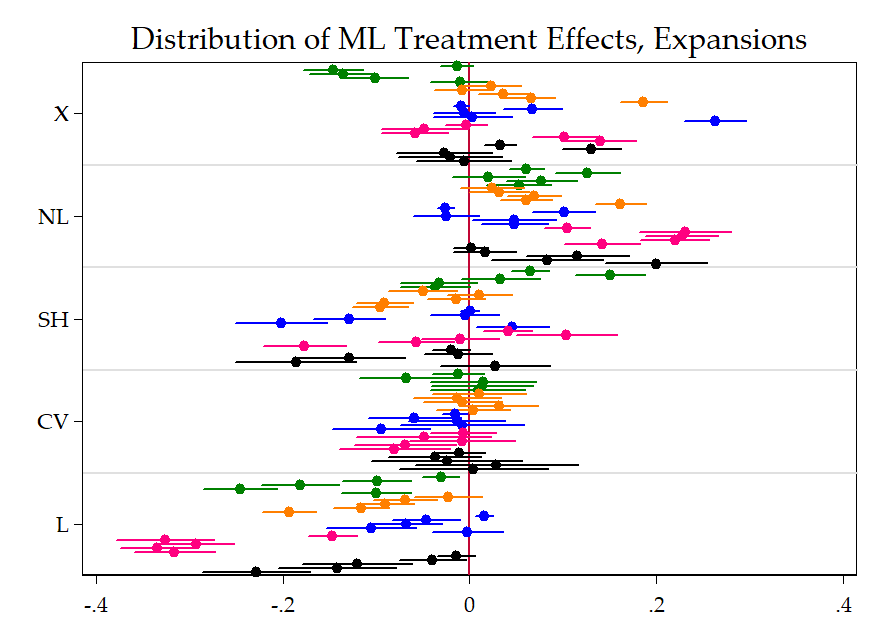}
\vspace{-0.5cm}
	\caption{\footnotesize{This figure plots the distribution of $\dot{\alpha}_F^{(h,v)}$ from equation \ref{r2_eq} done by $(h,v)$ subsets. The subsample under consideration here are \textbf{expansions}. The unit of the x-axis are improvements in OOS $R^2$ over the basis model. Variables are \textcolor{Green}{INDPRO}, \textcolor{orange}{UNRATE}, \textcolor{blue}{SPREAD}, \textcolor{magenta}{INF} and HOUST. Within a specific color block, the horizon increases from $h=1$ to $h=24$ as we are going down.  SEs are HAC. These are the 95\% confidence bands.}} 
	\label{dist_exp}
\end{figure}

\begin{figure}[!h]
\centering
\includegraphics[scale=.35]{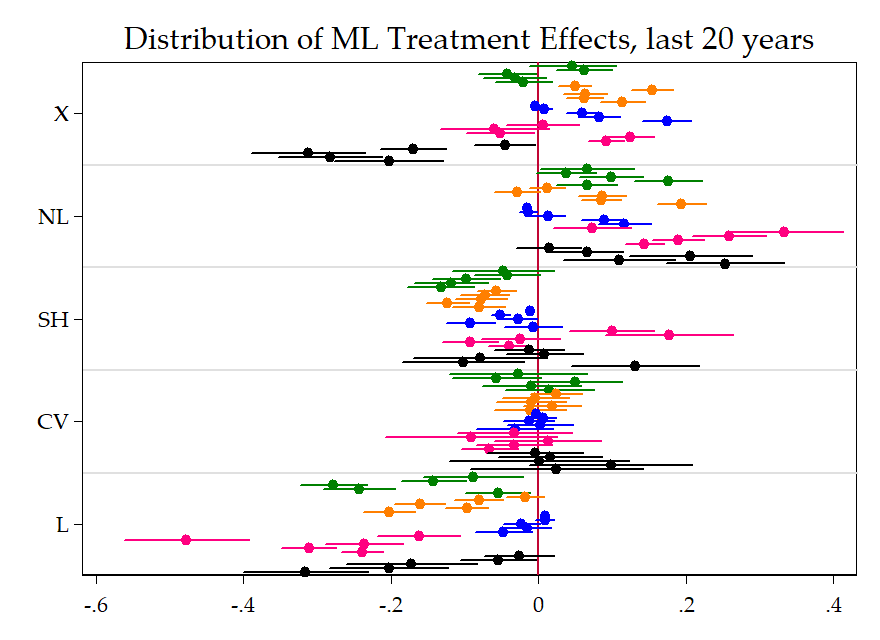}
\vspace{-0.5cm}
	\caption{\footnotesize{This figure plots the distribution of $\dot{\alpha}_F^{(h,v)}$ from equation \ref{r2_eq} done by $(h,v)$ subsets. The subsample under consideration here are \textbf{the last 20 years}. The unit of the x-axis are improvements in OOS $R^2$ over the basis model. Variables are \textcolor{Green}{INDPRO}, \textcolor{orange}{UNRATE}, \textcolor{blue}{SPREAD}, \textcolor{magenta}{INF} and HOUST. Within a specific color block, the horizon increases from $h=1$ to $h=24$ as we are going down.  SEs are HAC. These are the 95\% confidence bands.}} 
	\label{dist_last20yrs}
\end{figure}

\newpage
\section{Additional Results}\label{sec:addgraphs}

\begin{figure}
\centering
\includegraphics[scale=.6]{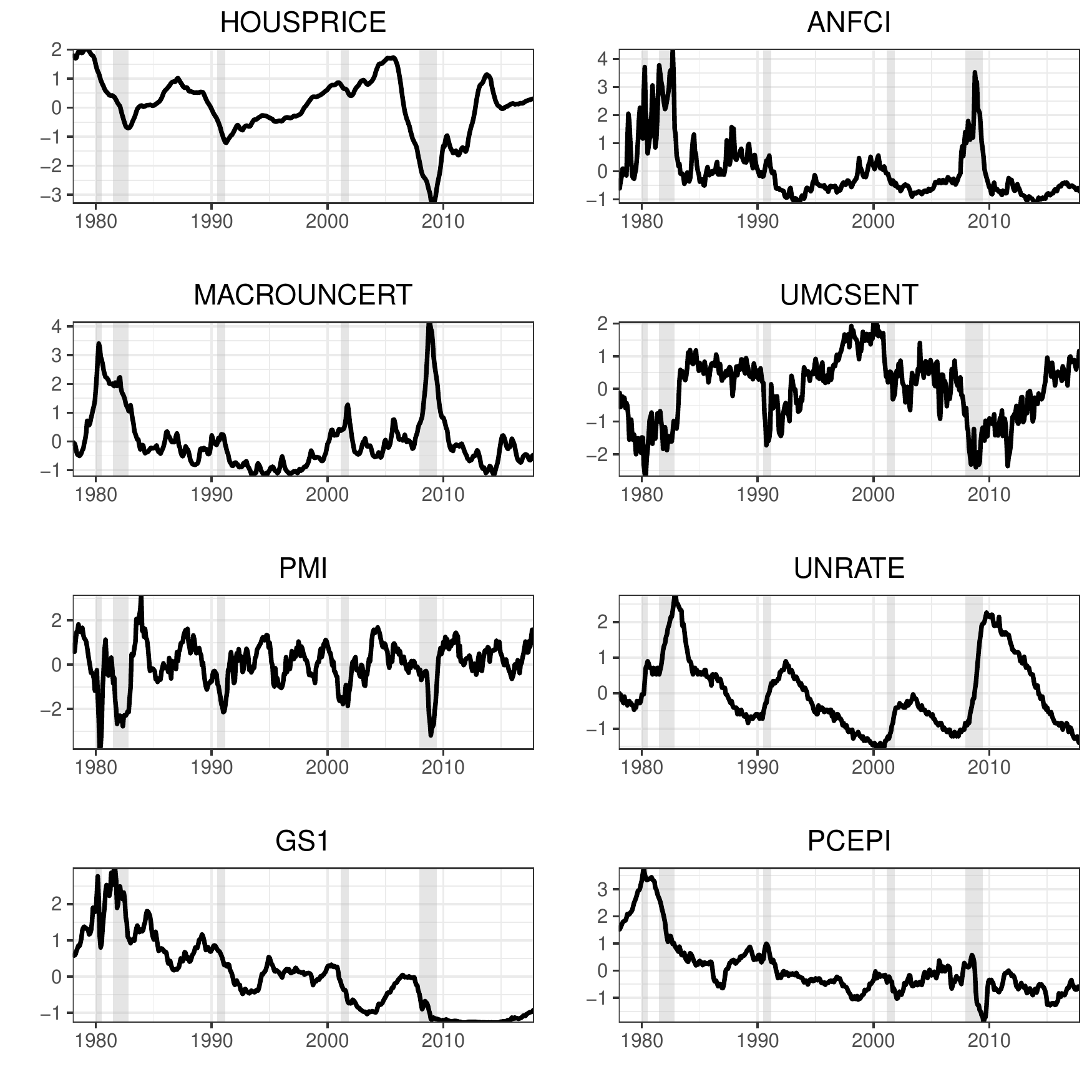}
	\caption{\footnotesize{This figure plots time series of variables explaining the heterogeneity of NL treatment effects in section \ref{BlackBox}.}}
	\label{xi_series}
\end{figure}

\begin{figure}
\centering
\vspace{-0.cm}\hspace*{-2.5cm} \includegraphics[width=\dimexpr\textwidth+4cm%
\relax,height=0.85%
\textheight]{{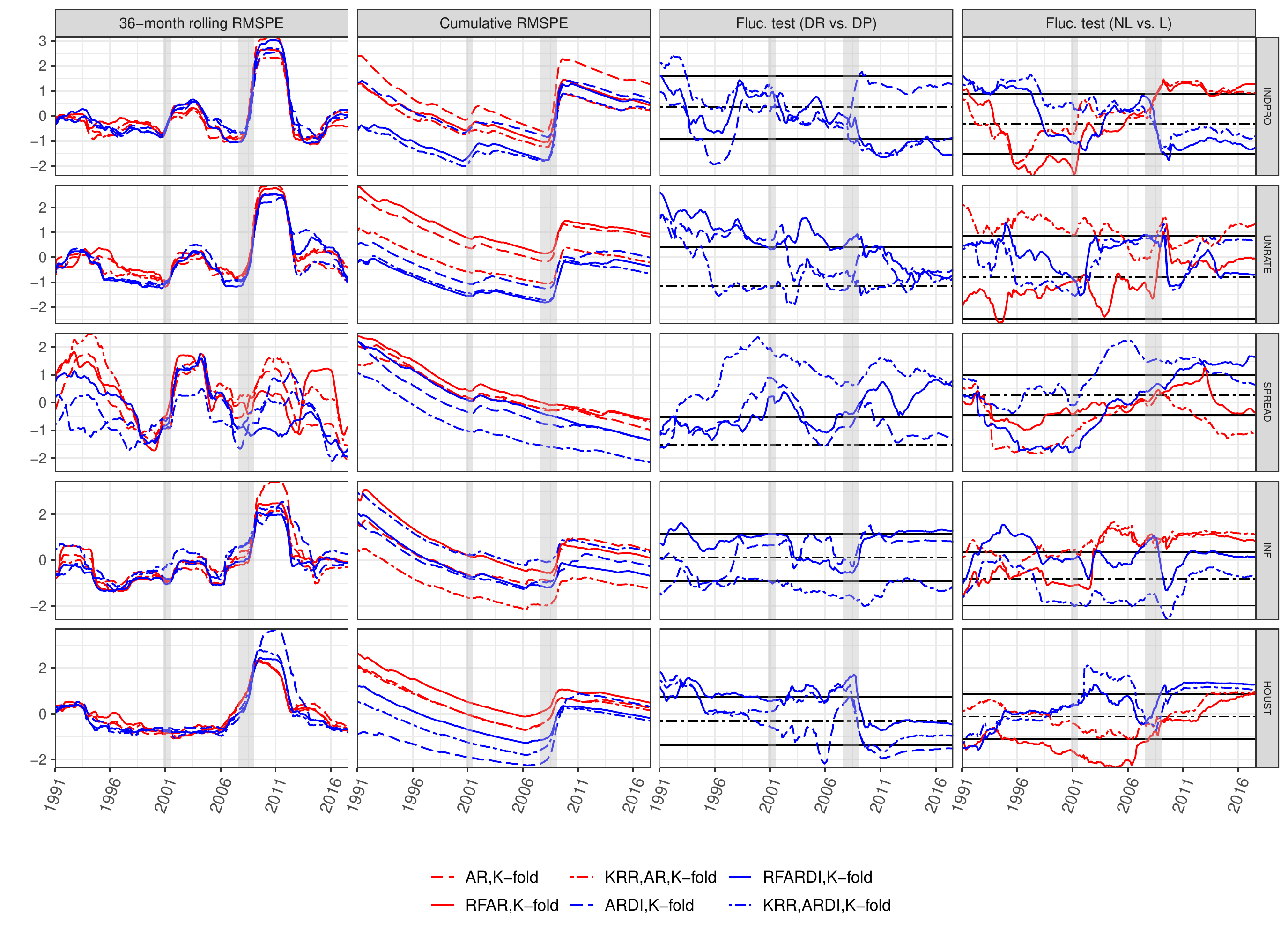}} 
\caption{\footnotesize{This figure shows the 3-year rolling window root MSPE, the cumulative root MSPE and \cite{Giacomini2010} fluctuation tests for linear and nonlinear 
data-poor and data-rich models, at 12-month horizon.}}
\label{TV_MSPE}
\end{figure}

\begin{figure}
\centering
\includegraphics[scale=.35]{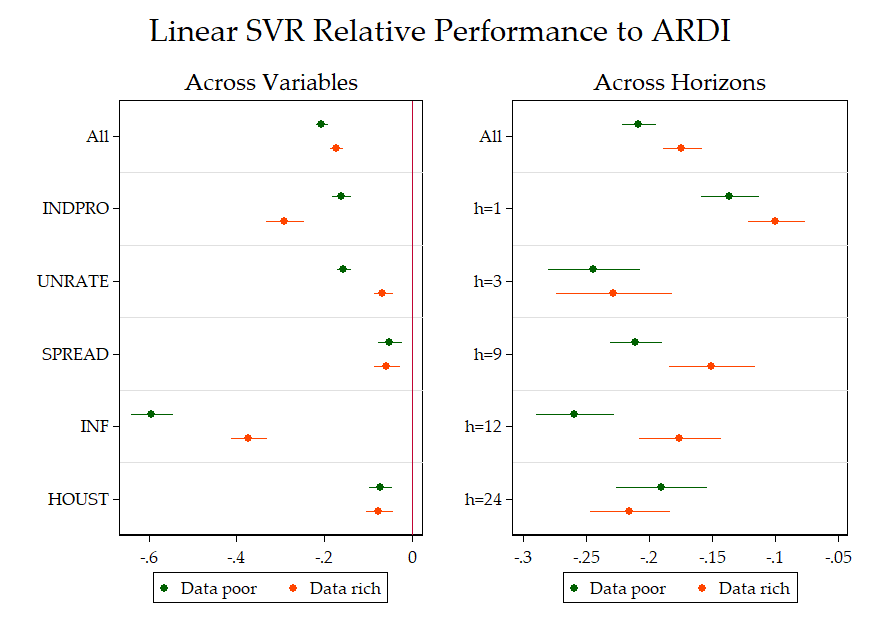}
	\caption{\footnotesize{This graph display the marginal (un)improvements by variables and horizons to opt for the SVR in-sample loss function in comparing the data-poor and data-rich environments for linear models. The unit of the x-axis are improvements in OOS $R^2$ over the basis model.  SEs are HAC. These are the 95\% confidence bands.}}
	\label{SVRlin_byX}
\end{figure}

\begin{figure}
\centering
\includegraphics[scale=.35]{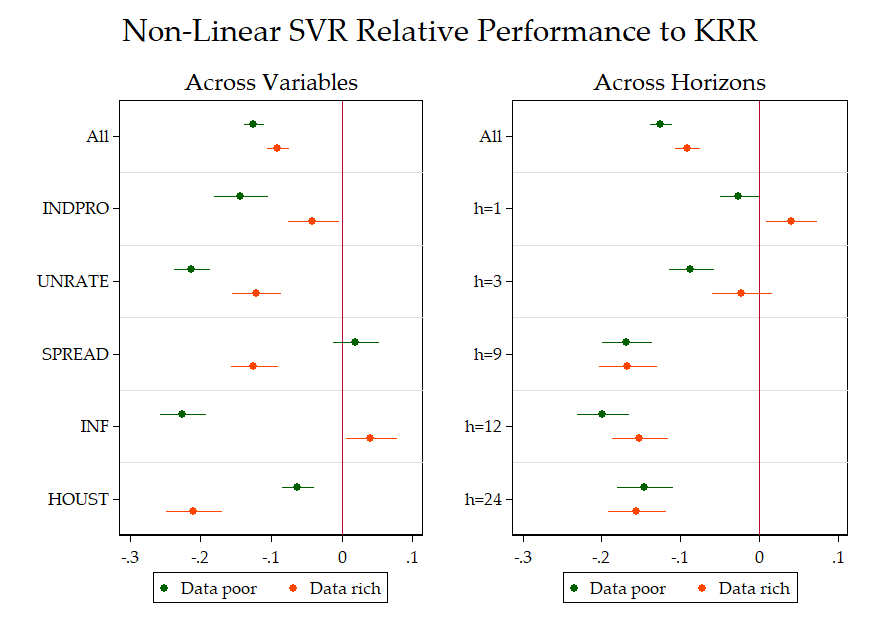}
	\caption{\footnotesize{This graph display the marginal (un)improvements by variables and horizons to opt for the SVR in-sample loss function in comparing the data-poor and data-rich environments for nonlinear models. The unit of the x-axis are improvements in OOS $R^2$ over the basis model.  SEs are HAC. These are the 95\% confidence bands.}}
	\label{SVRNL_byX}
\end{figure}

\clearpage
\section{Nonlinearites Matter -- A Robustness Check}\label{sec:NNBT}

In this appendix, we trade Random Forests for Boosted Trees and KRR for Neural Networks. First, we briefly introduce the newest addition to our nonlinear arsenal. Second, we demonstrate that very similar conclusions to that of section \ref{NLeffect} are reached using those. This further backs our claim that nonlinearities matter, whichever way they were obtained.

\subsection{Data-Poor}
	
	{\noindent \textbf{Boosted Trees AR (BTAR)}.} This algorithm provides an alternative means of approximating nonlinear functions by additively combining regression trees in a sequential fashion. Let $\eta \in [0,1]$ be the learning rate and $\hat{y}_{t+h}^{(n)}$ and $e_{t+h}^{(n)} := y_{t-h} - \eta \hat{y}_{t+h}^{(n)}$ be the step $n$ predicted value and pseudo-residuals, respectively. Then, the step $n+1$ prediction is obtained as \vspace{-1em}
	\begin{align*}
	\hat{y}_{t+h}^{(n+1)} = y_{t+h}^{(n)} + \rho_{n+1} f(Z_t, c_{n+1} )
	\end{align*}
	where $(c_{n+1}, \rho_{n+1}) := \text{arg}\underset{\rho, c}{\min} \sum_{t=1}^T \left( e_{t+h}^{(n)} - \rho_{n+1} f(Z_t, c_{n+1}) \right)^2$ and $c_{n+1} := \left( c_{n+1,m} \right)_{m=1}^M$ are the parameters of a regression tree. In other words, it recursively fits trees on pseudo-residuals. 
	The maximum depth of each tree is set to 10 and all features are considered at each split. We select the number of steps and $\eta \in [0,1]$ with Bayesian optimization. We impose $p_y = 12$.

	{\noindent \textbf{Neural Network AR (NNAR)}.} We opted for fully connected feed-forward neural networks. The value of the input vector $\left[ Z_{it} \right]_{i=1}^{N_0}$ is represented by a layer of input neurons, each taking on the value of a different element in the vector. Each neuron $j$ of the first hidden layer takes on a value $h_{jt}^{(n)}$ which is determined by applying a potentially nonlinear transformation to a weighted sum of the input value. The same is true of each subsequent hidden layer until we have reached the output layer which contains a single neuron whose value is the $h$ period ahead forecast of the model. Formally, our neural network models have the following form:
	\begin{align*}
		h_{jt}^{(n)} &=
		\begin{cases}
			f^{(1)} \left( \sum_{i=1}^{N_0} w_{ji}^{(1)} Z_{it} + w_{j0}^{(1)} \right) & n = 1 \\
			f^{(n)} \left( \sum_{i=1}^{N_k} w_{ji}^{(n)} h_{it}^{(n-1)} + w_{j0}^{(n)} \right) & n > 1
		\end{cases} \\
		\hat{y}_{t+h} &= \sum_{i=1}^{N_{N_h}} w_{i}^{(y)} h_{jt}^{(N_h)} + w_{0}^{(y)}.
	\end{align*}
	We restrict our attention to two fixed architectures: the first one uses a single hidden layer of 32 neurons ($(N_h,N_1)=(1,32)$) and the second one uses two hidden layers of 32 and 16 neurons, respectively ($(N_h,N_1,N_2) = (2,32,16)$). In all cases, we use rectified linear units (ReLU) as the activation functions, i.e.
	\begin{align*}
		f^{(n)}(z) = \max \{ 0, z \}, \forall n=1,...,N_h.
	\end{align*}
	The training is carried out by batch gradient descent using the Adam algorithm. This algorithm is initialized with a learning rate of 0.01 and we use an early stopping rule\footnote{If improvements in the performance metric doesn't exceed a tolerance threshold for 5 consecutive epochs, we stop the training.}. And, in an effort to mitigate the effects of overfitting and the impact of random initialization of weights, we train 5 neural networks with the same architecture and use their average output as our prediction value. In essence, those neural networks are simplified versions of the neural networks used in \cite{Gu2020a} where we got rid of the hyperparameter optimization and use 5 base learners instead of 10. For this algorithm, the input is a set of $p_y=12$ lagged values of the target variable. We do not make use of cross-validation, but we do estimate model weights recursively.

	\subsection{Data-Rich}
	
	{\noindent \textbf{Boosted Trees ARDI (BTARDI)}.} We consider a vanilla Boosted Trees where the maximum depth of each tree is set to 10 and all features are considered at each split. We select the number of steps and $\eta \in [0,1]$ with Bayesian optimization. We impose $p_y = 12$, $p_f = 12$ and $K = 8$.   
	
	{\noindent \textbf{Neural Network ARDI (NNARDI)}.} We opted for fully connected feed-forward neural network with the same architecture as the data-poor version, but we now use $(p_y,p_f,K)=(12,10,12)$ for the inputs.

	\begin{figure}
			\centering
\includegraphics[scale=.4]{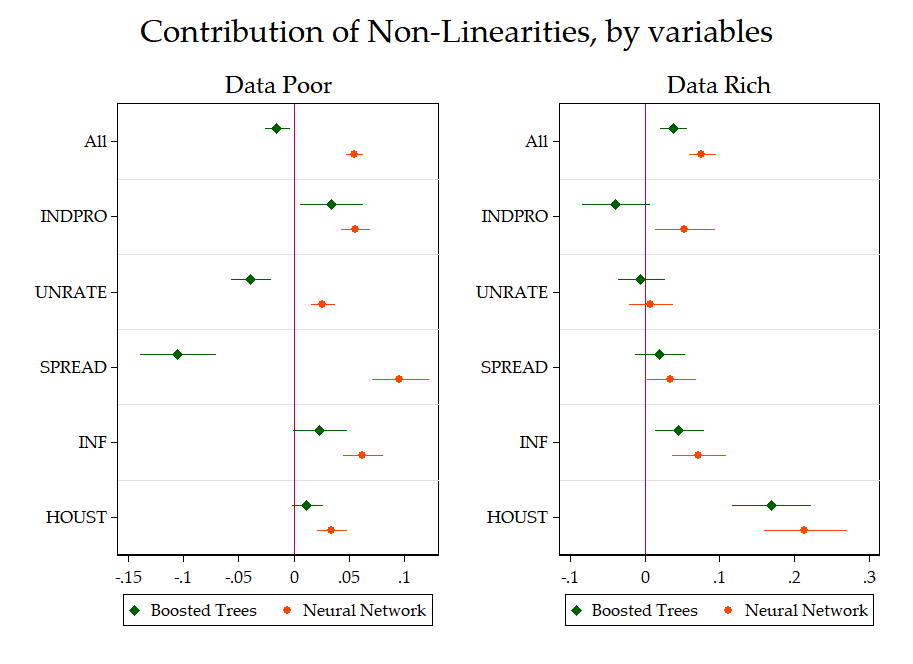}  
		\caption{\footnotesize{This figure compares the two alternative NL models averaged over all horizons. The unit of the x-axis are  improvements in OOS $R^2$ over the basis model. SEs are HAC. These are the 95\% confidence bands.}}
		\label{g_nl_v_NNBT}
		\end{figure}

		\begin{figure}
			\centering
\includegraphics[scale=.4]{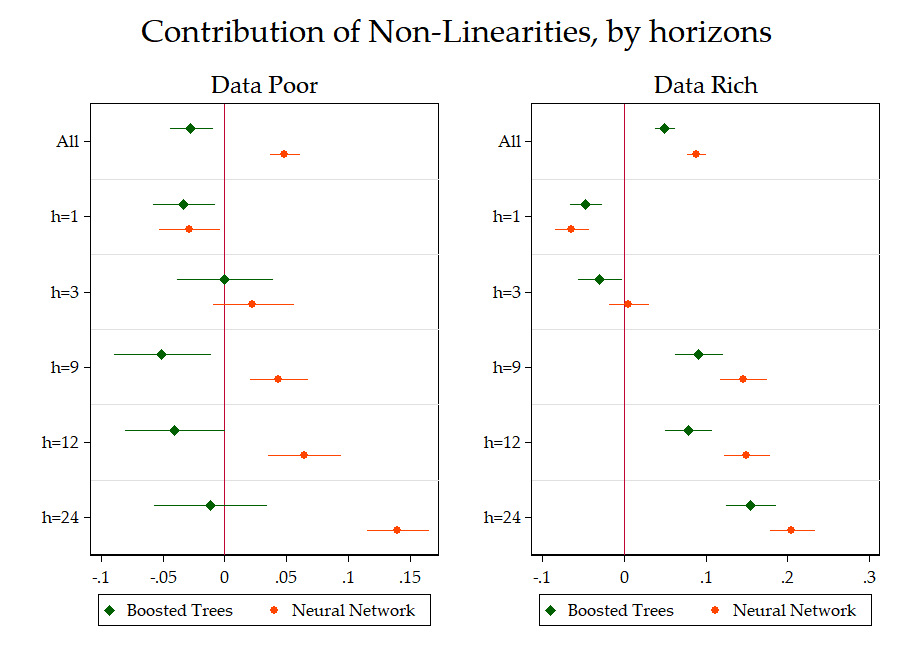}  
	\caption{\footnotesize{This figure compares the two alternative NL models averaged over all variables. The unit of the x-axis are improvements in OOS $R^2$ over the basis model.  SEs are HAC. These are the 95\% confidence bands.}}
	\label{g_nl_h_NNBT}
		\end{figure}

\subsection{Results}

In line with what reported in section \ref{NLeffect}, we find that NL's treatment effect is magnified for horizons 9, 12 and 24. Additionally, it is found that both algorithms give very homogeneous improvements in the data-rich environment, another finding detailed in the main text. Results for the data-poor environment are more scattered, as they were before. Targets benefiting most from NL in the data-rich environment are INF and HOUST, which is analogous to earlier findings. However, it was found that the real activity targets benefited more from NL in our main text configuration, which is the sole noticeable difference with results reported here.


\clearpage

\pagenumbering{arabic} \setcounter{section}{0} \setcounter{page}{1}
\onehalfspace

\includepdf[pages=-,pagecommand={},width=\textwidth]{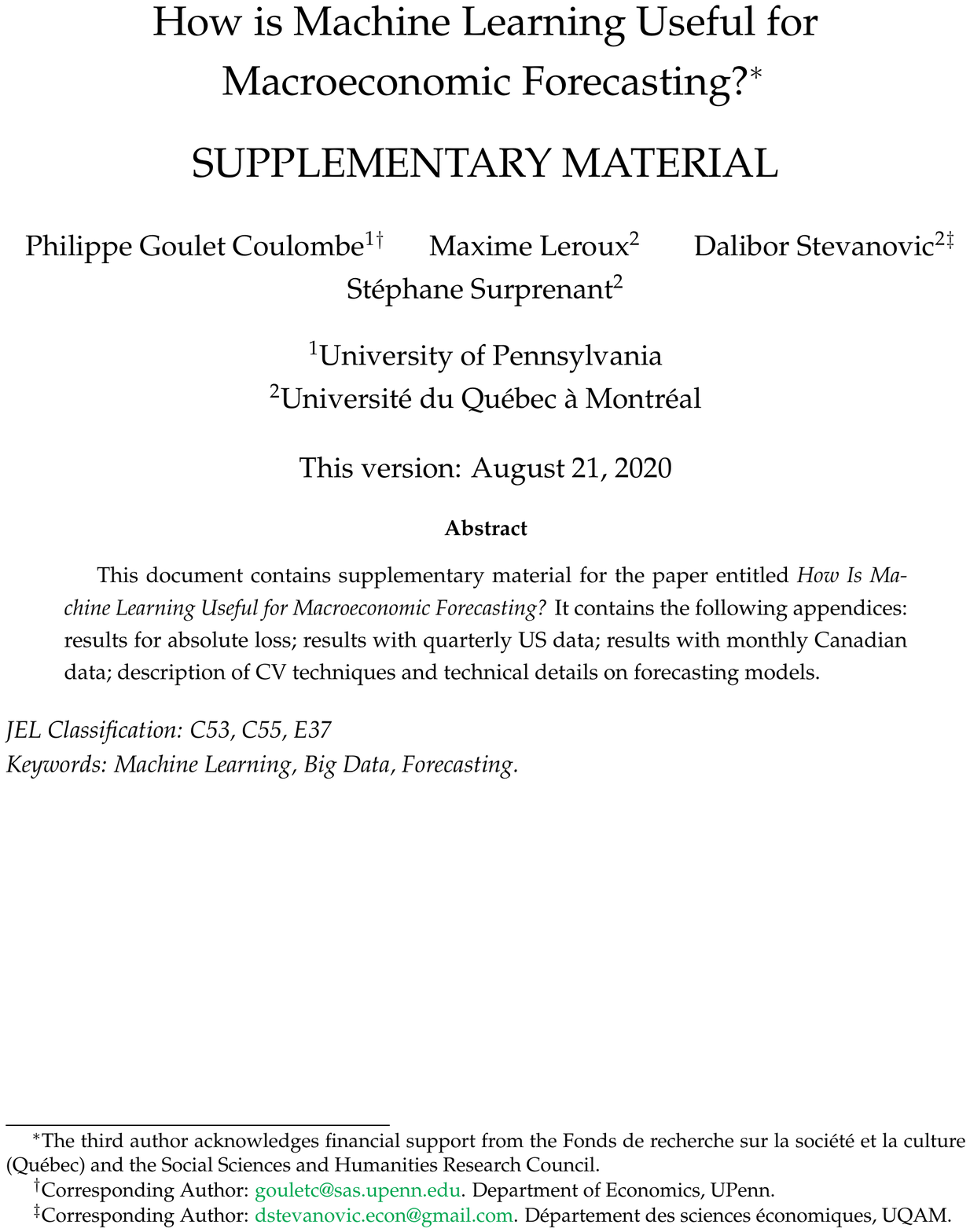}

\clearpage


\section{Results with Absolute Loss}\label{sec:mae}

In this section we present results for a different out-of-sample loss function that is often used in the literature: the absolute loss. Following 
\cite{Koenker1999}, we generate the pseudo-$R^1$  in order to perform regressions (\ref{r2_eq}) and (\ref{e_eq2}): 
$R^1_{t,h,v,m} \equiv 1 - \frac{|e_{t,h,v,m}|}{\frac{1}{T} \sum_{t=1}^T |y_{v,t+h} - \bar{y}_{v,h}|}$. Hence, the figure included in this section are exact replication of those included in the main text except that the target variable of all the regressions has been changed.

\begin{figure}
\centering
\includegraphics[scale=.75]{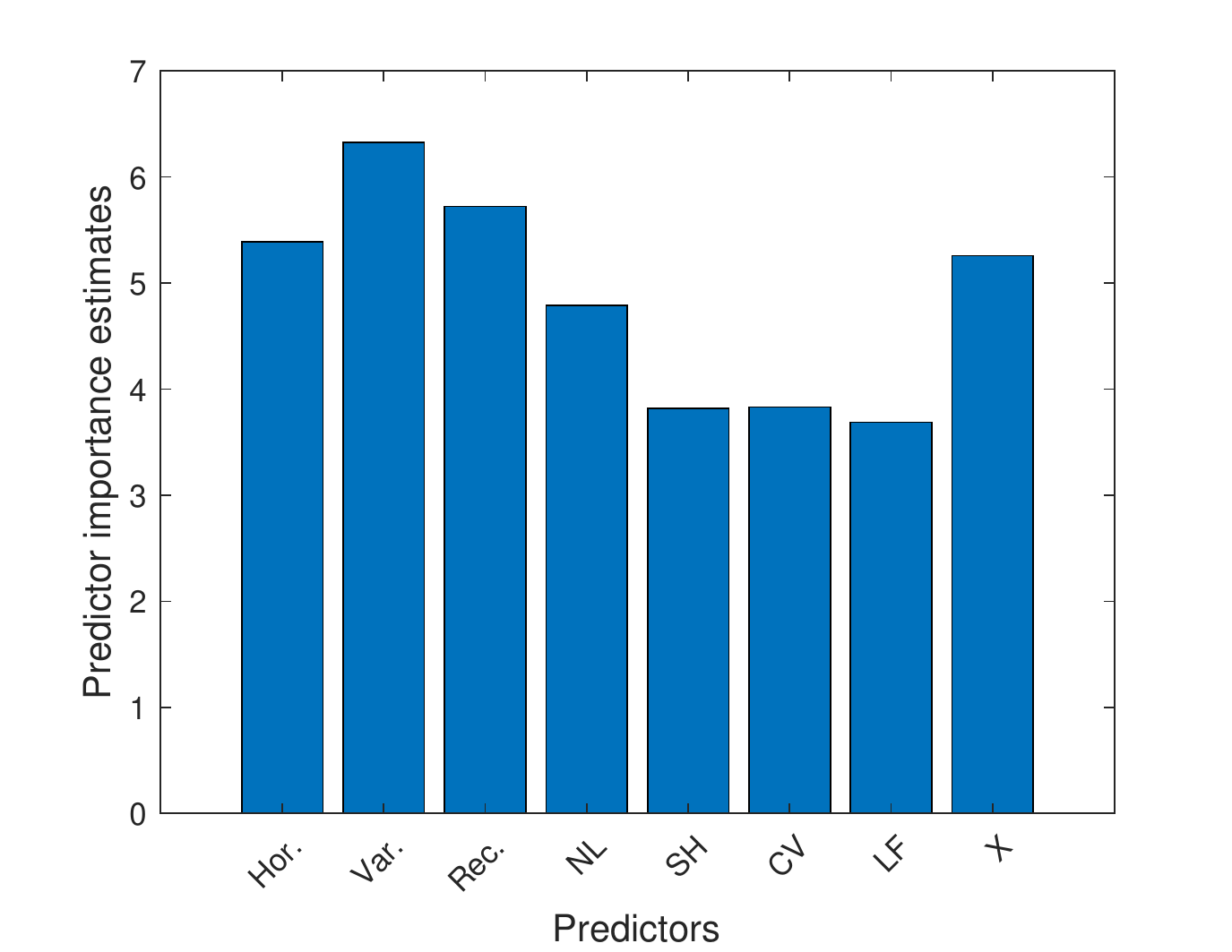}
	\caption{\footnotesize{This figure presents predictive importance estimates. Random forest is trained to predict $R^1_{t,h,v,m}$ defined in (\ref{r2_eq}) and 
	use out-of-bags observations to assess the performance of the model and compute features' importance. NL, SH, CV and LF stand for nonlinearity, shrinkage, 
	cross-validation and loss function features respectively. A dummy for $H_t^+$ models, $X$, is included as well.}}
	\label{Tree_absloss}
\end{figure}

\begin{figure}
\centering
\includegraphics[width=0.8\textwidth, height=0.38\textheight, trim= 0mm 10mm 0mm 0mm, clip]{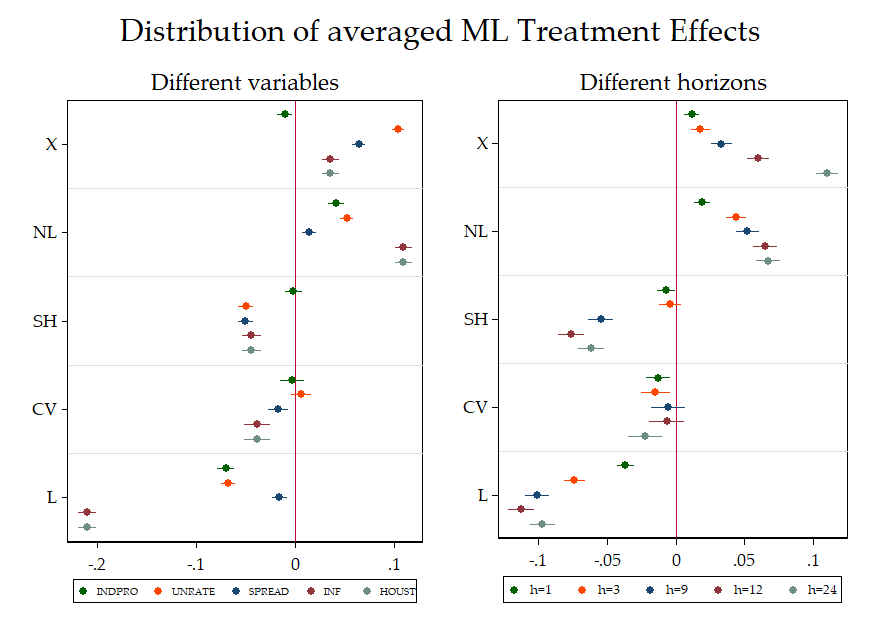}
	\caption{\footnotesize{This figure plots the distribution of $\dot{\alpha}_F^{(v)}$ and $\dot{\alpha}_F^{(h)}$ from equation (\ref{r2_eq}) done by $h$ and $v$ subsets. That is, we are looking at the average partial effect on the pseudo-OOS $R^1$ from augmenting the model with ML features, keeping everything else fixed. $X$ is making the switch from data-poor to data-rich. However, in this graph, $v-$specific heterogeneity and $h-$specific heterogeneity have been integrated out in turns. SEs are HAC. These are the 95\% confidence bands.}}
	\label{dist_all_hv_absloss}
\end{figure}

The main message here is that results obtained using the squared loss are very consistent with what one would obtain using the absolute loss. The importance of each feature, figure \ref{Tree_absloss}, and the way it behaves according to the variable/horizon pair is the same. Indeed, most of the heterogeneity is variable specific while there are clear horizon patterns emerging when we average out variables. For instance, we clearly see by comparing figures \ref{dist_all_hv_absloss} and \ref{dist_all_hv} that more data and nonlinearities usefulness increase linearly in $h$. CV is flat around the 0 line. Alternative shrinkage and loss function both are negative and follow a boomerang shape (they are not as bad for short and very long horizons, but quite bad in between).

The pertinence of nonlinearities and the impertinence of alternative shrinkage follow very similar behavior to what is obtained in the main body of this paper. However, for nonlinearities, the data-poor advantages are not robust to the choice of MSPE vs MAPE. Fortunately, besides that, the figures are all very much alike.

Results for the alternative in-sample loss function also seem to be independent of the proposed choices of out-of-sample loss function. Only for hyperparameters selection we do get slightly different results: CV-KF is now sometimes worse than BIC in a statistically significant way. However, the negative effect is again much stronger for POOS CV. CV-KF still outperforms any other model selection criteria on recessions.


\begin{figure}
\centering
\includegraphics[scale=.4, trim= 0mm 10mm 0mm 10mm, clip]{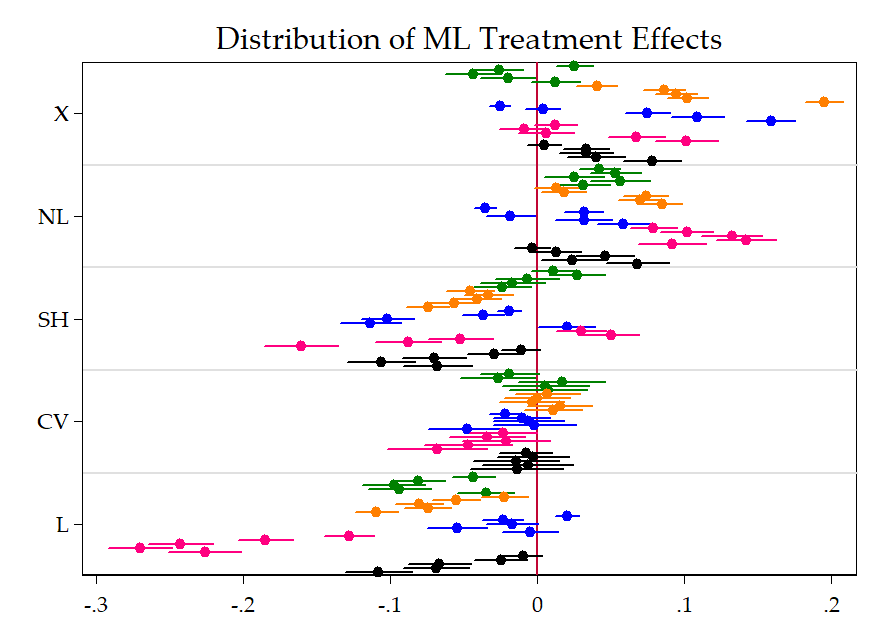}
\vspace{-0cm}
	\caption{\footnotesize{This figure plots the distribution of $\dot{\alpha}_F^{(h,v)}$ from equation (\ref{r2_eq}) done by $(h,v)$ subsets. That is, we are looking at the average partial effect on the pseudo-OOS $R^1$ from augmenting the model with ML features, keeping everything else fixed. $X$ is making the switch from data-poor to data-rich. Finally, variables are \textcolor{Green}{INDPRO}, \textcolor{orange}{UNRATE}, \textcolor{blue}{SPREAD}, \textcolor{magenta}{INF} and HOUST. Within a specific color block, the horizon increases from $h=1$ to $h=24$ as we are going down. As an example, we clearly see that the partial effect of $X$ on the $R^1$ of \textcolor{magenta}{INF} increases drastically with the forecasted horizon $h$. SEs are HAC. These are the 95\% confidence bands.}} 
	\label{dist_all_absloss}
\end{figure}

	\begin{figure}
			\centering
\includegraphics[scale=.4]{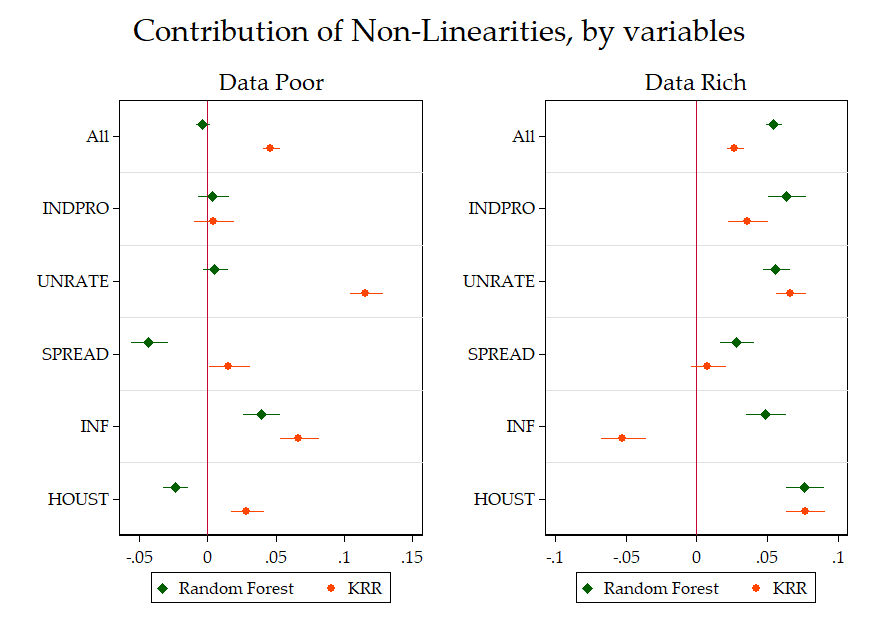}  
		\caption{\footnotesize{This compares the two NL models averaged over all horizons. The unit of the x-axis are  improvements in OOS $R^1$ over the basis model. SEs are HAC. These are the 95\% confidence bands.}}
		\label{g_nl_v_absloss}
		\end{figure}

		\begin{figure}
			\centering
\includegraphics[scale=.4]{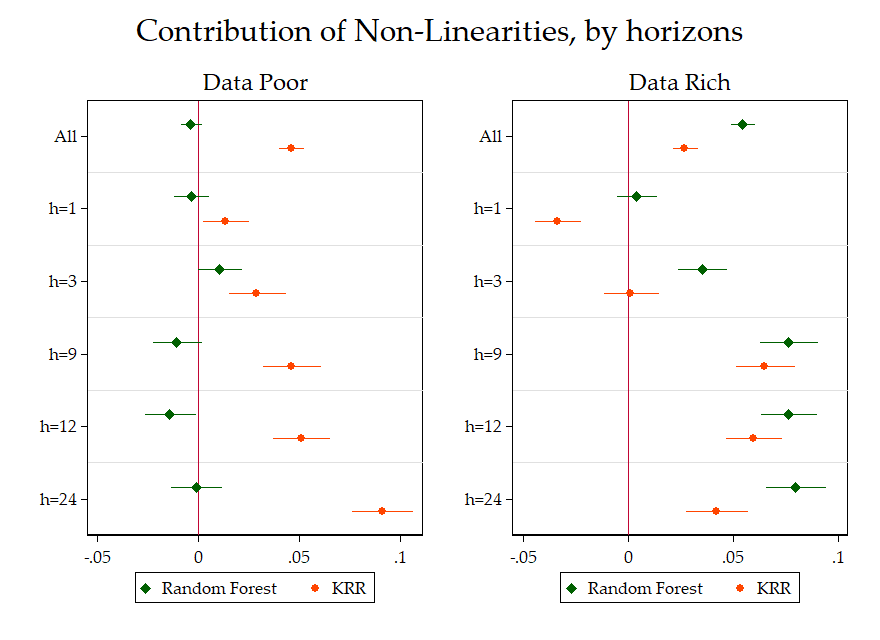}  
	\caption{\footnotesize{This compares the two NL models averaged over all variables. The unit of the x-axis are improvements in OOS $R^1$ over the basis model.  SEs are HAC. These are the 95\% confidence bands.}}
	\label{g_nl_h_absloss}
		\end{figure}


		\begin{figure}
			\centering
					
\includegraphics[scale=.35]{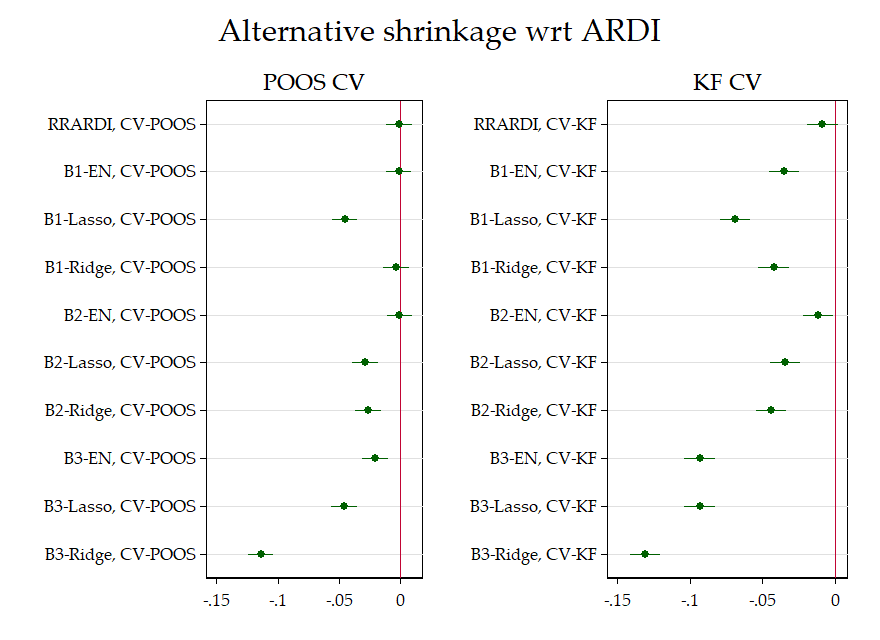}  
\caption{\footnotesize{This compares models of section \ref{sec:reg} averaged over all variables and horizons. The unit of the x-axis are improvements in OOS $R^1$ over the basis model.  The base models are ARDIs specified with POOS-CV and KF-CV respectively. SEs are HAC. These are the 95\% confidence bands.}}\label{SHgraph_absloss}
		\end{figure}


\begin{table}[htbp]\centering
\def\sym#1{\ifmmode^{#1}\else\(^{#1}\)\fi}
\caption{CV comparison}\label{CVreg_abs}
\begin{tabular}{l*{5}{c}}
\hline\hline
                    &\multicolumn{1}{c}{(1)}&\multicolumn{1}{c}{(2)}&\multicolumn{1}{c}{(3)}&\multicolumn{1}{c}{(4)}&\multicolumn{1}{c}{(5)}\\
                    &\multicolumn{1}{c}{All}&\multicolumn{1}{c}{Data-rich}&\multicolumn{1}{c}{Data-poor}&\multicolumn{1}{c}{Data-rich}&\multicolumn{1}{c}{Data-poor}\\
\hline
CV-KF               &      0.0114         &     -0.0233         &      0.0461         &      -0.221         &      -0.109         \\
                    &     (0.375)         &     (0.340)         &     (0.181)         &     (0.364)         &     (0.193)         \\
CV-POOS             &      -0.765\sym{*}  &      -0.762\sym{*}  &      -0.768\sym{***}&      -0.700         &      -0.859\sym{***}\\
                    &     (0.375)         &     (0.340)         &     (0.181)         &     (0.364)         &     (0.193)         \\
AIC                 &      -0.396         &      -0.516         &      -0.275         &      -0.507         &      -0.522\sym{**} \\
                    &     (0.375)         &     (0.340)         &     (0.181)         &     (0.364)         &     (0.193)         \\
CV-KF * Recessions  &                     &                     &                     &       1.609         &       1.264\sym{*}  \\
                    &                     &                     &                     &     (1.037)         &     (0.552)         \\
CV-POOS * Recessions&                     &                     &                     &      -0.506         &       0.747         \\
                    &                     &                     &                     &     (1.037)         &     (0.552)         \\
AIC * Recessions    &                     &                     &                     &     -0.0760         &       2.007\sym{***}\\
                    &                     &                     &                     &     (1.037)         &     (0.552)         \\
\hline
Observations        &       91200         &       45600         &       45600         &       45600         &       45600         \\
\hline\hline
\multicolumn{6}{l}{\footnotesize Standard errors in parentheses}\\
\multicolumn{6}{l}{\footnotesize \sym{*} \(p<0.05\), \sym{**} \(p<0.01\), \sym{***} \(p<0.001\)}\\
\end{tabular}
\end{table}

		\begin{figure}
			\centering
\includegraphics[scale=.4]{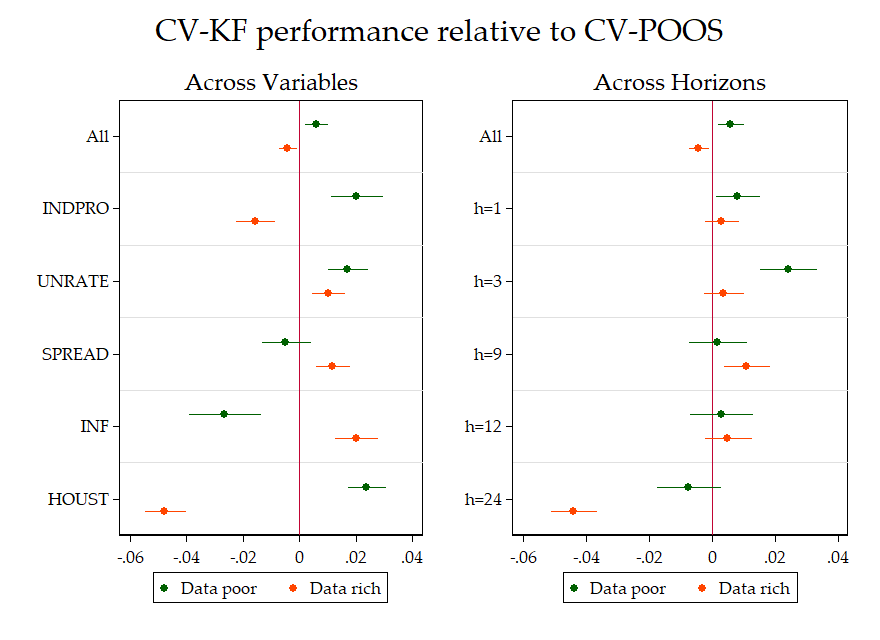}
		\caption{\footnotesize{This compares the two CVs procedure averaged over all the models that use them. The unit of the x-axis are improvements in OOS $R^1$ over the basis model.  SEs are HAC. These are the 95\% confidence bands.}}\label{CVbyX_absloss}
		\end{figure}
		
\begin{figure}
			\centering
\includegraphics[scale=.4]{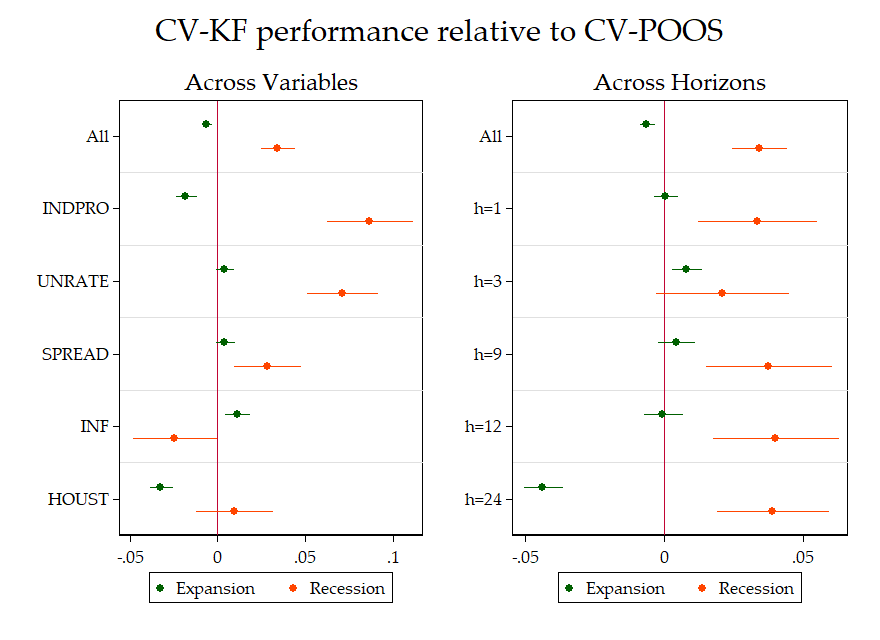}
		\caption{\footnotesize{This compares the two CVs procedure averaged over all the models that use them. The unit of the x-axis are improvements in OOS $R^1$ over the basis model.  SEs are HAC. These are the 95\% confidence bands.}}\label{CVbyrec_absloss}
		\end{figure}


		\begin{figure}
			\centering
\includegraphics[scale=.4]{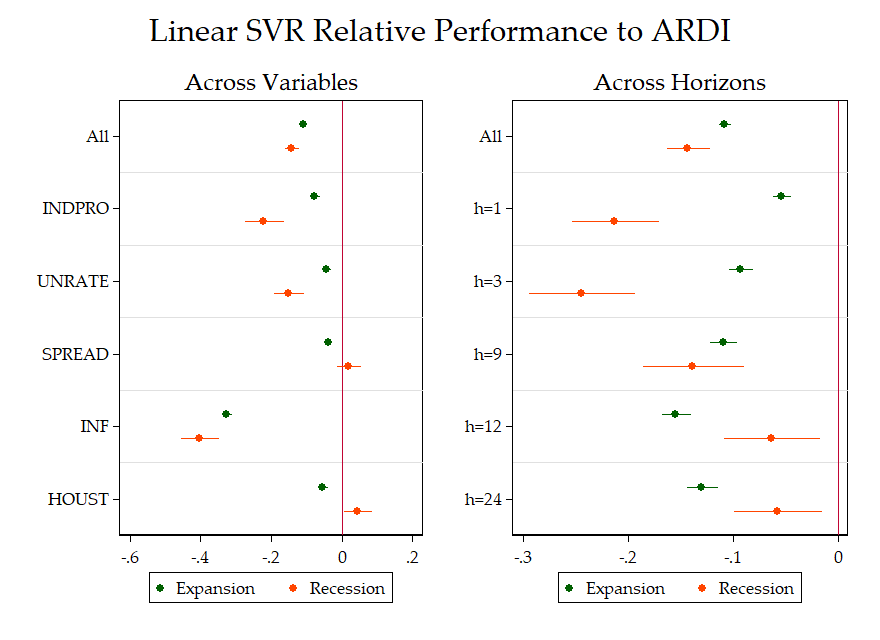}
		\caption{\footnotesize{This graph display the marginal (un)improvements by variables and horizons to opt for the SVR in-sample loss function in \textbf{both the data-poor and data-rich environments}. The unit of the x-axis are improvements in OOS $R^1$ over the basis model.  SEs are HAC. These are the 95\% confidence bands.}}\label{Leffectlin_absloss}
		\end{figure}

		\begin{figure}
			\centering
\includegraphics[scale=.4]{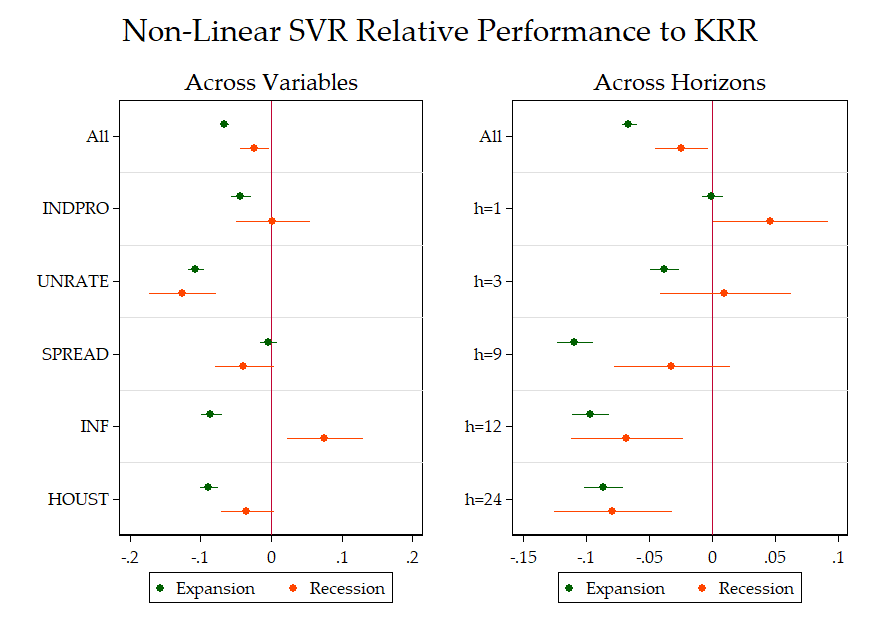}
		\caption{\footnotesize{This graph display the marginal (un)improvements by variables and horizons to opt for the SVR in-sample loss function in \textbf{both recession and expansion periods}. The unit of the x-axis are improvements in OOS $R^1$ over the basis model.  SEs are HAC. These are the 95\% confidence bands.}}\label{LeffectNL_absloss}
		\end{figure}

\clearpage


\clearpage

\section{Results with Quarterly Data}\label{sec:quart}

In this section we present results for quarterly frequency using the dataset FRED-QD, publicly available at the Federal Reserve of St-Louis's 
web site. This is the quarterly companion to FRED-MD monthly dataset used in the main part of paper. It contains 248 US macroeconomic and 
financial aggregates observed from 1960Q1 to 2018Q4. The series transformations to induce stationarity are the same as in \cite{Stock2012brookings}. 
The variables of interest are: real GDP, real personal consumption expenditures (CONS), real gross private investment (INV), 
real disposable personal income (INC) and the PCE deflator. All the targets are expressed in average growth rate 
over $h$ periods as in equation (\ref{fcst1}). Forecasting horizons are 1, 2, 3, 4 and 8 quarters.

The main message here is that results obtained using the quarterly data and predicting GDP components are consistent with those on monthly variables. Tables \ref{tab:GDP} - 
\ref{tab:PCE} summarize the overall predictive ability in terms of RMPSE relative to the reference AR,BIC model. GDP and consumption growths are best predicted at short run by the 
standard \cite{Stock2002a} ARDI,BIC model, while random forests dominate at longer horizons. Nonlinear models perform well for most horizons when 
predicting the disposable income growth. Finally, kernel ridge regressions (both data-poor and data-rich) are the best options to predict the PCE inflation. 

\begin{figure}
\centering
\includegraphics[scale=.75]{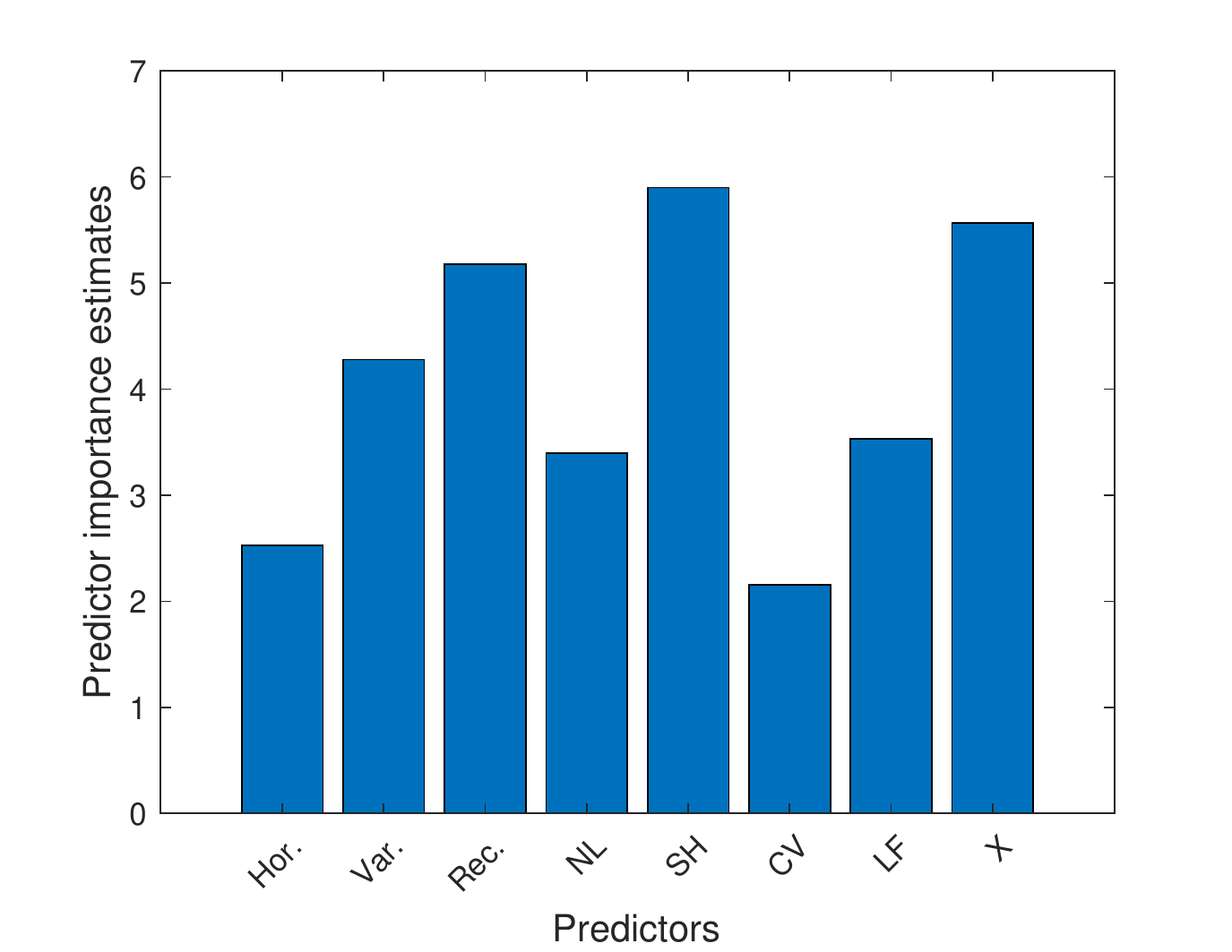}
	\caption{\footnotesize{This figure presents predictive importance estimates. Random forest is trained to predict $R^2_{t,h,v,m}$ defined in (\ref{r2_eq}) and 
	use out-of-bags observations to assess the performance of the model and compute features' importance. NL, SH, CV and LF stand for nonlinearity, shrinkage, 
	cross-validation and loss function features respectively. A dummy for $H_t^+$ models, $X$, is included as well.}}
	\label{Tree_q}
\end{figure}

\begin{figure}
\centering
\includegraphics[scale=.4, trim= 0mm 8mm 0mm 8mm, clip]{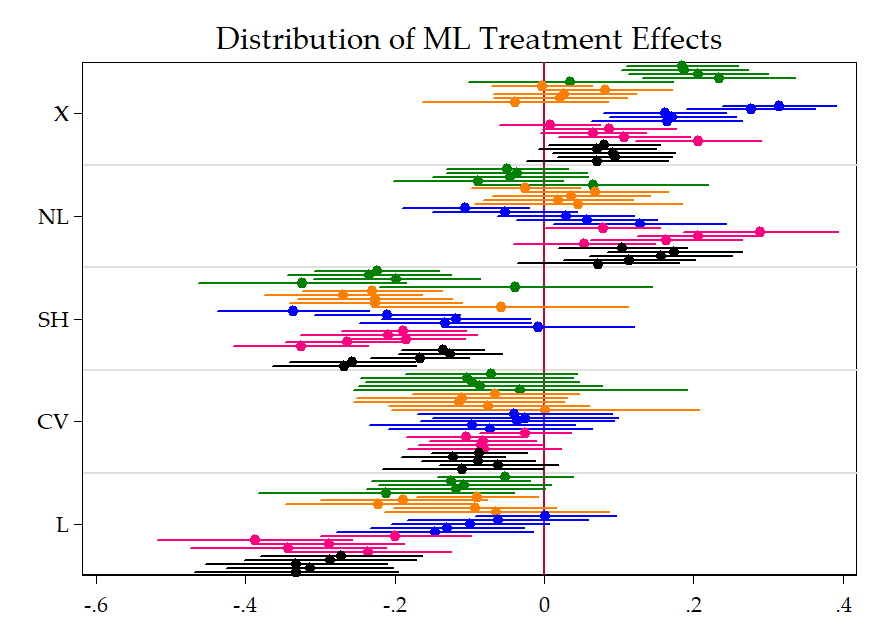}
\vspace{-0cm}
	\caption{\footnotesize{This figure plots the distribution of $\dot{\alpha}_F^{(h,v)}$ from equation (\ref{r2_eq}) done by $(h,v)$ subsets. That is, we are looking at the average partial effect on the pseudo-OOS $R^2$ from augmenting the model with ML features, keeping everything else fixed. $X$ is making the switch from data-poor to data-rich. Finally, variables are \textcolor{Green}{GDP}, \textcolor{orange}{CONS}, \textcolor{blue}{INV}, \textcolor{magenta}{INC} and PCE. Within a specific color block, the horizon increases from $h=1$ to $h=8$ as we are going down.  SEs are HAC. These are the 95\% confidence bands.}} 
	\label{dist_all_q}
\end{figure}

\begin{figure}
\centering
\includegraphics[width=0.8\textwidth, height=0.38\textheight, trim= 0mm 10mm 0mm 0mm, clip]{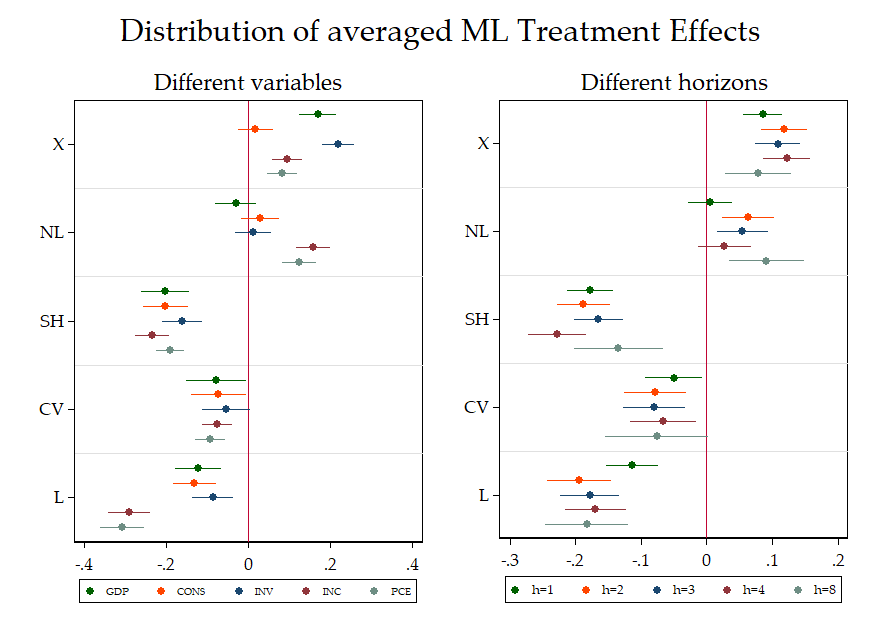}
	\caption{\footnotesize{This figure plots the distribution of $\dot{\alpha}_F^{(v)}$ and $\dot{\alpha}_F^{(h)}$ from equation (\ref{r2_eq}) done by $h$ and $v$ subsets. That is, we are looking at the average partial effect on the pseudo-OOS $R^2$ from augmenting the model with ML features, keeping everything else fixed. $X$ is making the switch from data-poor to data-rich. However, in this graph, $v-$specific heterogeneity and $h-$specific heterogeneity have been integrated out in turns. SEs are HAC. These are the 95\% confidence bands.}}
	\label{dist_all_hv_q}
\end{figure}

The ML features' importance is plotted in figure \ref{Tree_q}. Contrary to monthly data, horizons and variables fixed effects are much less important which 
is somehow expected because of relative smoothness of quarterly data and similar targets (4 out 5 are real activity series). Among ML treatments, shrinkage 
is the most important, followed by loss function and nonlinearity. As in the monthly application, CV is the least relevant, while the data-rich component remains 
very important. From figures \ref{dist_all_q} and \ref{dist_all_hv_q}, we see that: (i) the richness of predictors' set 
is very helpful for most of the targets; (ii) nonlinearity treatment has positive and significant effects for investment, income and PCE deflator, while it is not significant for GDP and 
CONS; (iii) the impertinence of alternative shrinkage follow very similar behavior to what is obtained in the main body of this paper; (iv) CV has in general negative but small and often insignificant effect; (v) SVR loss function decreases the predictive 
performance as in the monthly case, especially for income growth and inflation.

\clearpage 

\begin{table}[!h]
\caption{GDP: Relative Root MSPE}
\label{tab:GDP}\vspace{-0.3cm}
\par
\begin{center}
{\scriptsize
\begin{tabular}{llllll|lllll}
\hline\hline
 & \multicolumn{5}{c}{Full Out-of-Sample} & \multicolumn{5}{|c}{NBER Recessions Periods} \\
Models & h=1 & h=2 & h=3 & h=4 & h=8 & \multicolumn{1}{|l}{h=1} & h=2 & h=3 & h=4 & h=8 \\
\hline
\multicolumn{3}{l}{Data-poor ($H_t^-$) models} \\
AR,BIC (RMSPE) & \textbf{0.0752} & \textbf{0.0656} & \textbf{0.0619} & \textbf{0.0593} & \textbf{0.0521} & 0,1199 & 0,1347 & 0,1261 & 0,1285 & 0,1022 \\
AR,AIC & \textbf{1.004} & \textbf{0.994} & \textbf{0.999} & \textbf{1.000} & \textbf{1.000} & 1,034 & 0,995 & 1 & 1 & 1 \\
AR,POOS-CV & \textbf{0.984}** & \textbf{0.994} & \textbf{0.994} & \textbf{1.000} & \textbf{1.017} & \textbf{0.991} & 0,994 & 0,993 & 1 & 1,033 \\
AR,K-fold & \textbf{0.998} & \textbf{1.003} & \textbf{0.999} & 1,001 & \textbf{1.000} & 1,026 & 1,01 & 0,997 & 1,001 & 1 \\
RRAR,POOS-CV & \textbf{0.992} & \textbf{1.002} & \textbf{1.000} & 1,005 & \textbf{1.005} & 1,014 & 1 & 1,005 & 0,997 & 1,014 \\
RRAR,K-fold & \textbf{1.013} & \textbf{1.007} & \textbf{1.006} & 1,012 & \textbf{1.000} & 1.092* & 1.010* & 1.020*** & 1,02 & 0.999*** \\
RFAR,POOS-CV & 1.185*** & 1.104*** & 1.165*** & 1.129*** & 1.061** & 1.241** & 1.077* & 1.116** & 1.070** & 0.925*** \\
RFAR,K-fold & 1.082** & 1.124*** & 1.105** & 1.121*** & \textbf{1.064}** & 1.124* & 1,085 & 1,021 & 1.089** & 0,989 \\
KRR-AR,POOS-CV & 1,049 & 1,044 & 1,011 & 1.065* & \textbf{0.993} & 1.103** & 0,954 & 0.913* & 0.943* & 0.873*** \\
KRR,AR,K-fold & 1,044 & \textbf{1.033} & 1.051** & 1,013 & \textbf{0.995} & 1.172*** & 1,01 & 1,036 & 0,974 & 0.963*** \\
SVR-AR,Lin,POOS-CV & 1.161** & 1.136** & 1.129** & 1.143** & \textbf{1.045} & 1.233*** & 1.106** & 1.152*** & 1.061** & 1,071 \\
SVR-AR,Lin,K-fold & 1.082** & 1.092** & \textbf{1.054}* & 1.051** & \textbf{0.986} & 1.222*** & 1.110** & 1.088** & 1.054** & 0.964*** \\
SVR-AR,RBF,POOS-CV & 1,015 & \textbf{1.036}* & \textbf{1.026} & 1,051 & 1.095** & 1.038** & 1,01 & 1.037* & 0,991 & 1,016 \\
SVR-AR,RBF,K-fold & 1.043** & \textbf{1.032}* & \textbf{1.029}* & 1,018 & \textbf{1.011}* & 1.157*** & 1.032** & 1.041** & 0,986 & 1,002 \\
\hline
\multicolumn{3}{l}{Data-rich ($H_t^+$) models} \\
ARDI,BIC & \underline{\textbf{0.884}} & \underline{\textbf{0.811}}** & \underline{\textbf{0.824}}** & \textbf{0.817}** & \textbf{1.002} & \underline{\textbf{0.829}} & \underline{\textbf{0.649}}*** & \underline{\textbf{0.732}}** & \textbf{0.704}*** & 0.714*** \\
ARDI,AIC & \textbf{0.905} & \textbf{0.833}* & \textbf{0.844}* & \textbf{0.832}* & \textbf{0.989} & \textbf{0.931} & \textbf{0.652}*** & \textbf{0.741}** & \textbf{0.721}*** & 0.687*** \\
ARDI,POOS-CV & \textbf{0.913} & \textbf{0.861}* & \textbf{0.878} & \textbf{0.885} & \textbf{0.918} & \textbf{0.936} & \textbf{0.689}** & \textbf{0.742}** & \textbf{0.719}*** & 0.735*** \\
ARDI,K-fold & \textbf{0.978} & \textbf{0.881} & \textbf{0.871} & \textbf{0.815}* & \textbf{1.070} & 1,078 & 0.709** & 0.767** & \underline{\textbf{0.681}}*** & \textbf{0.595}*** \\
RRARDI,POOS-CV & \textbf{0.938} & \textbf{0.853}* & \textbf{0.846}* & \textbf{0.924} & \textbf{0.949} & 1,034 & 0.717*** & \textbf{0.742}** & 0.740*** & 0.770*** \\
RRARDI,K-fold & \textbf{0.906} & \textbf{0.839}* & \textbf{0.842}* & \underline{\textbf{0.810}}* & \textbf{1.021} & \textbf{0.924} & 0.720** & \textbf{0.755}** & \textbf{0.690}*** & \textbf{0.587}*** \\
RFARDI,POOS-CV & \textbf{0.938} & \textbf{0.929} & \textbf{0.876}* & \textbf{0.866}* & \textbf{0.887}* & 0,989 & 0.866* & 0.810** & 0.761*** & 0.739*** \\
RFARDI,K-fold & \textbf{0.941} & \textbf{0.908}* & \textbf{0.868}* & \textbf{0.856}* & \underline{\textbf{0.862}}** & 1,022 & 0.843** & 0.813* & 0.742*** & 0.692*** \\
KRR-ARDI,POOS-CV & 1,055 & \textbf{1.048} & 1.074** & 1,049 & \textbf{1.011} & 1.135* & 0,97 & 0,979 & 0.923* & 0.921* \\
KRR,ARDI,K-fold & 1,005 & 1,038 & 1,065 & 1,074 & \textbf{0.957} & 1 & 0,969 & 0,947 & 0,95 & 0.822*** \\
$(B_1,\alpha=\hat{\alpha})$,POOS-CV & 1.061* & \textbf{1.057} & \textbf{1.039} & 1.077** & \textbf{1.026} & 1.118** & 0,977 & 1,057 & 0,981 & 0.931** \\
$(B_1,\alpha=\hat{\alpha})$,K-fold & 1,015 & \textbf{0.964} & \textbf{1.016} & 1.079** & \textbf{1.010} & 1,041 & 0,955 & 0,98 & 0,972 & 0.907*** \\
$(B_1,\alpha=1)$,POOS-CV & 1.076** & 1.104* & \textbf{1.008} & 1.065* & \textbf{1.006} & 1.179*** & 1,007 & 1,003 & 0,954 & 0.937* \\
$(B_1,\alpha=1)$,K-fold & \textbf{0.994} & \textbf{1.018} & 1,033 & 1.079* & \textbf{0.971} & \textbf{0.989} & 0,989 & 1,013 & 0.947* & 0.890*** \\
$(B_1,\alpha=0)$,POOS-CV & 1.082* & \textbf{1.064} & 1.148*** & 1.145* & \textbf{0.992} & 1.242*** & 1,083 & 1.156*** & 1,033 & 0,979 \\
$(B_1,\alpha=0)$,K-fold & 1.191** & \textbf{1.079}* & 1,052 & 1.070* & \textbf{0.968} & 1.091** & 0,974 & 0,999 & 1,011 & 0.928* \\
$(B_2,\alpha=\hat{\alpha})$,POOS-CV & 1,043 & \textbf{1.022} & \textbf{1.021} & 1,032 & \textbf{1.015} & 1.083* & 1,01 & 1,007 & 0,907 & 0.900** \\
$(B_2,\alpha=\hat{\alpha})$,K-fold & \textbf{0.991} & \textbf{1.007} & \textbf{0.994} & \textbf{0.980} & \textbf{1.126} & 1,077 & 1,002 & 0,947 & \textbf{0.747}*** & 0.612*** \\
$(B_2,\alpha=1)$,POOS-CV & 1.110** & \textbf{1.072}* & \textbf{1.007} & \textbf{0.991} & \textbf{0.918} & 1.217** & 1.090* & 0,998 & 0,924 & 0.782*** \\
$(B_2,\alpha=1)$,K-fold & 1,039 & \textbf{1.027} & \textbf{1.003} & \textbf{0.961} & \textbf{1.069} & 1.136** & 1,029 & 0,957 & 0.777*** & \underline{\textbf{0.563}}*** \\
$(B_2,\alpha=0)$,POOS-CV & \textbf{1.000} & \textbf{1.000} & \textbf{1.001} & 0,989 & \textbf{0.978} & 1,106 & 0,959 & 0,976 & 0.852** & 0.772*** \\
$(B_2,\alpha=0)$,K-fold & \textbf{0.986} & \textbf{0.980} & \textbf{0.980} & 1,001 & 1,132 & 1,073 & 0,958 & 0,968 & 0.819** & 0.750*** \\
$(B_3,\alpha=\hat{\alpha})$,POOS-CV & 1,047 & 1,055 & 1.049* & 1,052 & \textbf{1.003} & 1,046 & 1,027 & 1.043* & 1,037 & 0.930*** \\
$(B_3,\alpha=\hat{\alpha})$,K-fold & 1,038 & \textbf{0.975} & \textbf{1.004} & 1,021 & \textbf{0.991} & 1,056 & 0,98 & 0,988 & 0.918*** & 0.839*** \\
$(B_3,\alpha=1)$,POOS-CV & 1.055* & 1.133** & \textbf{1.044} & 1.107** & \textbf{0.995} & 1,058 & 1.116* & 1,033 & 1,067 & 0.895** \\
$(B_3,\alpha=1)$,K-fold & 1,045 & \textbf{1.020} & \textbf{1.009} & 1,021 & \textbf{0.982} & 1,078 & 0,994 & 1,011 & 0.942* & 0.854*** \\
$(B_3,\alpha=0)$,POOS-CV & 1.142** & 1.153* & \textbf{0.979} & 1.217* & \textbf{0.992} & 1.124** & 1,046 & 0,976 & 1,162 & 0,973 \\
$(B_3,\alpha=0)$,K-fold & 1.225* & \textbf{1.105} & \textbf{0.994} & 1,139 & 1.068* & 1.197** & 1,021 & 0,987 & 1,098 & 0,979 \\
SVR-ARDI,Lin,POOS-CV & \textbf{1.014} & \textbf{1.088} & 1.130* & \textbf{0.966} & \textbf{1.073} & 0,972 & 0,984 & 1,016 & 0.806*** & 0.933* \\
SVR-ARDI,Lin,K-fold & 1,027 & \textbf{1.112} & 1,064 & 1,084 & 1.237** & \textbf{0.982} & 0,998 & 0,876 & 0,957 & 0.863*** \\
SVR-ARDI,RBF,POOS-CV & 1,033 & \textbf{1.015} & \textbf{0.924} & 1,013 & \textbf{1.034} & 1,201 & 1,001 & \textbf{0.779}** & 0.871* & 0.861** \\
SVR-ARDI,RBF,K-fold & \textbf{0.896} & \textbf{0.887} & \textbf{0.930} & 0,973 & 1,089 & \textbf{0.930} & 0.781** & 0.807* & 0.823** & 0.813*** \\
\hline\hline
\end{tabular}
}
\end{center}
\vspace{-0.4cm}
{\scriptsize \emph{%
\singlespacing{Note: The numbers represent the relative. with respect to AR,BIC model. root MSPE. Models retained in model confidence set are in bold. the
minimum values are underlined. while
$^{***}$. $^{**}$. $^{*}$ stand for 1\%. 5\% and 10\% significance of
Diebold-Mariano test.}}}
\end{table}

\clearpage

\begin{table}[!h]
\caption{Consumption: Relative Root MSPE}
\label{tab:CONS}\vspace{-0.3cm}
\par
\begin{center}
{\scriptsize
\begin{tabular}{llllll|lllll}
\hline\hline
 & \multicolumn{5}{c}{Full Out-of-Sample} & \multicolumn{5}{|c}{NBER Recessions Periods} \\
Models & h=1 & h=2 & h=3 & h=4 & h=8 & \multicolumn{1}{|l}{h=1} & h=2 & h=3 & h=4 & h=8 \\
\hline
\multicolumn{3}{l}{Data-poor ($H_t^-$) models} \\
AR,BIC (RMSPE) & 0,0604 & \textbf{0.0485} & 0,0451 & 0,0476 & \textbf{0.0480} & 0,0927 & 0,0848 & 0,0851 & 0,0947 & 0,0881 \\
AR,AIC & 0.982** & \textbf{0.993} & 1,001 & 0.979** & \textbf{1.000} & 0.961*** & 0,993 & 1,004 & 0.978* & 1 \\
AR,POOS-CV & 0.961** & \textbf{0.986}** & \textbf{0.998} & \textbf{0.974}** & \textbf{0.997} & 0.920* & 0,995 & 0,999 & 0.971** & 0,998 \\
AR,K-fold & 0.987* & \textbf{1.025} & 1,015 & 0.975** & \textbf{1.035} & 0.977*** & 1,026 & 1,014 & 0.974** & 1,062 \\
RRAR,POOS-CV & \textbf{0.944}** & \textbf{0.988}* & 1 & \textbf{0.968}** & \textbf{0.998} & 0.878** & 0,989 & 1 & 0.971* & 0,99 \\
RRAR,K-fold & 0.973** & \textbf{1.013} & 1.015** & 1 & \textbf{1.011}* & 0,947 & 1,013 & 1.017* & 1.015** & 1,014 \\
RFAR,POOS-CV & 0,989 & \textbf{1.036} & 1,02 & 1,01 & 1.065** & 0,977 & 0,987 & 0.929* & 0,965 & 1,035 \\
RFAR,K-fold & 1,015 & \textbf{1.008} & 1.044* & 1.052* & 1.067** & 0,951 & 0,897 & 0,959 & 1,002 & 0,979 \\
KRR-AR,POOS-CV & 0,986 & \textbf{0.995} & 1.072* & 1.064** & \textbf{1.010} & 0,994 & 0,946 & 0,953 & 0,973 & 0,951 \\
KRR,AR,K-fold & 1,012 & \textbf{0.980} & 1,031 & 1,003 & \textbf{0.994} & 1,017 & 0,924 & 0,943 & 0,95 & 0.946** \\
SVR-AR,Lin,POOS-CV & 1,013 & 1.339*** & 1.304*** & 1.166*** & \textbf{1.012} & 0,868 & 1.225* & 1.350*** & 1.150*** & 0.935* \\
SVR-AR,Lin,K-fold & 1,085 & 1.176** & 1.222*** & 1.117*** & \textbf{1.020}* & 1,101 & 1.234* & 1.251*** & 1.133*** & 0,989 \\
SVR-AR,RBF,POOS-CV & 1.081* & \textbf{1.098}** & 1.120*** & 1.052** & \textbf{1.005} & 1,06 & 1,07 & 1,003 & 0.937*** & 0.934* \\
SVR-AR,RBF,K-fold & 0,973 & \textbf{1.026} & 1.064*** & \textbf{0.956}** & 1.083** & 0.881* & 1 & 1.054* & 0.959** & 1.109** \\
\hline
\multicolumn{3}{l}{Data-rich ($H_t^+$) models} \\
ARDI,BIC & \underline{\textbf{0.897}}* & \underline{\textbf{0.879}} & \underline{\textbf{0.903}} & \textbf{0.938} & \textbf{1.017} & \underline{\textbf{0.782}}* & \underline{\textbf{0.729}}** & \textbf{0.782}** & \textbf{0.829}** & 0.809*** \\
ARDI,AIC & \textbf{0.916} & \textbf{0.939} & 0,983 & 0,988 & \textbf{1.094} & 0,857 & \textbf{0.752}* & 0.800* & \textbf{0.830}* & 0.761*** \\
ARDI,POOS-CV & 1,007 & \textbf{1.002} & 1,06 & 1,069 & \textbf{0.967} & 1,071 & 0,948 & 1,05 & 1,02 & 0.860* \\
ARDI,K-fold & 1,092 & \textbf{0.948} & \textbf{0.967} & \textbf{0.959} & 1,116 & 1,31 & \textbf{0.768}* & \textbf{0.764}** & \textbf{0.819}** & 0.769*** \\
RRARDI,POOS-CV & 1,009 & \textbf{1.005} & 1,018 & 1,018 & \textbf{1.049} & 1,151 & 0,965 & 1,023 & 0,976 & 0.802** \\
RRARDI,K-fold & 1,083 & \textbf{0.924} & \textbf{0.977} & 0,995 & \textbf{1.071} & 1,339 & \textbf{0.752}** & 0,889 & 0,853 & 0.682*** \\
RFARDI,POOS-CV & 0,976 & \textbf{0.946} & \textbf{0.969} & \textbf{0.928} & \textbf{0.982} & 0,895 & \textbf{0.853}* & 0.840** & \textbf{0.781}*** & 0.808*** \\
RFARDI,K-fold & \textbf{0.937}* & \textbf{0.961} & \textbf{0.979} & \underline{\textbf{0.913}} & \underline{\textbf{0.957}} & 0.872** & \textbf{0.785}** & \textbf{0.810}** & \underline{\textbf{0.775}}** & 0.757*** \\
KRR-ARDI,POOS-CV & 1.138** & \textbf{1.112}* & 1.181** & 1.141*** & \textbf{1.021} & 1,123 & 1,059 & 1,117 & 1,028 & 0,919 \\
KRR,ARDI,K-fold & 1,054 & \textbf{1.058} & 1.118** & 1,065 & \textbf{0.994} & 1,035 & 0,909 & 0,972 & 0,955 & 0.849** \\
$(B_1,\alpha=\hat{\alpha})$,POOS-CV & 1.153*** & 1.213*** & 1.168** & 1.107** & \textbf{1.038} & 1,134 & 1.238** & 1.191* & 1,009 & 0,926 \\
$(B_1,\alpha=\hat{\alpha})$,K-fold & 1,069 & 1.193*** & 1.186*** & 1.120** & 1.079* & 1,103 & 1,155 & 1.212*** & 1.151*** & 0.901* \\
$(B_1,\alpha=1)$,POOS-CV & 1.118** & 1.215*** & 1.184** & 1.153*** & \textbf{1.054} & 1,135 & 1,178 & 1.194* & 1,086 & 0,954 \\
$(B_1,\alpha=1)$,K-fold & 1,056 & 1.166*** & 1.122** & 1.079** & \textbf{1.016} & 1,048 & 1,151 & 1,078 & 1.117*** & 0.878** \\
$(B_1,\alpha=0)$,POOS-CV & 1.158*** & 1.281*** & 1.300*** & 1.171** & 1.062** & 1,119 & 1,163 & 1,172 & 1,049 & 1,012 \\
$(B_1,\alpha=0)$,K-fold & 1.453*** & \textbf{1.219}** & 1.288* & 1.103** & \textbf{1.039} & 1,325 & 0,947 & 1,069 & 1.072** & 0,966 \\
$(B_2,\alpha=\hat{\alpha})$,POOS-CV & 1.092* & \textbf{1.107}* & 1.140* & 1.105* & \textbf{1.082} & 0,98 & 1,143 & 1,14 & 0,997 & 0.826** \\
$(B_2,\alpha=\hat{\alpha})$,K-fold & 1,036 & \textbf{1.088}** & 1.167** & 1,082 & 1,129 & 1.080** & 1.139** & 1,119 & \textbf{0.814}** & \underline{\textbf{0.628}}*** \\
$(B_2,\alpha=1)$,POOS-CV & 1.158** & \textbf{1.136}* & 1.194** & 1.187*** & \textbf{1.027} & 1,051 & 1,188 & 1.223** & 1,005 & 0.839** \\
$(B_2,\alpha=1)$,K-fold & 1,057 & 1.179*** & 1.113* & 1,072 & 1,153 & 1,107 & 1.263*** & 1,056 & 0.872* & 0.672*** \\
$(B_2,\alpha=0)$,POOS-CV & 1.054* & \textbf{1.081}* & 1.194** & 1,049 & \textbf{1.079} & 1.084* & 1,1 & 1,056 & 0,883 & 0.865** \\
$(B_2,\alpha=0)$,K-fold & 1.072* & \textbf{1.088} & 1.133* & 1,083 & 1.255* & 1.133** & 1,135 & 1,13 & 0.853* & 0.791*** \\
$(B_3,\alpha=\hat{\alpha})$,POOS-CV & 1,061 & 1.128** & 1.165** & 1,055 & 1.052** & 1,05 & 1.164* & 1.183** & 1,027 & 1,003 \\
$(B_3,\alpha=\hat{\alpha})$,K-fold & 1.128** & \textbf{1.057} & 1.149** & 1.125*** & \textbf{1.005} & 1,091 & 1,049 & 1.093* & 1,023 & 0.764*** \\
$(B_3,\alpha=1)$,POOS-CV & 1.096* & 1.174** & 1.186** & 1.138** & 1.079*** & 1,095 & 1.202* & 1.192* & 1,05 & 1,006 \\
$(B_3,\alpha=1)$,K-fold & 1,065 & \textbf{1.106}** & 1.153** & 1.188*** & 1.129* & 1,052 & 1,107 & 1,149 & 1,04 & 0.825** \\
$(B_3,\alpha=0)$,POOS-CV & 1,063 & \textbf{1.100}* & 1.118*** & 1.168** & \textbf{1.015} & 1,012 & 1,14 & 1.144** & 1.166* & 1,001 \\
$(B_3,\alpha=0)$,K-fold & 1.441** & \textbf{1.188}* & 1.144*** & 1.152* & \textbf{1.049}* & 1.584** & 1,085 & 1.122*** & 1,104 & 0,986 \\
SVR-ARDI,Lin,POOS-CV & 1,046 & \textbf{1.201}* & 1,108 & 1,064 & 1.106* & 0,989 & 1,119 & 1,069 & 1,004 & 1,007 \\
SVR-ARDI,Lin,K-fold & 1,105 & \textbf{1.010} & 1.265** & 1,038 & 1,088 & 1,285 & 1,032 & 1,093 & 0,925 & 0.776*** \\
SVR-ARDI,RBF,POOS-CV & 1,053 & \textbf{1.021} & 1,118 & 1.080* & 1,441 & 1,077 & 1,043 & 1,069 & 0,999 & 1,754 \\
SVR-ARDI,RBF,K-fold & 0,986 & \textbf{0.987} & 1,058 & \textbf{0.981} & \textbf{1.016} & 0,932 & 0,873 & \underline{\textbf{0.755}}** & \textbf{0.830}* & \textbf{0.679}*** \\
\hline\hline
\end{tabular}
}
\end{center}
\vspace{-0.4cm}
{\scriptsize \emph{%
\singlespacing{Note: The numbers represent the relative. with respect to AR,BIC model. root MSPE. Models retained in model confidence set are in bold. the
minimum values are underlined. while
$^{***}$. $^{**}$. $^{*}$ stand for 1\%. 5\% and 10\% significance of
Diebold-Mariano test.}}}
\end{table}

\clearpage

\begin{table}[!h]
\caption{Investment: Relative Root MSPE}
\label{tab:INV}\vspace{-0.3cm}
\par
\begin{center}
{\scriptsize
\begin{tabular}{llllll|lllll}
\hline\hline
 & \multicolumn{5}{c}{Full Out-of-Sample} & \multicolumn{5}{|c}{NBER Recessions Periods} \\
Models & h=1 & h=2 & h=3 & h=4 & h=8 & \multicolumn{1}{|l}{h=1} & h=2 & h=3 & h=4 & h=8 \\
\hline
\multicolumn{3}{l}{Data-poor ($H_t^-$) models} \\
AR,BIC (RMSPE) & 0,4078 & 0,3385 & 0,2986 & 0,277 & 0,2036 & 0,7551 & 0,6866 & 0,5725 & 0,5482 & 0,3834 \\
AR,AIC & 1.015* & 1.011* & 1.007* & 1 & 0,996 & 1.023** & 1.015** & 1.010* & 1 & 0,991 \\
AR,POOS-CV & 0.995* & 1,004 & 1.007** & 1,004 & 1,007 & 1 & 1.008* & 1.006** & 1,008 & 1,03 \\
AR,K-fold & 1,007 & 1,004 & 1,009 & 1 & 1,021 & 1,002 & 1.018** & 1.024*** & 1,017 & 1.040* \\
RRAR,POOS-CV & 1,004 & 1,001 & 1.013*** & 1.007** & 1,001 & 1,01 & 1,002 & 1.016*** & 1.007* & 1,006 \\
RRAR,K-fold & 1.015** & 1.013* & 1.008* & 1 & 1,002 & 1.026*** & 1,012 & 1.016*** & 1.013*** & 0,998 \\
RFAR,POOS-CV & 1,055 & 1,013 & 0,979 & 0,985 & 1,046 & 1,024 & 0.905** & \textbf{0.880}*** & 0,978 & 1,022 \\
RFAR,K-fold & 1,036 & 1,016 & 1,019 & 1 & 0,977 & 0,992 & 0,942 & 1,007 & 0,934 & 0,957 \\
KRR-AR,POOS-CV & 1,036 & 1 & \textbf{0.979} & 1,001 & 0,953 & 1.079* & 0,937 & 0,989 & 1,003 & 0.947** \\
KRR,AR,K-fold & 0,996 & 1,008 & \textbf{0.961}* & 1 & 0.969** & 1,022 & 0,987 & 0,975 & 1,015 & 0.965*** \\
SVR-AR,Lin,POOS-CV & 1,033 & 1.097** & 1.096*** & 1.050* & 1.116** & 1,035 & 1,061 & 1.041** & 1 & 0,98 \\
SVR-AR,Lin,K-fold & 1.033* & 1.033* & 1.026** & 1.016* & 1,019 & 1.063** & 1,021 & 1.028* & 0,998 & 1,004 \\
SVR-AR,RBF,POOS-CV & 1.038*** & 1,13 & 1.062*** & 1.047** & 1.094*** & 1.050** & 1,145 & 1.069** & 1.008** & 1,006 \\
SVR-AR,RBF,K-fold & 1,03 & 1,026 & 1.039** & 1,01 & 0,986 & 1.066* & 1,018 & 1.040** & 0,994 & 0,995 \\
\hline
\multicolumn{3}{l}{Data-rich ($H_t^+$) models} \\
ARDI,BIC & \textbf{0.749}*** & \underline{\textbf{0.774}}** & \underline{\textbf{0.862}}* & \underline{\textbf{0.827}}** & \textbf{0.911}* & \textbf{0.603}*** & \underline{\textbf{0.665}}*** & \textbf{0.851} & \textbf{0.827}*** & 0,949 \\
ARDI,AIC & \textbf{0.757}*** & \textbf{0.894}* & \textbf{0.933} & \textbf{0.831}* & 0,948 & \textbf{0.601}*** & 0,847 & 0,936 & \underline{\textbf{0.773}}** & 0.849** \\
ARDI,POOS-CV & \textbf{0.745}*** & \textbf{0.801}** & \textbf{0.918} & 0,913 & 0,979 & \textbf{0.623}*** & \textbf{0.736}** & \textbf{0.939} & \textbf{0.809}*** & 0,924 \\
ARDI,K-fold & \textbf{0.765}*** & \textbf{0.905} & \textbf{0.944} & \textbf{0.854} & 1,009 & \underline{\textbf{0.584}}*** & 0,837 & 0,993 & \textbf{0.784}** & 0.811*** \\
RRARDI,POOS-CV & \textbf{0.776}*** & \textbf{0.858}** & \textbf{0.916} & 0,984 & 0,976 & \textbf{0.626}*** & 0,831 & \textbf{0.937} & 0,945 & 0,969 \\
RRARDI,K-fold & \underline{\textbf{0.742}}*** & \textbf{0.866}* & \textbf{0.912} & \textbf{0.925} & 0,985 & \textbf{0.603}*** & 0.810* & \textbf{0.931} & 0,923 & 0.828*** \\
RFARDI,POOS-CV & 0.907** & \textbf{0.910}** & \textbf{0.884}** & \textbf{0.833}** & \textbf{0.814}** & 0.917* & 0,898 & \textbf{0.885} & \textbf{0.790}*** & 0.750*** \\
RFARDI,K-fold & 0,951 & \textbf{0.927}* & \textbf{0.875}** & \textbf{0.830}** & \textbf{0.806}** & 0,966 & 0,92 & 0,922 & 0.830** & 0.735*** \\
KRR-ARDI,POOS-CV & 0,989 & \textbf{0.945} & \textbf{0.966} & 0,942 & 0.919* & 1,01 & 0,95 & 1,028 & 0,959 & 0,933 \\
KRR,ARDI,K-fold & 0,978 & \textbf{0.952} & 0,995 & 0.937* & 0.930* & 0,974 & 0.932* & 1,049 & 0,983 & 0,987 \\
$(B_1,\alpha=\hat{\alpha})$,POOS-CV & 1,036 & 0,976 & 1,014 & 1,007 & 0,939 & 0.884** & 0.916*** & 1,006 & 0.925* & 0,965 \\
$(B_1,\alpha=\hat{\alpha})$,K-fold & 1,046 & 0,967 & \textbf{0.939} & 0.915* & 1,012 & 1.076* & 0,964 & 0,951 & 0.894*** & 0,993 \\
$(B_1,\alpha=1)$,POOS-CV & 1,023 & 0,991 & 0,989 & 0,941 & 0,966 & 0.889* & 0,954 & 0,974 & 0.902* & 0,973 \\
$(B_1,\alpha=1)$,K-fold & 0,953 & \textbf{0.914}* & \textbf{0.918}* & \textbf{0.887}** & 1,018 & 0.905* & 0.941** & 0,959 & 0.899*** & 0,953 \\
$(B_1,\alpha=0)$,POOS-CV & 1,019 & 0,997 & 1.110** & 1,045 & 1,013 & 0,973 & 0,997 & 1.078*** & 1.071* & 1,008 \\
$(B_1,\alpha=0)$,K-fold & 1.117** & 0,98 & \textbf{0.977} & 0,971 & 0,93 & 1,012 & 0.931** & \textbf{0.897} & 0,914 & 0,912 \\
$(B_2,\alpha=\hat{\alpha})$,POOS-CV & 0,996 & 0,973 & 1,01 & 1,016 & \textbf{0.915} & 1,038 & 0,974 & 1,047 & 0,989 & 0.848** \\
$(B_2,\alpha=\hat{\alpha})$,K-fold & 0,974 & 0,975 & 0,958 & 1,005 & 0,956 & 1,026 & 0,965 & 0,94 & 0.886** & 0.662** \\
$(B_2,\alpha=1)$,POOS-CV & 0,988 & 0,961 & 1,076 & 1,069 & 1,003 & 1,008 & 0.959* & 1.150** & 1,067 & 0.874*** \\
$(B_2,\alpha=1)$,K-fold & 0,974 & 0,965 & 0,967 & 1,014 & \underline{\textbf{0.794}}** & 0,997 & 0,973 & 0,975 & \textbf{0.854}* & \underline{\textbf{0.615}}*** \\
$(B_2,\alpha=0)$,POOS-CV & 1,033 & 0,975 & 1,048 & 1,057 & \textbf{0.904}* & 1,056 & 0,991 & 1,102 & 1,031 & 0.871** \\
$(B_2,\alpha=0)$,K-fold & 1,023 & 0,923 & 0,966 & 0,996 & 0,966 & 1,025 & 0.892** & 0,993 & 0,946 & 0,894 \\
$(B_3,\alpha=\hat{\alpha})$,POOS-CV & 0,961 & 0,982 & 1,006 & 0,988 & \textbf{0.920}** & 0.901* & 0,991 & 1,058 & 0,996 & 0.929*** \\
$(B_3,\alpha=\hat{\alpha})$,K-fold & 0.948* & 0,976 & \textbf{0.921} & \textbf{0.884}** & 0,941 & 0.928* & 0.967* & \textbf{0.913} & 0.845** & 0.888*** \\
$(B_3,\alpha=1)$,POOS-CV & 0,946 & 0,985 & \textbf{0.957} & 0,977 & 0.939* & 0,916 & 0,993 & 1,037 & 0,975 & 0.941** \\
$(B_3,\alpha=1)$,K-fold & 0,956 & 0,966 & \textbf{0.891}** & \textbf{0.894}** & 0,954 & 0,937 & 0,973 & \textbf{0.894}** & 0.881*** & 0.880*** \\
$(B_3,\alpha=0)$,POOS-CV & 1.110* & 1.036* & 1,027 & 1,027 & 1 & 1,011 & 0,97 & 1,004 & 1,011 & 1,001 \\
$(B_3,\alpha=0)$,K-fold & 1,151 & 0,989 & 0,982 & 1,136 & 1,023 & 0,99 & 0,965 & 0,974 & 1,089 & 0,968 \\
SVR-ARDI,Lin,POOS-CV & 0,975 & 0,995 & 1,077 & 1,013 & 1,013 & 1,042 & 0,974 & 1,086 & 0,986 & 0,938 \\
SVR-ARDI,Lin,K-fold & \textbf{0.758}*** & \textbf{0.805}** & \textbf{0.908} & 1,094 & 1,098 & \textbf{0.623}*** & \textbf{0.739}*** & \underline{\textbf{0.808}}* & 0,975 & 0,964 \\
SVR-ARDI,RBF,POOS-CV & \textbf{0.791}*** & \textbf{0.909} & \textbf{0.969} & 0,956 & 0,948 & \textbf{0.711}*** & 0,856 & \textbf{0.876} & 0,934 & 0.904** \\
SVR-ARDI,RBF,K-fold & \textbf{0.804}*** & \textbf{0.836}* & \textbf{0.913} & 0,962 & 0,979 & 0.737*** & \textbf{0.728}** & \textbf{0.852} & 0,965 & 0.812** \\
\hline\hline
\end{tabular}
}
\end{center}
\vspace{-0.4cm}
{\scriptsize \emph{%
\singlespacing{Note: The numbers represent the relative. with respect to AR,BIC model. root MSPE. Models retained in model confidence set are in bold. the
minimum values are underlined. while
$^{***}$. $^{**}$. $^{*}$ stand for 1\%. 5\% and 10\% significance of
Diebold-Mariano test.}}}
\end{table}

\clearpage

\begin{table}[!h]
\caption{Income: Relative Root MSPE}
\label{tab:INC}\vspace{-0.3cm}
\par
\begin{center}
{\scriptsize
\begin{tabular}{llllll|lllll}
\hline\hline
 & \multicolumn{5}{c}{Full Out-of-Sample} & \multicolumn{5}{|c}{NBER Recessions Periods} \\
Models & h=1 & h=2 & h=3 & h=4 & h=8 & \multicolumn{1}{|l}{h=1} & h=2 & h=3 & h=4 & h=8 \\
\hline
\multicolumn{3}{l}{Data-poor ($H_t^-$) models} \\
AR,BIC (RMSPE) & \textbf{0.1011} & \textbf{0.0669} & \textbf{0.0581} & \textbf{0.0528} & 0,0417 & 0,1336 & 0,088 & 0,0803 & 0,0772 & 0,0683 \\
AR,AIC & \textbf{0.995} & \textbf{0.991} & \textbf{0.998} & \textbf{1.000} & 1 & 1 & 0.969* & 1 & 1 & 1 \\
AR,POOS-CV & \textbf{0.985}* & \textbf{0.996} & \textbf{1.002} & \textbf{0.999} & \textbf{0.991} & \textbf{0.938}** & 0.980** & 0.992* & 0,998 & 0,993 \\
AR,K-fold & \textbf{0.987} & \textbf{0.992} & \textbf{0.994} & \textbf{0.998} & 1,002 & \textbf{0.947}** & 0.963** & 0.969** & 1 & 0,999 \\
RRAR,POOS-CV & \textbf{0.987} & \textbf{0.996} & \textbf{1.002} & 1.006*** & 0,995 & \textbf{0.939}** & 0.976** & 0,994 & 1.006*** & 0.991** \\
RRAR,K-fold & \textbf{0.988} & \textbf{0.991} & \textbf{1.000} & \textbf{1.003}* & 1 & \textbf{0.945}** & 0.972*** & 1 & 1.008*** & 0.999** \\
RFAR,POOS-CV & \textbf{1.028} & 1.068** & 1.075** & 1,016 & 1,008 & 1,072 & 1.103* & 0.939* & 0,975 & 0,975 \\
RFAR,K-fold & 1.132*** & \textbf{1.024} & 1.056* & 1,01 & 1,036 & 1.124** & 0,976 & 0,989 & 0,985 & 0,957 \\
KRR-AR,POOS-CV & \textbf{0.990} & \textbf{1.000} & 1,033 & 1.070** & \textbf{0.967} & \textbf{0.923}** & 0.905** & 0,959 & 0,979 & 0.908* \\
KRR,AR,K-fold & \textbf{0.988} & \textbf{0.991} & \textbf{1.004} & 1.049* & 1,037 & \textbf{0.964} & 0.897*** & 0,956 & 0,978 & 0.913** \\
SVR-AR,Lin,POOS-CV & \textbf{1.000} & 1,056 & \textbf{1.009} & 1,881 & 1.165** & 0,976 & 0,954 & 0,97 & 0,993 & 1,111 \\
SVR-AR,Lin,K-fold & \textbf{0.993} & \textbf{0.995} & \textbf{0.996} & \textbf{0.988} & \textbf{0.962}*** & 0,976 & 0,996 & 1,015 & 1.016** & 0.965*** \\
SVR-AR,RBF,POOS-CV & \textbf{0.975} & 1,049 & 1,022 & 1.066* & \textbf{0.969} & \textbf{0.939}** & 0,959 & 0,973 & 1,01 & 0.928*** \\
SVR-AR,RBF,K-fold & 1.012* & \textbf{0.996} & \textbf{1.009} & 1,012 & 1.018* & 1,01 & 1 & 1.026* & 1.036*** & 1.029** \\
\hline
\multicolumn{3}{l}{Data-rich ($H_t^+$) models} \\
ARDI,BIC & \textbf{1.059} & \textbf{0.981} & \textbf{0.913}** & \textbf{0.939} & \textbf{0.963} & 1,257 & \textbf{0.773}** & \underline{\textbf{0.726}}*** & \textbf{0.777}*** & 0.769*** \\
ARDI,AIC & \textbf{1.016} & \textbf{0.940} & \underline{\textbf{0.911}}* & \textbf{0.966} & \textbf{0.992} & 1,05 & \underline{\textbf{0.611}}*** & \textbf{0.757}** & 0,886 & 0.721*** \\
ARDI,POOS-CV & \textbf{1.040} & \textbf{0.975} & \textbf{0.945} & \textbf{0.933} & 1,128 & 1,149 & \textbf{0.757}** & \textbf{0.753}*** & \textbf{0.757}*** & 0.770** \\
ARDI,K-fold & 1,065 & \textbf{0.946} & \textbf{0.953} & \textbf{0.974} & 1,028 & 1,175 & \textbf{0.664}** & 0.796** & 0,898 & 0.689*** \\
RRARDI,POOS-CV & \textbf{1.038} & \textbf{1.007} & \textbf{0.971} & \underline{\textbf{0.917}} & 1,058 & 1,12 & \textbf{0.796}* & 0,869 & \textbf{0.767}*** & 0.743*** \\
RRARDI,K-fold & 1,06 & \textbf{0.973} & \textbf{0.925} & \textbf{0.919} & \textbf{0.999} & 1,197 & 0,82 & 0.830* & 0,871 & \underline{\textbf{0.627}}*** \\
RFARDI,POOS-CV & \underline{\textbf{0.954}}* & \underline{\textbf{0.932}}** & \textbf{0.936}* & \textbf{0.919}* & \textbf{0.910}* & \underline{\textbf{0.916}} & \textbf{0.807}*** & 0.822** & \textbf{0.762}*** & 0.678*** \\
RFARDI,K-fold & \textbf{0.977} & \textbf{0.957} & \textbf{0.929}** & \textbf{0.925}** & \underline{\textbf{0.886}}* & \textbf{0.931} & 0.821** & 0.802** & 0.795*** & \textbf{0.675}*** \\
KRR-ARDI,POOS-CV & 1,026 & 1.069*** & 1,025 & 1.090* & 0,985 & \textbf{0.948} & 0,991 & 0.936** & 0,954 & 0.894** \\
KRR,ARDI,K-fold & \textbf{0.969} & 1,012 & 1,075 & 1.084* & 0,991 & \textbf{0.947} & 0.925** & 0,942 & 0.929* & 0.849*** \\
$(B_1,\alpha=\hat{\alpha})$,POOS-CV & \textbf{1.010} & 1.045* & \textbf{0.997} & 1,016 & 1,015 & \textbf{0.948}*** & 0,993 & 1,018 & 1,034 & 0.922* \\
$(B_1,\alpha=\hat{\alpha})$,K-fold & \textbf{1.008} & 1,02 & \textbf{1.031} & 1,025 & 1,055 & 0,988 & 1,063 & 0.882*** & 0,972 & 0,903 \\
$(B_1,\alpha=1)$,POOS-CV & \textbf{1.010} & 1.105** & 1.070* & 1.035* & 1,016 & 0,998 & 0,963 & 0,985 & 1.067** & 0.914** \\
$(B_1,\alpha=1)$,K-fold & 1,017 & \textbf{1.020} & 1,014 & 1,015 & 1,091 & 1,036 & 1,066 & 0,974 & 0,958 & 0.895* \\
$(B_1,\alpha=0)$,POOS-CV & 1.030* & 1,034 & 1.050** & 1.075*** & 1,014 & \textbf{0.942}*** & 1,021 & 1,034 & 1,031 & 1.120* \\
$(B_1,\alpha=0)$,K-fold & 1.023* & \textbf{0.996} & \textbf{1.032} & \textbf{1.010} & \textbf{0.953} & 0.972* & 0.921* & 0.904*** & 0,964 & 0,936 \\
$(B_2,\alpha=\hat{\alpha})$,POOS-CV & \textbf{1.001} & \textbf{0.976} & \textbf{0.989} & 1,027 & \textbf{0.972} & 0,994 & 0.874** & 0,998 & 1.043** & 0.772** \\
$(B_2,\alpha=\hat{\alpha})$,K-fold & \textbf{1.020} & \textbf{0.979} & \textbf{0.975} & \textbf{0.988} & 1.220** & 1.054* & 0,934 & 0,931 & 0,897 & 0.790** \\
$(B_2,\alpha=1)$,POOS-CV & \textbf{0.992} & \textbf{0.988} & \textbf{0.991} & \textbf{1.005} & \textbf{0.947} & 0,978 & 1,003 & 0,991 & 1,002 & 0.877*** \\
$(B_2,\alpha=1)$,K-fold & 1.080* & \textbf{0.971} & \textbf{0.958} & \textbf{0.966} & 1.262** & 1.253* & 0.872** & 0.848** & 0.838** & \textbf{0.691}** \\
$(B_2,\alpha=0)$,POOS-CV & \textbf{1.022} & \textbf{0.978} & \textbf{0.958} & \textbf{0.993} & \textbf{0.964} & 1,061 & 0.844*** & 0,924 & 0,931 & 0.722*** \\
$(B_2,\alpha=0)$,K-fold & 1,028 & \textbf{1.000} & \textbf{0.990} & \textbf{0.997} & 1,158 & 1,051 & 0,955 & 0,983 & 0,921 & 0.830** \\
$(B_3,\alpha=\hat{\alpha})$,POOS-CV & \textbf{1.009} & \textbf{1.010} & 1,013 & 1,032 & 1,015 & 0.953* & 0,993 & 1.047** & 1,027 & 0.935** \\
$(B_3,\alpha=\hat{\alpha})$,K-fold & \textbf{0.990} & \textbf{0.995} & \textbf{0.997} & 1,024 & 1.085* & 0,962 & 0,924 & 0,969 & 1.051* & 0.882*** \\
$(B_3,\alpha=1)$,POOS-CV & \textbf{0.995} & \textbf{1.005} & \textbf{1.006} & 1,035 & 1.040** & 0,978 & 0,984 & 1.056** & 1,047 & 0.991* \\
$(B_3,\alpha=1)$,K-fold & \textbf{1.003} & \textbf{1.006} & \textbf{1.005} & \textbf{0.999} & 1.171*** & 1,001 & 0.931* & 0,999 & 1,002 & 0.862*** \\
$(B_3,\alpha=0)$,POOS-CV & \textbf{0.985} & \textbf{0.987} & \textbf{0.986} & 1,04 & \textbf{0.984} & \textbf{0.941}** & 0,954 & 0,987 & 1,145 & 0.959** \\
$(B_3,\alpha=0)$,K-fold & \textbf{0.993} & 1,132 & \textbf{1.000} & 1,078 & 1.166** & \textbf{0.947}** & 0.906** & 0,991 & 1,134 & 1,001 \\
SVR-ARDI,Lin,POOS-CV & 1,06 & 1,081 & \textbf{1.005} & \textbf{0.982} & 1,082 & \textbf{0.958} & 1,019 & 0,906 & 0.863* & 0.888** \\
SVR-ARDI,Lin,K-fold & 1.170* & \textbf{0.968} & 1,042 & \textbf{0.984} & 1,144 & 1.512* & 0,852 & 0.821* & \underline{\textbf{0.736}}** & 0,988 \\
SVR-ARDI,RBF,POOS-CV & 1.147** & 1,097 & \textbf{0.975} & \textbf{0.972} & 1,025 & 1.311* & 1,069 & 0,97 & 0,992 & 0,931 \\
SVR-ARDI,RBF,K-fold & \textbf{1.008} & 1,117 & \textbf{0.985} & \textbf{0.998} & 1,191 & \textbf{0.943} & 1,286 & 0.827** & 0.843** & 0.770*** \\
\hline\hline
\end{tabular}
}
\end{center}
\vspace{-0.4cm}
{\scriptsize \emph{%
\singlespacing{Note: The numbers represent the relative. with respect to AR,BIC model. root MSPE. Models retained in model confidence set are in bold. the
minimum values are underlined. while
$^{***}$. $^{**}$. $^{*}$ stand for 1\%. 5\% and 10\% significance of
Diebold-Mariano test.}}}
\end{table}

\clearpage

\begin{table}[!h]
\caption{PCE Deflator: Relative Root MSPE}
\label{tab:PCE}\vspace{-0.3cm}
\par
\begin{center}
{\scriptsize
\begin{tabular}{llllll|lllll}
\hline\hline
 & \multicolumn{5}{c}{Full Out-of-Sample} & \multicolumn{5}{|c}{NBER Recessions Periods} \\
Models & h=1 & h=2 & h=3 & h=4 & h=8 & \multicolumn{1}{|l}{h=1} & h=2 & h=3 & h=4 & h=8 \\
\hline
\multicolumn{3}{l}{Data-poor ($H_t^-$) models} \\
AR,BIC (RMSPE) & \textbf{0.0442} & 0,0421 & 0,0395 & \textbf{0.0387} & \textbf{0.0418} & \textbf{0.0798} & 0,0827 & 0,078 & 0,069 & 0,0644 \\
AR,AIC & \textbf{1.000} & 0,999 & 0.992** & \textbf{0.991}** & \textbf{0.976}* & 1.033* & 1,018 & 0,997 & 1 & 0.976* \\
AR,POOS-CV & \textbf{0.991} & \textbf{0.969}** & \textbf{0.990}* & \textbf{0.968}** & \textbf{0.968}** & 1,025 & 0,976 & 0,998 & 0.984** & 0.974* \\
AR,K-fold & \textbf{0.992} & \textbf{0.984} & 0,998 & \textbf{0.984}** & \textbf{0.988} & 1,032 & 1,007 & 0,997 & 0,993 & 0,989 \\
RRAR,POOS-CV & \textbf{0.974}** & \textbf{0.953}** & \textbf{0.964}** & \textbf{0.967}* & \textbf{0.958}** & 1.019* & 0,965 & 0,968 & 0,981 & 0.938*** \\
RRAR,K-fold & \textbf{1.000} & \textbf{0.983} & \textbf{0.988}*** & \textbf{0.992}* & \textbf{0.976}* & 1,025 & 1,005 & 0.994** & 0,993 & 0.955** \\
RFAR,POOS-CV & \textbf{0.981} & \textbf{0.917}** & \textbf{0.917}* & \textbf{0.936} & 1,053 & 1,059 & 0,937 & 0,94 & 1,022 & 0,896 \\
RFAR,K-fold & \textbf{0.969} & \textbf{0.921}** & \textbf{0.923}* & \textbf{0.917}* & 1,025 & \textbf{1.030} & 0,936 & 0,947 & 1,013 & 0.795** \\
KRR-AR,POOS-CV & \textbf{1.042} & \underline{\textbf{0.894}}** & \textbf{0.867}* & \textbf{0.891} & \textbf{0.903}* & 1,178 & \textbf{0.873}* & 0,817 & \textbf{0.760}** & 0.775** \\
KRR,AR,K-fold & \textbf{0.997} & \textbf{0.908} & \underline{\textbf{0.860}}* & \underline{\textbf{0.870}}* & 1,009 & \textbf{1.021} & \textbf{0.855} & \textbf{0.770}* & \textbf{0.768}** & 0.783** \\
SVR-AR,Lin,POOS-CV & \textbf{1.011} & 1.198*** & 1.075* & 1.488** & 1.410*** & 1,04 & 1.084** & 1,001 & 1,202 & 1.300* \\
SVR-AR,Lin,K-fold & 1.563*** & 1.950*** & 1.914*** & 1.805*** & 1.662*** & 1.329* & 1.622*** & 1.293** & 1,116 & 0,948 \\
SVR-AR,RBF,POOS-CV & \textbf{0.990} & 1,007 & 1,04 & \textbf{1.058} & 1.188** & 1,009 & 0,933 & 1,017 & 1,114 & 1,002 \\
SVR-AR,RBF,K-fold & \textbf{1.083}** & 1.040** & 1,059 & 1.222** & 1.189** & 1.019** & 0,992 & 0.931*** & 1,032 & 0,865 \\
\hline
\multicolumn{3}{l}{Data-rich ($H_t^+$) models} \\
ARDI,BIC & \textbf{1.016} & \textbf{0.978} & \textbf{0.994} & \textbf{0.990} & \textbf{0.986} & 1,048 & \textbf{0.949} & 0,939 & \underline{\textbf{0.714}}** & 0.731** \\
ARDI,AIC & \textbf{1.043} & 1,027 & 1,052 & \textbf{1.050} & 1,068 & 1,104 & 0,99 & 0,924 & \textbf{0.844} & 0.806** \\
ARDI,POOS-CV & \textbf{1.091} & 1,055 & 1,084 & \textbf{1.013} & \textbf{0.918} & 1.221** & 1,113 & 1,015 & \textbf{0.751}* & 0.686** \\
ARDI,K-fold & \textbf{1.037} & 1,027 & 1.092* & \textbf{1.069} & 1,047 & 1,107 & 1,007 & 0,926 & 0,853 & 0.816** \\
RRARDI,POOS-CV & \textbf{1.010} & \textbf{1.041} & \textbf{1.037} & \textbf{1.000} & \textbf{0.990} & 1,058 & 1,063 & 0,977 & \textbf{0.720}** & \textbf{0.639}** \\
RRARDI,K-fold & \textbf{0.988} & 1,014 & 1.117* & \textbf{1.073} & 1,167 & 1,023 & 0,972 & 0,976 & 0,857 & 0.681*** \\
RFARDI,POOS-CV & \underline{\textbf{0.963}} & \textbf{0.900}** & \textbf{0.895}* & \textbf{0.914} & 1,088 & 1,032 & 0,944 & 0,906 & 0,956 & 0.786*** \\
RFARDI,K-fold & \textbf{0.970} & \textbf{0.904}** & \textbf{0.931} & \textbf{0.946} & \textbf{1.040} & 1,046 & 0,932 & 0,924 & 1,026 & 0.786*** \\
KRR-ARDI,POOS-CV & \textbf{1.017} & \textbf{0.914} & \textbf{0.924} & \textbf{0.958} & \textbf{0.948} & \textbf{0.996} & \textbf{0.850}* & \textbf{0.783}* & 0.835* & 0,902 \\
KRR,ARDI,K-fold & \textbf{0.988} & \textbf{0.925} & \textbf{0.893}* & \textbf{0.904}* & \underline{\textbf{0.835}}** & 1,045 & \textbf{0.858} & 0.842* & \textbf{0.822}** & 0.668** \\
$(B_1,\alpha=\hat{\alpha})$,POOS-CV & \textbf{1.133}** & 1.200*** & 1.195** & 1.310*** & 1.267** & \textbf{0.967} & 1,018 & \textbf{0.778}* & 1,005 & 0.833** \\
$(B_1,\alpha=\hat{\alpha})$,K-fold & 1.123** & 1.221*** & 1.187* & 1.316*** & 1.179* & 1,029 & \textbf{0.871} & \textbf{0.749}** & 0,905 & 0.766*** \\
$(B_1,\alpha=1)$,POOS-CV & 1.251*** & 1.276*** & 1.208** & 1.221** & 1.403*** & 1,137 & 1,01 & \textbf{0.828} & 0,973 & 1,015 \\
$(B_1,\alpha=1)$,K-fold & 1.368*** & 1.340*** & 1.412*** & 1.409*** & 1.270** & 1.280** & 0,91 & 0,957 & 0,903 & 0.726** \\
$(B_1,\alpha=0)$,POOS-CV & 1.488** & 1.562** & 1.269* & 1.396** & 1.431*** & 1.153* & 0,961 & 0,979 & \textbf{0.793} & 1,307 \\
$(B_1,\alpha=0)$,K-fold & 1.540** & 1.493** & 1.489** & 1.429** & 1.317** & 1.125* & \underline{\textbf{0.815}} & \underline{\textbf{0.706}}* & \textbf{0.738} & 1,074 \\
$(B_2,\alpha=\hat{\alpha})$,POOS-CV & 1.131*** & 1.249** & 1.152** & 1.193** & 1,111 & 1,051 & 1,268 & 0.903* & 0.843** & \textbf{0.637}** \\
$(B_2,\alpha=\hat{\alpha})$,K-fold & 1.111** & 1,266 & 1.103* & \textbf{1.142}* & 1,079 & 1,115 & 1,387 & 0,925 & \textbf{0.823}* & 0,749 \\
$(B_2,\alpha=1)$,POOS-CV & \textbf{1.075}** & 1.078** & 1.095* & 1.233** & 1.259** & 1,026 & 0,974 & 0.912** & 0,884 & \textbf{0.606}** \\
$(B_2,\alpha=1)$,K-fold & 1.078* & 1,315 & 1.098* & \textbf{1.130}* & 1,172 & 1,11 & 1,449 & 0,933 & \textbf{0.798}** & 0.679* \\
$(B_2,\alpha=0)$,POOS-CV & 1.316** & 1.332** & 1.418*** & 1.393*** & 1.169* & 1,373 & 1.345* & 1,298 & 0,948 & \textbf{0.629}*** \\
$(B_2,\alpha=0)$,K-fold & 1.358** & 1.291** & 1.388** & 1.313** & 1,13 & 1,487 & 1,263 & 1,339 & 1,016 & \underline{\textbf{0.597}}*** \\
$(B_3,\alpha=\hat{\alpha})$,POOS-CV & \textbf{1.033}* & 1,009 & 1.063* & 1.092** & 1,102 & 1,016 & 0.945* & 0,972 & 0.885* & 0.854** \\
$(B_3,\alpha=\hat{\alpha})$,K-fold & \textbf{1.009} & 1,033 & 1.094*** & 1,056 & 1,101 & \textbf{1.000} & 1,001 & 0.946* & 0.936* & 0.790*** \\
$(B_3,\alpha=1)$,POOS-CV & \textbf{1.010} & 1.042* & 1.086** & 1.101** & 1,12 & \underline{\textbf{0.955}}* & 0.953* & 0,993 & 0,923 & 0.824** \\
$(B_3,\alpha=1)$,K-fold & \textbf{0.995} & 1,032 & 1.048** & \textbf{1.042} & 1.209** & \textbf{0.965}** & 1,007 & 0,997 & 0,947 & 0.907* \\
$(B_3,\alpha=0)$,POOS-CV & 1.084** & \textbf{1.001} & 1,017 & \textbf{1.016} & 1.117* & 1.067* & \textbf{0.910} & 0,904 & 0,917 & 0,885 \\
$(B_3,\alpha=0)$,K-fold & \textbf{1.071}* & 1.198* & 1,12 & 1.133* & 1.127* & 1.085* & 1,149 & 0,979 & 0,948 & 0,923 \\
SVR-ARDI,Lin,POOS-CV & \textbf{1.086}* & 1.271*** & 1.292*** & 1.228** & 1.220** & \textbf{1.009} & 1,13 & 1,081 & 0,945 & 0,97 \\
SVR-ARDI,Lin,K-fold & 1.136* & 1.161* & 1.351* & 1.301** & 1.169* & 1.228* & \textbf{0.881} & 1,173 & 1,145 & 1,026 \\
SVR-ARDI,RBF,POOS-CV & \textbf{1.236} & 1,019 & 1,017 & \textbf{0.958} & \textbf{0.991} & 1,47 & 0,968 & 0,939 & \textbf{0.768}*** & 0.798** \\
SVR-ARDI,RBF,K-fold & \textbf{1.054} & 1,062 & 1,063 & 1.236*** & 1,075 & 1,096 & 1,048 & 0,909 & 0,985 & 0,891 \\
\hline\hline
\end{tabular}
}
\end{center}
\vspace{-0.4cm}
{\scriptsize \emph{%
\singlespacing{Note: The numbers represent the relative. with respect to AR,BIC model. root MSPE. Models retained in model confidence set are in bold. the
minimum values are underlined. while
$^{***}$. $^{**}$. $^{*}$ stand for 1\%. 5\% and 10\% significance of
Diebold-Mariano test.}}}
\end{table}

	\begin{figure}
			\centering
\includegraphics[scale=.4]{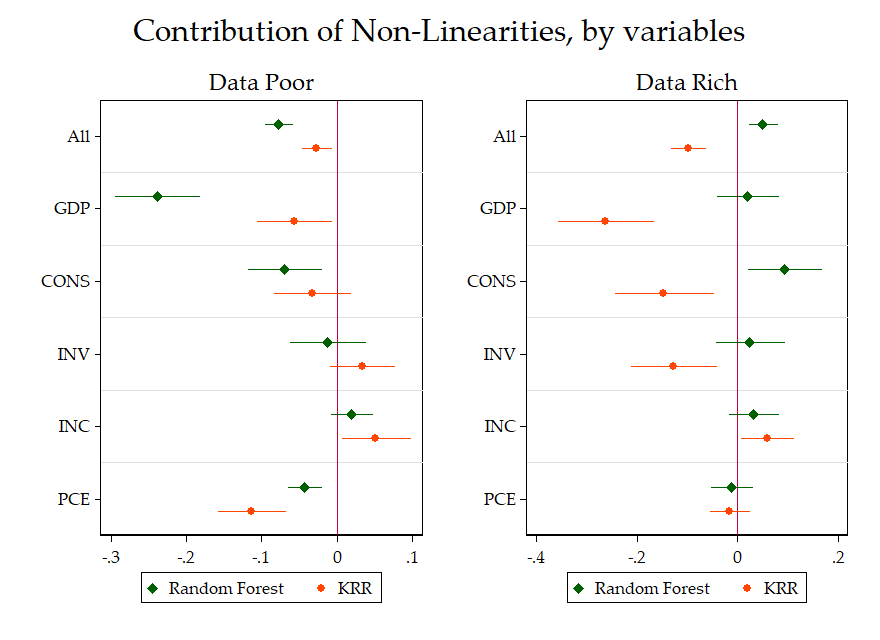}  
		\caption{\footnotesize{This compares the two NL models averaged over all horizons. The unit of the x-axis are  improvements in OOS $R^2$ over the basis model. SEs are HAC. These are the 95\% confidence bands.}}
		\label{g_nl_v_q}
		\end{figure}

		\begin{figure}
			\centering
\includegraphics[scale=.4]{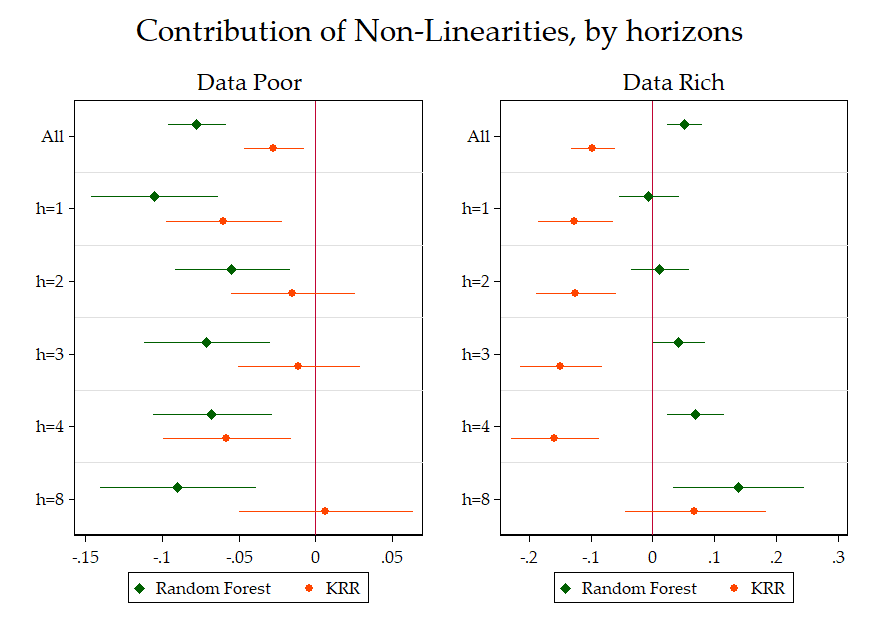}  
	\caption{\footnotesize{This compares the two NL models averaged over all variables. The unit of the x-axis are improvements in OOS $R^2$ over the basis model.  SEs are HAC. These are the 95\% confidence bands.}}
	\label{g_nl_h_q}
		\end{figure}


		\begin{figure}
			\centering
					
\includegraphics[scale=.35]{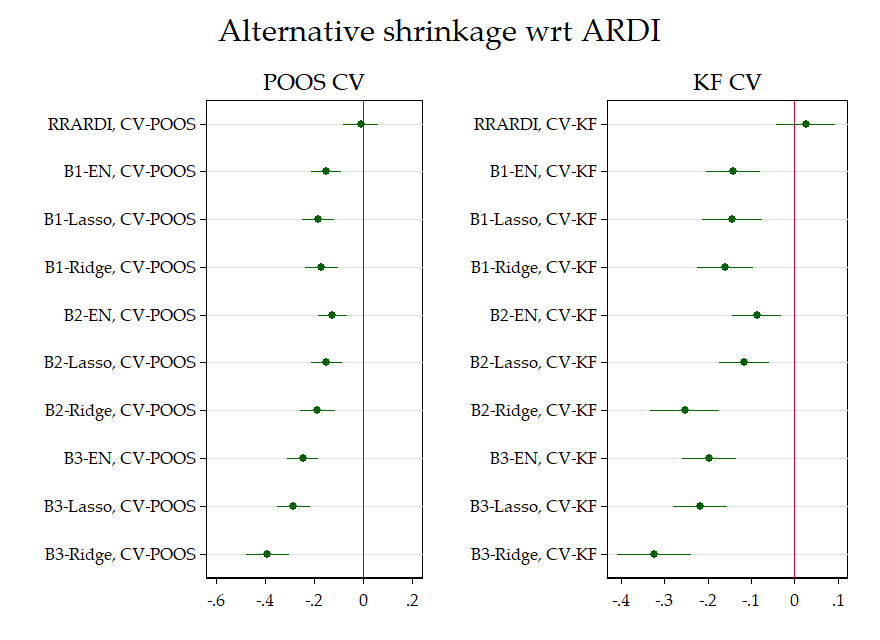}  
\caption{\footnotesize{This compares models of section \ref{sec:reg} averaged over all variables and horizons. The unit of the x-axis are improvements in OOS $R^2$ over the basis model.  The base models are ARDIs specified with POOS-CV and KF-CV respectively. SEs are HAC. These are the 95\% confidence bands.}}\label{SHgraph_q}
		\end{figure}


\begin{table}[htbp]\centering
\def\sym#1{\ifmmode^{#1}\else\(^{#1}\)\fi}
\caption{CV comparison}
\begin{tabular}{l*{5}{c}}
\hline\hline
                    &\multicolumn{1}{c}{(1)}&\multicolumn{1}{c}{(2)}&\multicolumn{1}{c}{(3)}&\multicolumn{1}{c}{(4)}&\multicolumn{1}{c}{(5)}\\
                    &\multicolumn{1}{c}{All}&\multicolumn{1}{c}{Data-rich}&\multicolumn{1}{c}{Data-poor}&\multicolumn{1}{c}{Data-rich}&\multicolumn{1}{c}{Data-poor}\\
\hline
CV-KF               &      -4.248\sym{*}  &      -8.304\sym{***}&      -0.192         &      -9.651\sym{***}&       0.114         \\
                    &     (1.940)         &     (1.787)         &     (0.424)         &     (1.886)         &     (0.386)         \\
CV-POOS             &      -2.852         &      -6.690\sym{**} &       0.985\sym{**} &      -6.772\sym{**} &       0.917\sym{*}  \\
                    &     (1.887)         &     (2.163)         &     (0.382)         &     (2.270)         &     (0.386)         \\
AIC                 &      -2.182         &      -4.722\sym{**} &       0.358         &      -5.557\sym{**} &       0.373         \\
                    &     (1.816)         &     (1.598)         &     (0.320)         &     (1.694)         &     (0.303)         \\
CV-KF * Recessions  &                     &                     &                     &       13.21\sym{**} &      -2.956\sym{*}  \\
                    &                     &                     &                     &     (4.893)         &     (1.500)         \\
CV-POOS * Recessions&                     &                     &                     &       1.002         &       0.683         \\
                    &                     &                     &                     &     (5.345)         &     (1.125)         \\
AIC * Recessions    &                     &                     &                     &       8.421         &      -0.127         \\
                    &                     &                     &                     &     (4.643)         &     (1.101)         \\
\hline
Observations        &       36960         &       18480         &       18480         &       18360         &       18360         \\
\hline\hline
\multicolumn{6}{l}{\footnotesize Standard errors in parentheses}\\
\multicolumn{6}{l}{\footnotesize \sym{*} \(p<0.05\), \sym{**} \(p<0.01\), \sym{***} \(p<0.001\)}\\
\end{tabular}
\end{table}
\label{CVreg_q}

		\begin{figure}
			\centering
\includegraphics[scale=.4]{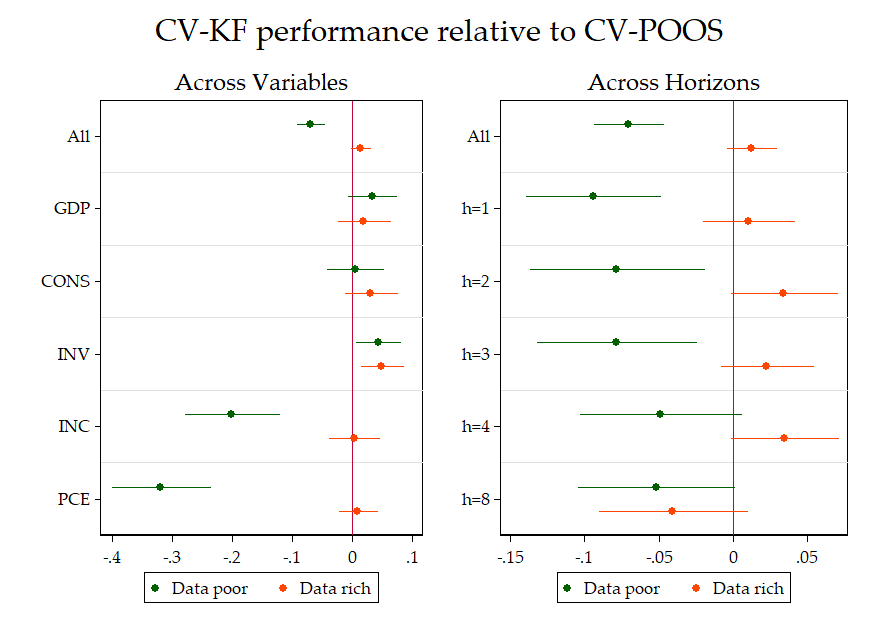}
		\caption{\footnotesize{This compares the two CVs procedure averaged over all the models that use them. The unit of the x-axis are improvements in OOS $R^2$ over the basis model.  SEs are HAC. These are the 95\% confidence bands.}}\label{CVbyX_q}
		\end{figure}
		
\begin{figure}
			\centering
\includegraphics[scale=.4]{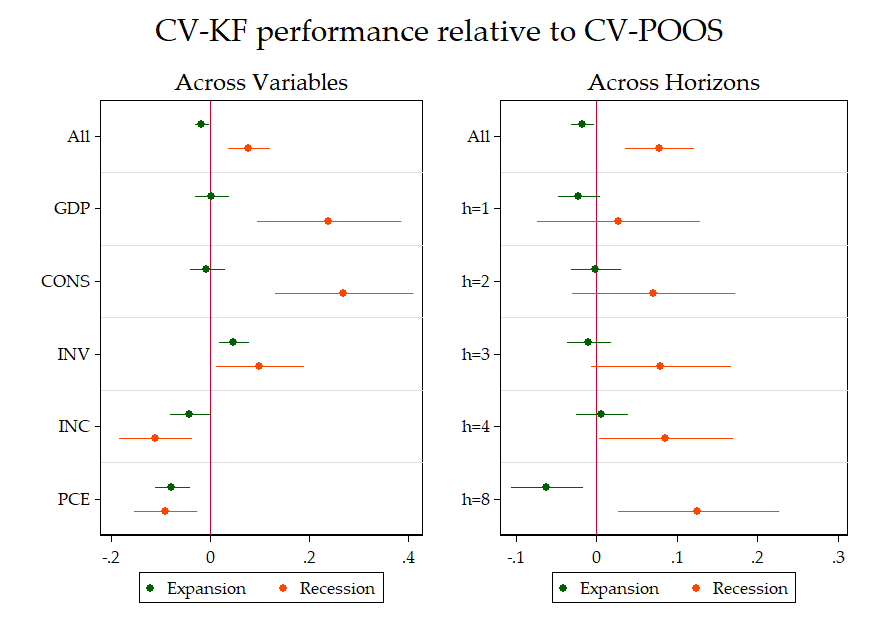}
		\caption{\footnotesize{This compares the two CVs procedure averaged over all the models that use them. The unit of the x-axis are improvements in OOS $R^2$ over the basis model.  SEs are HAC. These are the 95\% confidence bands.}}\label{CVbyrec_q}
		\end{figure}


		\begin{figure}
			\centering
\includegraphics[scale=.4]{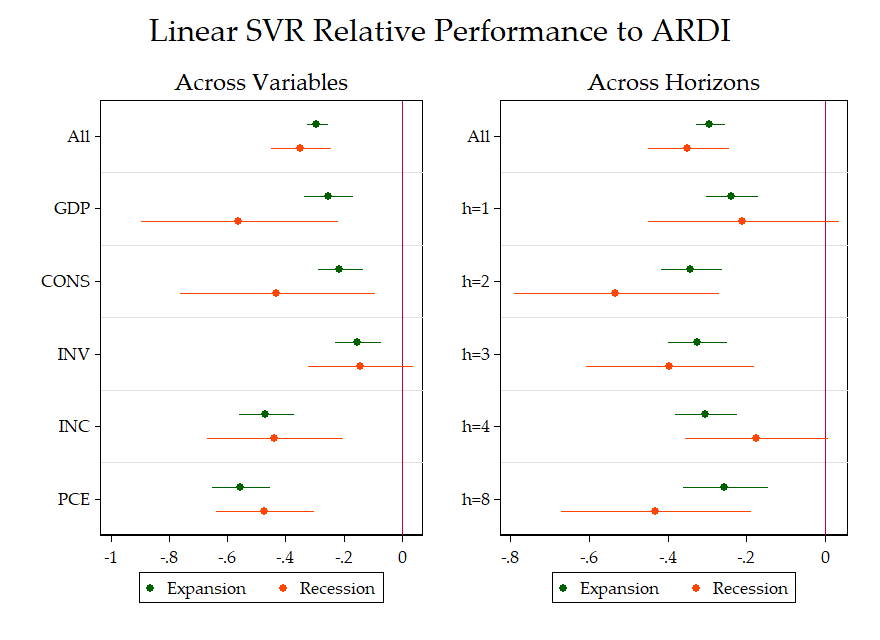}
		\caption{\footnotesize{This graph display the marginal (un)improvements by variables and horizons to opt for the SVR in-sample loss function in \textbf{both the data-poor and data-rich environments}. The unit of the x-axis are improvements in OOS $R^2$ over the basis model.  SEs are HAC. These are the 95\% confidence bands.}}\label{Leffectlin_q}
		\end{figure}

		\begin{figure}
			\centering
\includegraphics[scale=.4]{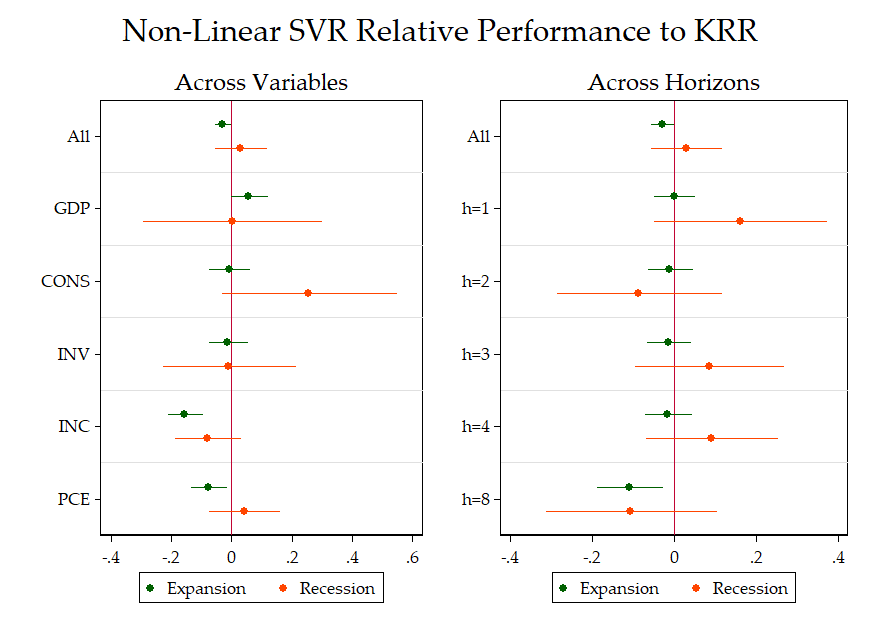}
		\caption{\footnotesize{This graph display the marginal (un)improvements by variables and horizons to opt for the SVR in-sample loss function in \textbf{both recession and expansion periods}. The unit of the x-axis are improvements in OOS $R^2$ over the basis model.  SEs are HAC. These are the 95\% confidence bands.}}\label{LeffectNL_q}
		\end{figure}

\clearpage

\section{Results with Canadian data}\label{sec:can}

In this section we present results obtained with Canadian data from \cite{FLSS2020}. It is a monthly dataset of 139 macroeconomic and financial variables, with 
categories similar to those from \cite{McCracken2016}, except that it contains much more international trade indicators to take into account the openness of Canadian 
economy. Data starts on 1981M01 and ends on 2017M12. The out-of-sample starts on 2000M01. The variables of interest are the same as in US application: industrial 
growth, unemployment rate change, term spread, CPI inflation and housing starts growth. Forecasting horizons are 1, 3, 9, 12 and 24 months. We do not compute results 
for recession periods separately since Canada has experienced only one downturn in the evaluation period. 

The results with Canadian data are overall similar to those in the paper. The main difference is a smaller NL treatment effect. That can be potentially explained through lenses of the 
analysis in section \ref{BlackBox}. The pseudo-out-of-sample covers 2000-2017 period during which Canadian financial system did not experience a dramatic nonfinancial cycle as in the US., and 
the housing bubble did not burst. The main reason for this discrepancy being more concentrated and strictly regulated (since 80's) Canadian financial system \citep{Bordo2015}. Hence, the nonlinearities associated to financial frictions found in the US case were probably less important and nonlinear methods did not have a 
significant effect on predicting real activity series on average. However, NL treatment is very important for inflation and housing. Shrinkage is still not a good idea for industrial production 
and unemployment rate,  but can be very helpful other variables at some specific horizons. Cross-validation does not have a big impact and the SVR loss function 
is still harmful.

\begin{figure}
\centering
\includegraphics[scale=.75]{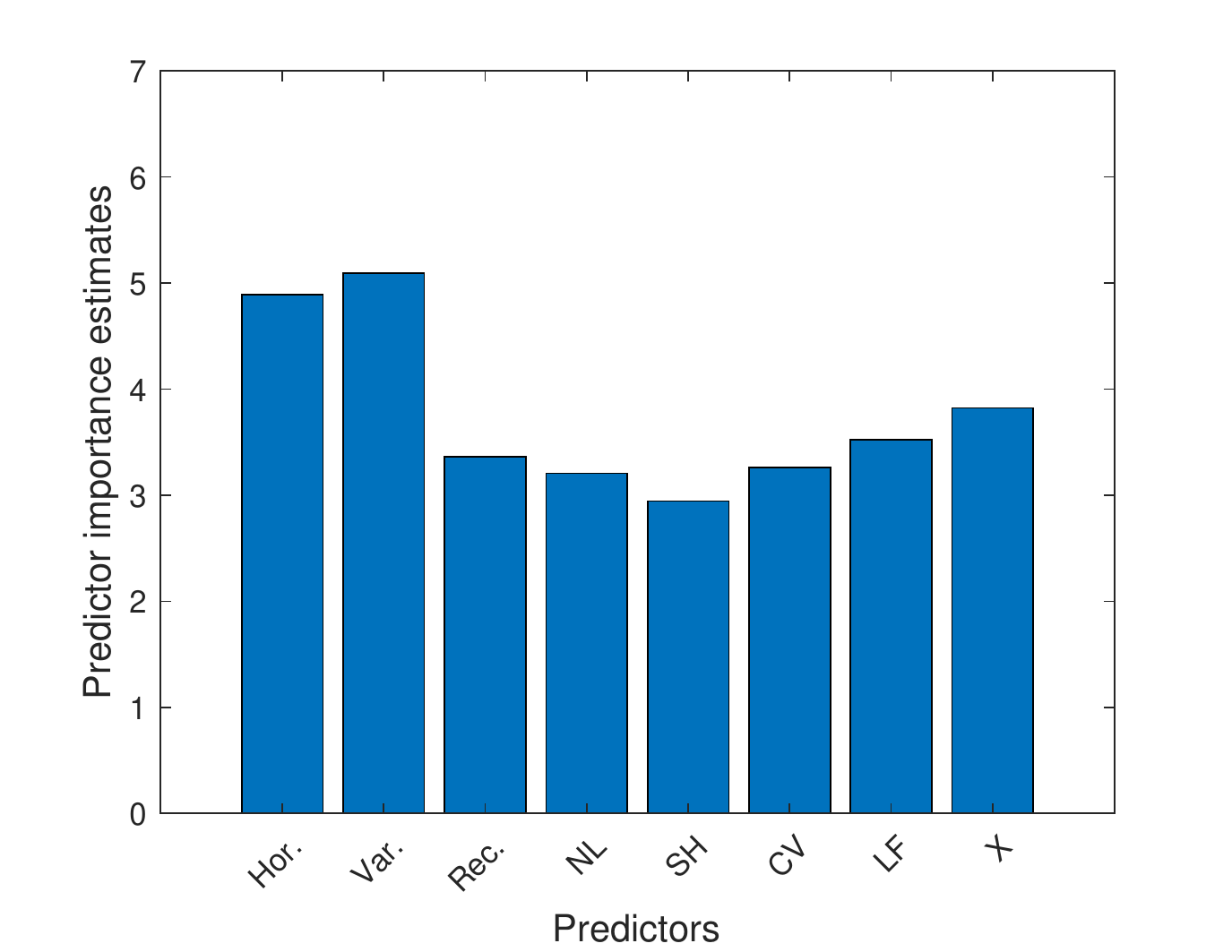}
	\caption{\footnotesize{This figure presents predictive importance estimates. Random forest is trained to predict $R^2_{t,h,v,m}$ defined in (\ref{r2_eq}) and 
	use out-of-bags observations to assess the performance of the model and compute features' importance. NL, SH, CV and LF stand for nonlinearity, shrinkage, 
	cross-validation and loss function features respectively. A dummy for $H_t^+$ models, $X$, is included as well.}}
	\label{Tree_can}
\end{figure}

\begin{figure}
\centering
\includegraphics[scale=.4, trim= 0mm 8mm 0mm 8mm, clip]{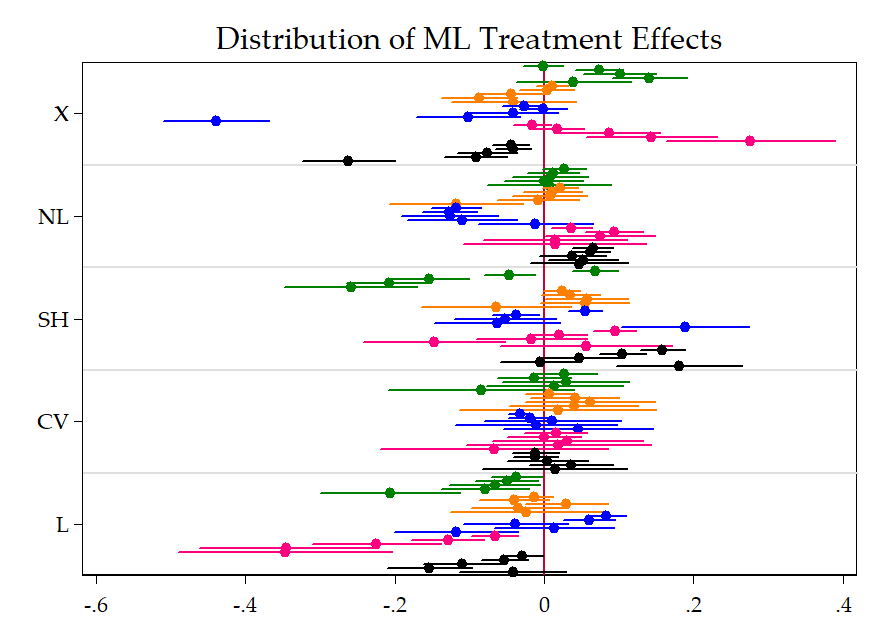}
\vspace{-0cm}
	\caption{\footnotesize{This figure plots the distribution of $\dot{\alpha}_F^{(h,v)}$ from equation (\ref{r2_eq}) done by $(h,v)$ subsets. That is, we are looking at the average partial effect on the pseudo-OOS $R^2$ from augmenting the model with ML features, keeping everything else fixed. $X$ is making the switch from data-poor to data-rich. Finally, variables are \textcolor{Green}{INDPRO}, \textcolor{orange}{UNRATE}, \textcolor{blue}{SPREAD}, \textcolor{magenta}{INF} and HOUS. Within a specific color block, the horizon increases from $h=1$ to $h=24$ as we are going down.  SEs are HAC. These are the 95\% confidence bands.}} 
	\label{dist_all_can}
\end{figure}

\clearpage

\section{Detailed Implementation of Cross-validations}\label{CVdetails}

All of our models involve some kind of hyperparameter selection prior to estimation. To curb the overfitting problem, we use two distinct methods that we refer to loosely as cross-validation methods. To make it feasible, we optimize hyperparameters every 24 months as the expanding window grows our in-sample set. The resulting optimization points are the same across all models, variables and horizons considered. In all other periods, hyperparameter values are frozen to the previous values and models are estimated using the expanded in-sample set to generate forecasts.

\begin{center}
	\begin{figure}[H]
		\begin{subfigure}{3.5in}
			\caption*{POOS}
			\includegraphics[width=3.2in]{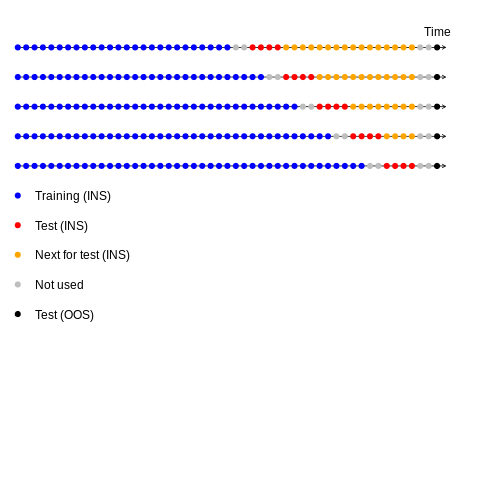}
		\end{subfigure}%
		\begin{subfigure}{3.5in}
			\caption*{K folds}
			\includegraphics[width=3.2in]{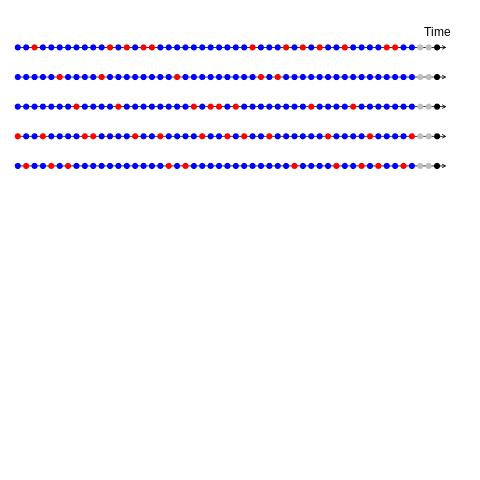}
		\end{subfigure}
	\vspace{-5em}
	\caption{Illustration of cross-validation methods} \label{CV_methods}
	\end{figure}
\end{center}
\vspace{-1.5cm}
\begin{footnotesize}
	\flushleft
	Notes: Figures are drawn for 3 months forecasting horizon and depict the splits performed in the in-sample set. The pseudo-out-of-sample observation to be forecasted here is shown in black.
\end{footnotesize} \\

The first cross-validation method we consider mimics in-sample the pseudo-out-of-sample comparison we perform across models. For each set of hyperparameters considered, we keep the last 25\% of the in-sample set as a comparison window. Models are estimated every 12 months, but the training set is gradually expanded to keep the forecasting horizon intact. This exercise is thus repeated 5 times. Figure \ref{CV_methods} shows a toy example with smaller jumps, a smaller comparison window and a forecasting horizon of 3 months, hence the gaps. Once hyperparameters have been selected, the model is estimated using the whole in-sample set and used to make a forecast in the pseudo-out-of-sample window that we use to compare all models (the black dot in the figure). This approach is a compromise between two methods used to evaluate time series models detailed in \cite{Tashman2000}, rolling-origin recalibration and rolling-origin updating.\footnote{In both cases, the last observation (the origin of the forecast) of the training set is rolled forward. However, in the first case, hyperparameters are recalibrated and, in the second, only the information set is updated.} For a simulation study of various cross-validation methods in a time series context, including the rolling-origin recalibration method, the reader is referred to \cite{Bergmeir2012}. We stress again that the compromise is made to bring down computation time. 

The second cross-validation method, K-fold cross-validation, is based on a re-sampling scheme \citep{Bergmeir2018}. We chose to use 5 folds, meaning the in-sample set is randomly split into five disjoint subsets, each accounting on average for 20 \% of the in-sample observations. For each one of the 5 subsets and each set of hyperparameters considered, 4 subsets are used for estimation and the remaining corresponding observations of the in-sample set used as a test subset to generate forecasting errors. This is illustrated in figure \ref{CV_methods} where each subset is illustrated by red dots on different arrows. 

Note that the average mean squared error in the test subset is used as the performance metric for both cross-validation methods to perform hyperparameter selection.

\section{Forecasting models in detail}\label{sec:models}

\subsection{Data-poor ($H_t^-$) models}
	
	In this section we describe forecasting models that contain only lagged values
	of the dependent variable, and hence use a small amount of predictors, $H^{-}_t$.
	
	\paragraph{Autoregressive Direct (AR)}
	
	The first univariate model is the so-called \textit{autoregressive direct}
	(AR)\ model, which is specified as:%
	\begin{equation*}
	y_{t+h}^{(h)}= c + \rho(L)y_{t}+e_{t+h},\quad t=1,\ldots ,T, 
	\end{equation*}%
	where $h\geq 1$ is the forecasting horizon. The only hyperparameter in
	this model is $p_y$, the order of the lag polynomial  $\rho(L)$.
	The optimal $p$ is selected in four ways: (i) Bayesian Information Criterion (AR,BIC);
	(ii) Akaike Information Criterion (AR,AIC); (iii) Pseudo-out-of-sample cross validation (AR,POOS-CV);
	and (iv) K-fold cross validation (AR,K-fold). The lag order is selected from the following subset 
	$p_y \in \{1,3,6,12\}$.
	Hence, this model enters the following categories: linear $g$ function,
	no regularization, in-sample and cross-validation selection of hyperparameters and
	quadratic loss function.
	
	\paragraph{Ridge Regression AR (RRAR)}
	
	The second specification is a penalized version of the previous AR model that allows
	potentially more lagged predictors by using Ridge regression. The model is written as in (\ref{ard}), 
	and the parameters are estimated using Ridge penalty. The Ridge hyperparameter is
	selected with two cross validation strategies, which gives two models: RRAR,POOS-CV and
	RRAR,K-fold. The lag order is selected from the following subset 
	$p_y \in \{1,3,6,12\}$ and
	for each of these value we choose the Ridge hyperparameter. This model creates variation on following axes: linear $g$,
	Ridge regularization,
	cross-validation for tuning parameters and quadratic loss function.
	
	\paragraph{Random Forests AR (RFAR)}
	
	A popular way to introduce nonlinearities in the predictive function $g$ is to use a tree method
	that splits the predictors space in a collection of dummy variables and their interactions. Since
	a standard tree regression is prompt to the overfit, we use instead the random forest approach
	described in Section \ref{RF}. We adopt the default value in the literature of one third for 'mtry', the share of randomly selected predictors that are candidates for splits in each tree. Observations in each set are sampled with replacement to get
	as many observations in the trees as in the full sample.
	The number of lags of 
	$y_t$, is chosen from the subset $p_y \in \{1,3,6,12\}$ with cross-validation while the number of trees is 
	selected internally with out-of-bag observations. This model generates nonlinear
	approximation of the optimal forecast, without regularization, using both CV techniques with the
	quadratic loss function: RFAR,K-fold and RFAR,POOS-CV.
	
	\paragraph{Kernel Ridge Regression AR (KRRAR)}
	
	This specification adds a nonlinear approximation of the function $g$ by using the Kernel trick
	as in Section \ref{KTregression}. The model is written as in (\ref{KT1}) and (\ref{KT2}) but with the
	autoregressive part only
	\begin{align*}
	y_{t+h} &= c + g(Z_t) + \varepsilon_{t+h}, \\
	Z_t &= \left[{\lbrace y_{t-0} \rbrace}_{j=0}^{p_y} \right],
	\end{align*}
	and the forecast is obtained using the equation (\ref{KT4}). The hyperparameters of Ridge and of its kernel are selected by two cross-validation procedures, which gives two forecasting
	specifications: (i) KRRAR,POOS-CV, (ii) KRRAR,K-fold. $Z_t$ consists of $y_t$ and its $p_y$ lags, 
	$p_y \in \{1,3,6,12\}$.
	This model is representative of a nonlinear
	$g$ function, Ridge regularization, cross-validation to select $\tau$ and quadratic $\hat L$.

	\paragraph{Support Vector Regression AR (SVR-AR)}
	
	We use the SVR model to create variation along the loss function dimension. In the data-poor
	version the predictors set $Z_t$ contains $y_t$ and a number of lags chosen from $p_y \in \{1,3,6,12\}$. The hyperparameters are
	selected with both cross-validation techniques, and we consider 2 kernels to approximate
	basis functions, linear and RBF. Hence, there are 4 versions: (i) SVR-AR,Lin,POOS-CV, (ii) SVR-AR,Lin,K-fold, (iii) SVR-AR,RBF,POOS-CV and (iv) SVR-AR,RBF,K-fold. The forecasts are generated
	using (\ref{svr1}).
	
	\subsection{Data-rich ($H_t^+$) models}
	
	We now describe forecasting models that use a large dataset of
	predictors, including the autoregressive components, $H^{+}_t$.
	
	\paragraph{Diffusion Indices (ARDI)}
	
	The reference model in the case of large predictor set is the autoregression augmented with diffusion
	indices from \cite{Stock2002}:
	\begin{eqnarray}
	y_{t+h}^{(h)} &=& c+ \rho(L)y_{t} + \beta(L)F_{t} +e_{t+h},\quad t=1,\ldots ,T   \\
	X_{t} &=&\Lambda F_{t}+u_{t} 
	\end{eqnarray}%
	where $F_{t}$ are $K$ \emph{consecutive} \emph{static} factors, and $\rho(L)$ and $\beta(L)$
	are lag polynomials of orders $p_y$ and $p_f$ respectively. The feasible procedure requires an
	estimate of $F_t$ that is usually done by PCA.
	The optimal values of hyperparamters $p$, $K$ and $m$ are selected in four ways:
	(i) Bayesian Information Criterion (ARDI,BIC);
	(ii) Akaike Information Criterion (ARDI,AIC); (iii) Pseudo-out-of-sample cross validation (ARDI,POOS-CV);
	and (iv) K-fold cross validation (ARDI,K-fold). These are selected from following subsets: 
	$p_y \in \{1,3,6,12\}$, $K \in \{3,6,10 \}$, $p_f \in \{1,3,6,12\}$.
	Hence, this model following features: linear $g$ function,
	PCA regularization, in-sample and cross-validation selection of hyperparameters and $L^2$.
	
	\paragraph{Ridge Regression Diffusion Indices (RRARDI)}
	As for the small data case, we explore how a regularization  affects the predictive performance of the
	reference model ARDI above.  The predictive regression is written as in (\ref{swardi1}) and
	$p_y$, $p_f$ and $K$ are selected from the same subsets of 
	values as for the ARDI case above. The parameters  are estimated
	using Ridge penalty. All the hyperparameters are
	selected with two cross validation strategies, giving two models: RRARDI,POOS-CV and
	RRARDI,K-fold. This model creates variation on following axes: linear $g$, Ridge regularization,
	CV for tuning parameters and $L^2$.
	
	\paragraph{Random Forest Diffusion Indices (RFARDI)}
	We also explore how nonlinearities affect the predictive performance of the ARDI model. The model is as in (\ref{swardi1})
	but a Random Forest of regression trees is used.  
	The ARDI hyperparameters  are chosen from the grid as in the linear case, while the number of trees is selected with out-of-bag observations. Both 
	POOS and K-fold CV are used to generate two forecasting models: RFARDI,POOS-CV and RFARDI,K-fold. This model 
	generates nonlinear treatment, with PCA regularization, using both CV techniques with the quadratic loss function.
	
	\paragraph{Kernel Ridge Regression Diffusion Indices (KRRARDI)}
	
	As for the autoregressive case, we can use the KT to generate nonlinear predictive functions $g$.
	The model is represented by equations (\ref{KT1}) - (\ref{KT3}) and the forecast is obtained using the
	equation (\ref{KT4}). The hyperparameters of Ridge and of its kernel, as well as $p_y$, $K$ and $p_f$ are selected by two cross-validation procedures, 
	which gives two forecasting specifications: (i) KRRARDI,POOS-CV, (ii) KRRARDI,K-fold. We use the same grid as in ARDI case for discrete hyperparameters. This model is representative of a nonlinear $g$ function, Ridge regularization with PCA,
	cross-validation to select $\tau$ and quadratic $\hat L$.

	\paragraph{Support Vector Regression ARDI (SVR-ARDI)}
	
	We use four versions of the SVR model: (i) SVR-ARDI,Lin,POOS-CV, (ii) SVR-ARDI,Lin,K-fold, (iii) SVR-ARDI,RBF,POOS-CV and (iv) SVR-ARDI,RBF,K-fold. The SVR hyperparameters are chosen by cross-validation and the ARDI hyperparameters are chosen using a grid that search in the same subsets 
	as the ARDI model. The forecasts are generated from equation (\ref{svr1}). This model creates variations in all categories: 
	nonlinear $g$, PCA regularization,
	CV and $\bar{\epsilon}$-insensitive loss function.
	
	\subsubsection{Generating shrinkage schemes}
	
	The rest of the forecasting models relies on using different $B$ operators to generate variations across shrinkage
	schemes, as depicted in section \ref{sec:reg}.
	
	\paragraph{$B_1$: taking all observables $H_t^+$}
	
	When $B$ is identity mapping, we consider $Z_t = H_t^+$ in the Elastic Net problem (\ref{en1}), where $H_t^+$ is 
	defined by (\ref{Hs}). The following 
	lag structures for $y_t$ and $X_t$ are considered, $p_y \in \{1, 3, 6, 12 \}$ $p_f \in \{1, 3, 6, 12 \}$, and the exact number is cross-validated. 
	The hyperparameter $\lambda$ is
	always selected by two cross validation procedures, while we consider three cases for $\alpha$: $\hat \alpha$, $\alpha=1$ and
	$\alpha=0$, which correspond to EN, Ridge and Lasso specifications respectively. In case of EN, $\alpha$ is
	also cross-validated. This gives six combinations:  ($B_1,\alpha=\hat{\alpha}$),POOS-CV; ($B_1,\alpha=\hat{\alpha}$),K-fold;
	($B_1,\alpha=1$),POOS-CV; ($B_1,\alpha=1$),K-fold; ($B_1,\alpha=0$),POOS-CV and ($B_1,\alpha=0$),K-fold. They create variations
	within regularization and hyperparameters' optimization.
	
	\paragraph{$B_2$: taking all principal components of $X_t$}
	
	Here $B_2()$ rotates $X_t$ into $N$ factors, $F_t$, estimated by principal components, which then constitute $Z_t$ to be used in (\ref{en1}). 
	Same lag structures and hyperparameters' 
	optimization from the $B_1$ case are used to generate the following six specifications:  ($B_2,\alpha=\hat{\alpha}$),POOS-CV; ($B_2,\alpha=\hat{\alpha}$),K-fold;
	($B_2,\alpha=1$),POOS-CV; ($B_2,\alpha=1$),K-fold; ($B_2,\alpha=0$),POOS-CV and ($B_2,\alpha=0$),K-fold.
	
	\paragraph{$B_3$: taking all principal components of $H_t^+$}
	
	Finally, $B_3()$ rotates $H_t^+$ by taking all principal components, where $H_t^+$ lag structure is to be selected as 
	in the $B_1$ case. Same variations and hyperparameters' selection are used to generate
	the following six specifications:  ($B_3,\alpha=\hat{\alpha}$),POOS-CV; ($B_3,\alpha=\hat{\alpha}$),K-fold;
	($B_3,\alpha=1$),POOS-CV; ($B_3,\alpha=1$),K-fold; ($B_3,\alpha=0$),POOS-CV and ($B_3,\alpha=0$),K-fold.

\end{document}